\documentclass{aa}  
\usepackage[varg]{txfonts}

\usepackage{bm} 

\usepackage[T1]{fontenc}
\usepackage{ae,aecompl}
\usepackage{graphicx}   
\usepackage{amsmath}    
\usepackage{amssymb}    

\usepackage[x11names]{xcolor}

\usepackage{amsmath}	
\usepackage{mathtools}
\usepackage{amssymb}	
\usepackage{natbib}
\bibpunct{(}{)}{;}{a}{}{,} 

\usepackage{float}

\usepackage[draft]{hyperref}

\hypersetup{
   colorlinks=true,
   linkcolor=blue,
   citecolor=blue,
   urlcolor=blue,
}

\usepackage{enumerate}
\usepackage{enumitem}

\newcommand{\itemfont}{
  \usefont{T1}{phv}{m}{it}\fontsize{9pt}{9pt}\selectfont}

\newcommand{\mstar}{$M_{\rm star}$}

\newcommand{\mgas}{$M_{\rm gas}$}
\newcommand{\fgas}{$f_{\rm gas}$}
\newcommand{\fhtwo}{$f_{\rm H2}$}
\newcommand{\fhi}{$f_{\rm HI}$}
\newcommand{\logoh}{12$+$log(O/H)}
\newcommand{\ohsol}{(O/H)$_\odot$}
\newcommand{\lco}{$L^\prime_{\rm CO}$}
\newcommand{\mhi}{$M_{\rm HI}$}
\newcommand{\mhtwo}{$M_{\rm H2}$}
\newcommand{\aco}{$\alpha_{\rm CO}$}
\newcommand{\acosun}{$\alpha_{\rm CO}^{\odot}$}

\newcommand{\lcounits}{${\rm K\,km\,s^{-1}\,pc^{2}}$}

\newcommand{\nuvw}{NUV$-$W1}
\newcommand{\msunpc}{$M_\odot\,{\mathrm pc}^{-2}$}
\newcommand{\rhi}{$R_\mathrm{HI}$}
\newcommand{\rwise}{$R_\mathrm{W1}$}

\newcommand{\micron}{$\mu$m}
\newcommand{\zzsun}{$Z/Z_\odot$}
\newcommand{\zsun}{$Z_\odot$}
\newcommand{\msun}{M$_\odot$}

\newcommand{\msunyr}{M$_\odot$\,yr$^{-1}$}
\newcommand{\htwo}{H$_2$}
\newcommand{\hi}{H{\sc i}}

\newcommand{\sfegas}{SFE$_{\rm gas}$}
\newcommand{\sfehtwo}{SFE$_{\rm H2}$}
\newcommand{\sfehi}{SFE$_{\rm HI}$}
\newcommand{\sfedmh}{SFE$_{\rm DMH}$}
\newcommand{\taudep}{$\tau_{\rm H2}$}
\newcommand{\taudephi}{$\tau_{\rm HI}$}

\newcommand{\hii}{H{\sc ii}}

\newcommand{\av}{$A_{\rm V}$}

\newcommand{\nii}{[N{\sc ii}]}
\newcommand{\te}{T$_{\rm e}$}

\newcommand{\lfir}{L$_{\rm FIR}$}

\newcommand{\fuvnuv}{FUV$-$NUV}

\newcommand{\hers}{{\it Herschel}}

\newcommand{\pone}{Paper\,I}

\begin{document}

\title{Scaling~relations~and~baryonic~cycling~in~local~star-forming~galaxies: II. Gas content and star-formation efficiency}
\titlerunning{Scaling~relations~and~baryonic~cycling~in~local~star-forming~galaxies: II. Gas content and SFE}

\author{L. K. Hunt\inst{\ref{inst1}}
	\and
	C. Tortora\inst{\ref{inst1},\ref{inst2}}
	\and
	M. Ginolfi\inst{\ref{inst3}}
	\and
	R. Schneider\inst{\ref{inst4}}
}

\institute{INAF -- Osservatorio Astrofisico di Arcetri, Largo Enrico Fermi 5, I-50125 Firenze, Italy
	\email{leslie.hunt@inaf.it}\label{inst1}
	\and
{INAF -- Osservatorio Astronomico di Capodimonte, Salita Moiariello, 16, I-80131 Napoli, Italy
\label{inst2}}
\and
{Observatoire de Gen\`eve, Universit\`e de Gen\`eve, 51 Ch. des Maillettes, 1290 Versoix, Switzerland \label{inst3}}
\and
{Universit\`a\ degli Studi di Roma ``La Sapienza", Roma, Italy \label{inst4}}
}

\date{Received XXX; accepted YYY}

\abstract{Assessments of the cold-gas reservoir in galaxies are a cornerstone for understanding star-formation processes 
and the role of feedback and baryonic cycling in galaxy evolution.
Here we exploit a sample of 392 galaxies (dubbed MAGMA, Metallicity and Gas for Mass Assembly), presented
in a recent paper, to quantify molecular and atomic gas properties across a broad range in stellar mass, \mstar, 
from $\sim\,10^7 - 10^{11}$\,\msun.
First, we find the metallicity ($Z$) dependence of \aco\ to be shallower than previous estimates, with
\aco$\,\propto\,(Z/Z_\odot)^{-1.55}$.
Second, molecular gas mass \mhtwo\ is found to be strongly correlated with \mstar\ and star-formation rate (SFR), enabling predictions of
\mhtwo\ good to within $\sim$0.2\,dex; analogous relations for atomic gas mass \mhi\ and total gas mass \mgas\ are
less accurate, $\sim$0.4\,dex and $\sim$0.3\,dex, respectively.
Indeed, the behavior of atomic gas mass \mhi\ in MAGMA scaling relations
suggests that it may be a third, independent variable that encapsulates information about the circumgalactic
environment and 
gas accretion.
If \mgas\ is considered to depend on \mhi, together with \mstar\ and SFR, we obtain a relation that predicts
\mgas\ to within $\sim$0.05\,dex.
Finally, the analysis of depletion times and the scaling of \mhi/\mstar\ and \mhtwo/\mstar\ over three different
mass bins suggests that the partition of gas 
and the regulation of star formation through gas content depends on the mass regime.
Dwarf galaxies
(\mstar $\,\lesssim\,3\times 10^{9}$\,\msun) tend to be overwhelmed by (\hi) accretion, and 
despite short \taudep\ (and thus presumably high star-formation efficiency),
star formation is unable to keep up with the gas supply.
For galaxies in the intermediate \mstar\ ``gas-equilibrium'' bin ($3\times10^9$\,\msun\,$\lesssim$ \mstar\,$\lesssim 3\times10^{10}$\,\msun), 
star formation proceeds apace with gas availability, and \hi\ and \htwo\ are both proportional to SFR.
In the most massive ``gas-poor, bimodality'' regime (\mstar\,$\gtrsim\,3\times10^{10}$\,\msun), 
\hi\ does not apparently participate in star formation, although it generally dominates in mass over \htwo.
Our results confirm that atomic gas plays a key role in baryonic cycling, and 
is a fundamental ingredient for current and future star formation, especially in dwarf galaxies.
}

\maketitle

\section{Introduction}
\label{sec:introduction}

Galaxy evolution is inevitably regulated by the 
gas reservoir available for star formation.
In its molecular form, this gas fuels star formation that, during stellar evolution,
enriches the interstellar medium (ISM) in a galaxy with metals and dust.
Over the course of a galaxy's lifetime, stars exploding as supernovae (SNe)
impart mechanical and radiative energy to the surrounding environment, and sow the seeds for dust
grain formation in SNe ejecta and the circumnuclear envelopes of evolved stars.
Metals produced by massive stars and other stellar events enrich the surrounding ISM,
changing its properties and enhancing its ability to cool.
Gas may also be expelled from the galaxy, becoming part of 
the circumgalactic medium (CGM) surrounding the galaxy, where it may be reaccreted \citep[see recent review by][]{tumlinson17}. 
Baryonic cycling in the galaxy eco-system comprises this combination of
gas accretion, the conversion of gas into stars, the injection of energy, and the ISM enrichment by metals
and dust grains. 
The manifestation of baryonic cycling can be found in scaling relations among star-formation rate (SFR) and stellar, gas, and metal content 
which are becoming powerful diagnostics for understanding how these processes 
depend on their mass and environment.

Two key concepts in baryonic cycling are the overall efficiency with which stars are ultimately produced
from the initial conditions in the host dark-matter halo (DMH) and the efficiency with which
stars are formed from the available gas supply.
Unlike more massive galaxies, such as the Milky Way (MW), 
dwarf galaxies seem to be highly inefficient at producing stars
\citep[e.g.,][]{behroozi13,moster13,graziani15,graziani17}.
This is a consequence of the relation between parent dark-matter halos (DMHs) and stars, 
and is thought to be regulated by baryonic physics, namely the energy released into the ISM
and intergalactic medium (IGM) by young stellar populations or active galactic nuclei (AGN)
in the more massive galaxies.

These feedback processes are arguably the most important drivers of galaxy evolution. 
Many fundamental relations are thought to be manifestations of feedback, resulting
in self-regulated balances between gas cooling and AGN-(black-hole) driven and supernova-driven outflows.
These range from  
galaxy stellar mass functions \citep[e.g.,][]{baldry12,ilbert13},
the mass-metallicity relation \citep[MZR, e.g.,][]{tremonti04}, 
gas scaling with SFR \citep[e.g.,][]{hopkins14},
to the baryonic Tully-Fisher relation \citep[e.g.,][]{mcgaugh00,mcgaugh12}, 
and the cosmic star-formation history of the Universe \citep[e.g.,][]{madau14}.

Dwarf galaxies\footnote{With stellar masses \mstar\ $\la 10^9$\,\msun.}
are particularly susceptible to feedback, because winds from massive stars and SNe can efficiently eject gas 
and metals from a shallow potential well, thus suppressing star formation \citep[e.g.,][]{gnedin09,gnedin10,graziani20}. 
The low star-formation efficiency (SFE) for dwarfs, predicted by the stellar-DMH relation, 
contrasts with observational evidence in which
some local metal-poor star-forming dwarf galaxies show short molecular depletion times,
a signature of efficient star formation \citep{hunt15,amorin16}. 
However, the low SFE for halos [\sfedmh$\,=\,M_{*}/(f_B\,M_{\rm halo})$] is relative to the
cosmic baryon fraction $f_B$; 
the low SFE for SFR is the ``observer's definition'' relative to the gas, where \sfegas\
($=\,{\rm SFR}/M_{\rm gas}$) and \sfehtwo\ ($=\,{\rm SFR}/M_{\rm H2}$)
are the inverses of gas depletion times.

The stochastic star-formation histories (SFHs) common in dwarf galaxies \citep[e.g.,][]{tolstoy09,mcquinn10,weisz11,gallart15} may act over long time
scales to produce a net low \sfedmh, as expected theoretically from the DMH relation.
Short depletion times (high \sfehtwo) in dwarf galaxies could be a ``smoking gun'' for 
a typical, bursty SFH \citep[e.g.,][]{madau14reversal}.
On the other hand, they could be a signature of galaxies with current higher \sfedmh\ such
as those found recently for galaxies in the Epoch of Reionization by the ``Universe Machine'' \citep{behroozi19}.

From an observational point of view, feedback is not straightforward to quantify and many clues are indirect.
The inefficiency of low-mass halos \sfedmh\ to form stars depends on at least three processes: 
\textit{(P1)}~preventive feedback, the (lack of) availability of cold baryons from the host halo; 
\textit{(P2)}~inefficiency of the star-formation process, the conversion of the available gas into stars; and
\textit{(P3)}~ejective feedback, the outright expulsion of energy, momentum, and material including gas, dust, and metals.

In this paper, we focus on two of these mechanisms: the 
availability of cold gas to form stars \textit{(P1)}, and the inefficiency of the 
star-formation process itself \textit{(P2)}.
In a companion paper (Tortora et al. 2020, in prep.), we examine the remaining
mechanism, \textit{(P3)}, namely the efficiency of metal enrichment and ejective feedback. 
To perform our analysis, we need a sample of galaxies, with homogeneously-determined parameters,
that contains a significant quantity of dwarf galaxies. 
In \citet[][hereafter \pone]{ginolfi20},
we have constructed such a sample, Metallicity And Gas for Mass Assembly (MAGMA), 
as described in Sect. \ref{sec:magma}. 

The rest of the paper is organized as follows:
we first reexamine the conversion of CO luminosity \lco\
to molecular gas mass \mhtwo\ in Sect. \ref{sec:alphaco}.
This new formulation of \aco\ is used in Sect. \ref{sec:gasscaling} to
establish scaling relations for the total gas content, and the separate molecular and atomic components.
In Sect. \ref{sec:gasphases}, we assess the interplay between gas and star formation, focusing in particular
on the efficiency of star formation with \htwo\ (Sect. \ref{sec:sfe}); 
\hi\ properties (Sect. \ref{sec:hi}); 
and the link between molecular and atomic gas (Sect. \ref{sec:h2hi}).
We discuss the ramifications of our results in Sect. \ref{sec:discussion}, and give our conclusions in Sect. \ref{sec:conclusions}.
Throughout the paper we assume that solar metallicity \ohsol\ is given by
\logoh\,=\,8.69 \citep{Asplund2009}.

\section{The MAGMA sample}
\label{sec:magma}

We have recently compiled a sample of 392 galaxies (dubbed MAGMA) with uniformly-derived stellar masses,
\mstar, SFRs, metallicity [\logoh], atomic gas mass (\mhi), and molecular gas mass (\mhtwo);
more than 30\% of the MAGMA galaxies are dwarfs with \mstar\,$\la\,3\times10^9$\,\msun\
(\pone).
MAGMA comprises more than 9 separate studies in the literature with measurements of \hi\ mass
and CO luminosity, \lco; 
these include xCOLDGASS \citep{Saintonge2017,Catinella2018};
ALLSMOG \citep[APEX Low-redshift Legacy Survey of MOlecular Gas:][]{Cicone2017};
HRS \citep[\hers\ Reference Sample:][]{Boselli2010,Boselli2014a,Boselli2015};
KINGFISH \citep[Key Insights into Nearby Galaxies, a Far-Infrared Survey with \hers:][]{Kennicutt2011,Aniano2020};
NFGS \citep[Nearby Field Galaxy Survey:][]{Kannappan2013};
Virgo star-forming dwarf galaxies \citep{Grossi2016};
BCDs (Blue Compact Dwarfs) from \citet[][]{hunt15} and Hunt et al. (2020, in preparation);
DGS \citep[Dwarf Galaxy Survey:][]{Cormier2014};
and other single objects from various papers (see \pone\ for more details).
The main requirement
for MAGMA was the existence of both CO and \hi\ detections, together with spectroscopic measurements of O/H.
As described in \pone, galaxies with spectroscopic evidence for active galactic nuclei (AGN) or a composite
nature were eliminated from the sample.
In addition, when possible (for HRS and KINGFISH), galaxies showing an \hi\ deficiency were not included.

Thus, MAGMA is a representative sample of star-forming, isolated galaxies in the Local Universe.
MAGMA contains fewer low-mass galaxies than a purely mass-limited sample, because
of the difficulty in obtaining CO detections and because of the selection criteria in some of the parent surveys
(e.g., xCOLDGASS, HRS).
It is rather representative of the location of galaxies along the local star-formation main sequence (SFMS) over a vast range in stellar mass.

\begin{figure*}[!ht]
\centering
\includegraphics[width=0.85\linewidth]{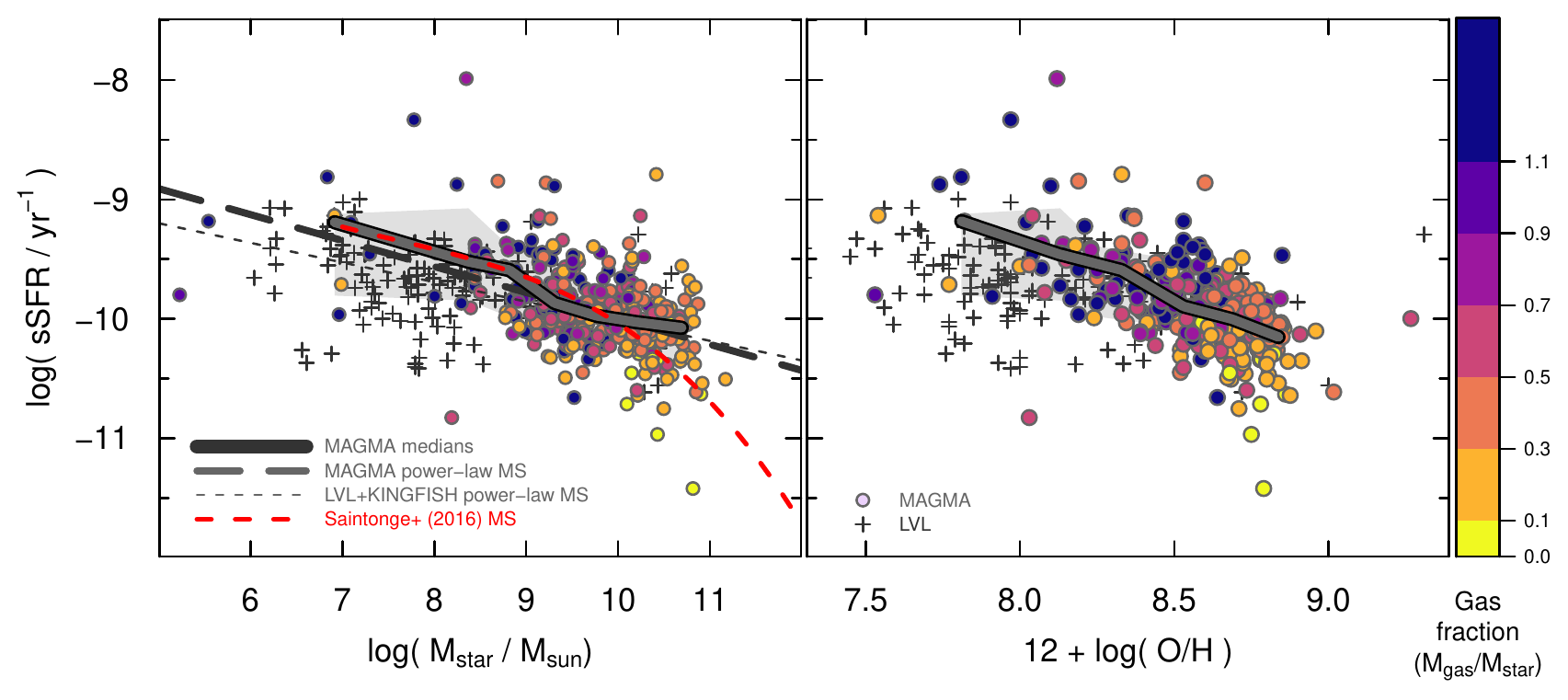}
\caption{sSFR plotted against \mstar\ in logarithmic units (left panel) and \logoh\ (right) for the MAGMA sample. 
The heavy dark-gray curves correspond to the medians of the binned MAGMA data, and the underlying gray region to
a standard deviation in either direction.
The dark-gray plus signs report galaxies from the LVL sample as described in the text.
The regression shown by the long-dashed line indicates the best (power-law) fit for the SFMS to MAGMA,
as given by Eqn. (\ref{eqn:ms}),
and the short-dashed line to fit of LVL$+$KINGFISH galaxies as described in the text.
The heavy dashed (red) curve corresponds to the [cubic in log(\mstar)] SFMS found by \citet{Saintonge2016}.
Galaxies are color-coded by gas fraction (\fgas, here defined as \mgas/\mstar) as indicated in the color wedge in the right-most panel.
}
	\label{fig:magma_ms}
\end{figure*}

Stellar masses and SFRs were uniformly calculated for the entire MAGMA sample based on 
photometry from the Wide-field Infrared Survey Explorer \citep[WISE,][]{Wright2010} and 
the Galaxy Evolution Explorer \citep[GALEX,][]{Morrissey2007}.
For \mstar, we used the formulation by \citet{Hunt2019} given for the CIGALE calibration \citep[see also][]{Boquien2019}.
This formulation is based on WISE W1 (3.4\,\micron) or IRAC (3.6\,\micron) luminosities,
after correcting for free-free nebular continuum emission associated with ionized gas.
SFR was calculated according to the precepts of \citet{Leroy2019}, based on WISE (W3, W4) and GALEX 
photometry. 
These are calibrated on the GALEX-SDSS-WISE Legacy Catalog(GSWLC) presented by \citet{Salim2016,Salim2018},
that relies on CIGALE fits of $\sim$700\,000 low-redshift galaxies.

Both \mstar\ and SFR are based on the underlying CIGALE calibration, and on the
Initial Mass Function (IMF) by \citet{Chabrier2003}.
In addition to the IMF, any \mstar\ and SFR calibration also depends on the underlying SFH.
\citet{Salim2018} fit the SFH to an old stellar population, added to an exponentially decaying SFH, and a young burst with nearly constant SFR.
The CIGALE calibration presented by \citet{Hunt2019} is based on a ``delayed'' parameterization SFH, 
coupled with an additional star-formation episode that could correspond to an old population or a younger one.
Such a functional form gives a nearly linear increase of the SFR from the onset of star formation rather than an abrupt one 
\citep[see][for more details]{Boquien2019}.
The main ingredients, namely an exponential (or approximately) declining SFH coupled with an additional stellar population, is the same in both calibrations, 
and should be similar enough to ensure consistency in our estimates of \mstar\ and SFR.

Figure \ref{fig:magma_ms} shows for the MAGMA sample the relation of specific SFR (sSFR\,$\equiv$\,SFR/\mstar)
and \mstar, where we have framed the SFMS 
in terms of sSFR rather than SFR.
The underlying SFMS is shown as a dashed line:
{\small
\begin{equation}
\log(\mathrm{sSFR/yr^{-1}})\,=\,(-0.22\,\pm\,0.02)\,\log(M_\mathrm{star}/\mathrm{M}_\odot) - (7.82\,\pm\,0.17)\ .
\label{eqn:ms}
\end{equation}
}
\noindent
This SFMS corresponds to the robust least-squares power-law best fit\footnote{For all statistical analysis, we rely on the
{\it R} statistical package: 
R Core Team (2018), R: A language and environment for statistical
  computing, R Foundation for Statistical Computing, Vienna, Austria
  (\url{https://www.R-project.org/}).},
for the MAGMA galaxies.
Also shown in Fig. \ref{fig:magma_ms} is the cubic form for the SFMS found by \citet{Saintonge2016};
the formulation pertains formally only to galaxies with \mstar$\ga\,10^{8}$\,\msun.
In any case, it corresponds perfectly down to \mstar\,$\la 10^7$\,\msun\ in MAGMA galaxies,
and also traces the downward curvature beyond \mstar\,$\ga 3\times10^{10}$\,\msun.
Also shown in Fig. \ref{fig:magma_ms}, is
the fit to galaxies from the Local Volume Legacy \citep[LVL, see][]{Dale2009} 
and KINGFISH taken from \citet{Hunt2019}, with the parameters computed in the same way as for the MAGMA galaxies.
It has a slightly shallower power-law slope, $\sim -0.16\,\pm\,0.02$.
The power-law slope for MAGMA is consistent
with earlier estimates of the local SFMS \citep[e.g.,][]{elbaz07} with slopes of $\sim -0.23$. 
Finally, the MAGMA medians are virtually identical to the curved SFMS found by \citet{Saintonge2016}, 
even at low \mstar.
All this suggests that overall, the MAGMA galaxies are fundamentally a main-sequence sample; 
there are very few galaxies having deviations larger than $\sim 1\sigma$ from the SFMS trend,
except possibly at low \mstar\ (see \pone).

Metallicities in the form of nebular oxygen abundances, O/H,
were taken from the original works in the (linear) \nii\ calibration by \citet[][hereafter PP04N2]{Pettini2004},
or according to the ``direct'' electron-temperature \te\ method;
the PP04N2 calibration is the closest strong-line approximation to the \te\ technique
\citep[see, e.g.,][for further discussion]{Hunt2016a}.
When neither of these were available, the original calibration was converted to PP04N2 according
to the expressions given by \citet{Kewley2008}.
\citet{ginolfi20} performed careful checks to ensure that the various methodologies for metallicity
determinations from the original samples were not unduly affected by metallicity gradients or aperture effects.

Atomic gas masses \mhi\ were taken from the original works (see \pone), and were added
to the molecular gas mass to obtain total gas mass, \mgas.
When possible, the \hi\ mass was defined within the optical radius of the galaxies
\citep[see][]{hunt15}, but in the majority of galaxies, the measurements were global.
This point will be discussed further in Sect. \ref{sec:lowmass}.
Molecular gas masses \mhtwo\ were inferred from \lco, using a metallicity-dependent
conversion factor \aco.
In \pone, we used an \aco\ with an approximately quadratic metallicity dependence below Solar O/H, 
according to the expression given by \citet[][]{hunt15}.
Such a dependence is consistent with the \aco\ dependence obtained from dust-to-gas ratios
\citep[e.g.,][]{leroy11,sandstrom13} and with independent calibrations of molecular gas mass
\citep{accurso17}.
In this paper, we recalculate molecular and total gas masses, revisiting
the \aco\ conversion factor by exploiting the MAGMA sample itself as the fiducial reference calibration.

\section{The conversion of CO luminosity to molecular gas mass: A new assessment}
\label{sec:alphaco}

Under typical physical conditions, CO emission is optically thick, 
but by comparing virial masses of molecular clouds with their
CO luminosity, it has been possible to associate a given \lco\ with 
\mhtwo\ through a simple conversion factor \aco:
\begin{equation}
M_\mathrm{H2}\,=\,\alpha_\mathrm{CO}\ L^\prime_\mathrm{CO}\quad .
\label{eqn:mh2}
\end{equation}
CO can be photodissociated in an intense UV radiation field \citep[e.g.,][]{vandishoeck88,wolfire10},
so that self-shielding is necessary for significant CO emission to emerge.
The implication is that \aco\ needs to be larger in low-metallicity environments with low dust content
and with harder and stronger UV fields.
It also means that carbon resides in atomic or ionized form when CO is photodissociated,
and that CO may cease to be a reliable tracer of \mhtwo\ in these conditions
\citep[e.g.,][]{wolfire10,bolatto13}.

Much effort has been devoted over the last few decades to quantifying the use of CO luminosity \lco\
as a tracer of molecular gas mass \mhtwo\
\citep[e.g.,][]{wilson95,israel97,bolatto08,bolatto13,leroy09,leroy11,schruba12,genzel12,sandstrom13,genzel15,hunt15,amorin16,accurso17}.
While most of the variation in \aco\ is usually attributed to metallicity, O/H, there are also indications
that other parameters are at play such as sSFR \citep[e.g.,][]{genzel15,accurso17}.

Here we use MAGMA to explore the limits above which CO can trace molecular gas, and 
the dependence of \aco\ on O/H, sSFR, and other available quantities.
A caveat with such an analysis is that the observers' definition of 
\sfehtwo\ is the inverse of the molecular depletion time \taudep:
\begin{equation}
\tau_\mathrm{H2}\,=\,M_\mathrm{H2}/\mathrm{SFR}\,=\,\alpha_\mathrm{CO}\ L^\prime_\mathrm{CO}/\mathrm{SFR}\,=\,(\mathrm{SFE}_\mathrm{H2})^{-1} \ .
\label{eqn:taudep}
\end{equation}
Thus it is virtually impossible to disentangle the effects of \sfehtwo, \taudep\ and \aco\ varying with metallicity or other
parameters;
in fact, only the ratio of \taudep/\aco\ is constrained by the observables, \lco\ and SFR.
Moreover, the common definition of \sfehtwo\ is really an inverse timescale, 
making a physical discussion of \sfehtwo\ difficult, if not impossible, with the observables at hand.

\begin{figure}[!h]
\centering
\includegraphics[width=0.98\linewidth]{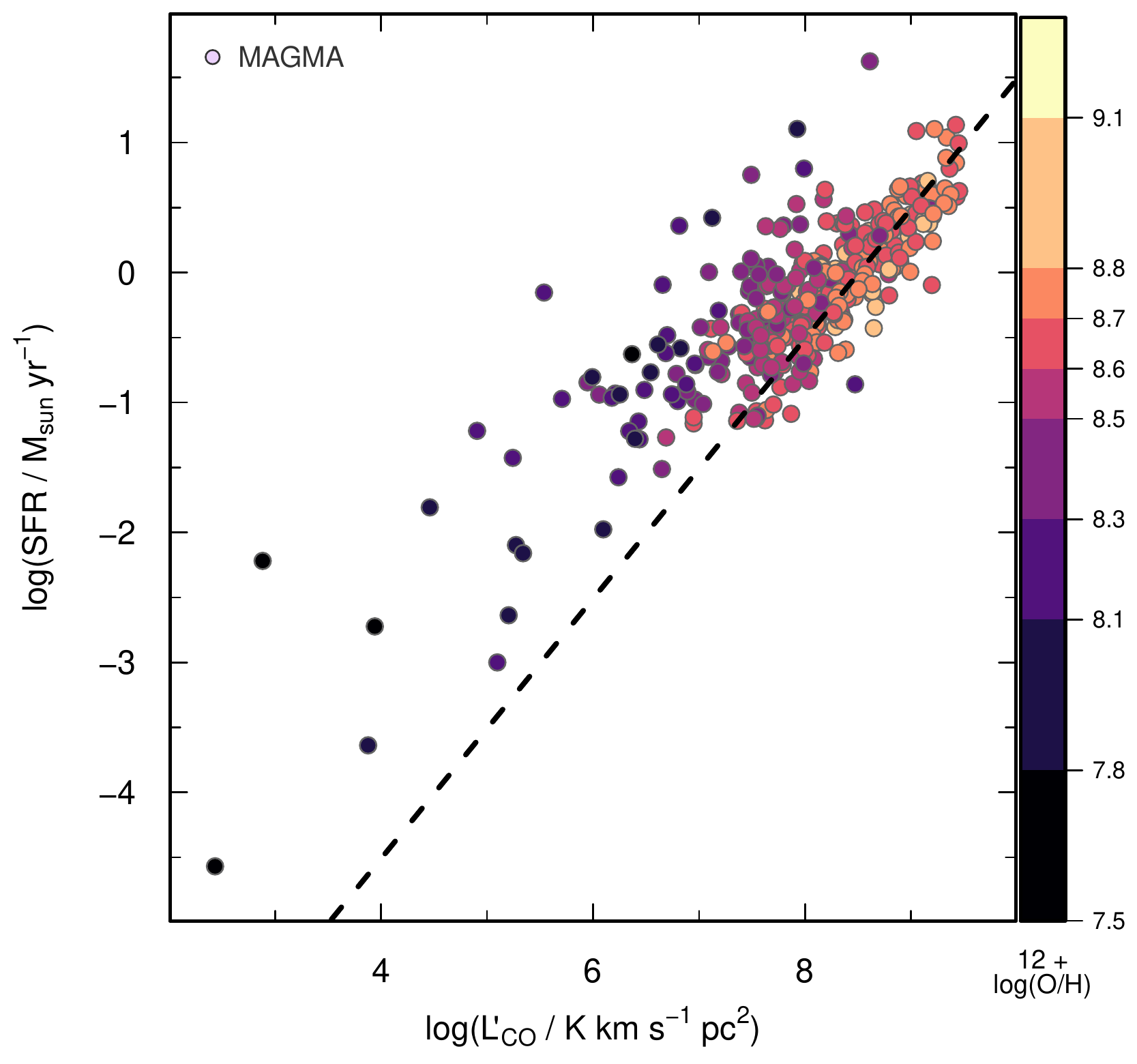}
\caption{SFR plotted against CO luminosity \lco. 
The regression shown by a long-dashed line indicates an imposed unit slope and a best-fit offset of
$-8.51\,\pm\,0.02$, roughly consistent with the correlation found by \citet{Gao2004} as described in the text.
Galaxies are color-coded by \logoh\ as indicated in the color wedge in the far-right panel.
}
	\label{fig:lcosfr}
\end{figure}

The correlation between SFR and CO luminosity \lco\ for MAGMA galaxies is shown in Fig. \ref{fig:lcosfr}.
The dashed line corresponds to an imposed unit slope and a fitted offset of $-8.51\,\pm\,0.02$,
roughly consistent with the offset of $-8.42$ found by \citet{Gao2004}, converted from far-infrared luminosity \lfir\
to SFR by \citet{hunt15}.
Part of this difference can be attributed to the different IMFs;
the conversion of \lfir\ to SFR by \citet{hunt15} assumed a \citet{Kroupa2001} IMF  
that has an offset of $-0.03$\,dex relative to the \citet{Chabrier2003} IMF used here.
At a given metallicity, the MAGMA data are best fit with a slightly sublinear slope (in logarithmic space) between SFR and \lco,
but overall are consistent with the unit slope reported by \citet{Gao2004}.
Figure \ref{fig:lcosfr} shows that at a given SFR, the more metal-poor galaxies are shifted to the
left, with smaller \lco, as also found by \citet{hunt15}.
This shift corresponds to a variation of \aco\ with metallicity that we explore below.

Assuming that SFR and \lco\ are approximately linearly related enables the exploitation of
results by \citet{Saintonge2011a,Saintonge2011b}.
They found that \taudep\ is correlated with sSFR, in the sense that shorter depletion times are
associated with larger sSFR \citep[see also][]{Huang2014,huang15,genzel15}.
However, their sample was mass-selected with little metallicity variation, so they could not explore
trends of \taudep\ and metallicity.
In fact, it is well established that sSFR and O/H are inversely correlated \citep[e.g.,][]{Hunt2012,salim14,Hunt2016a,ginolfi20},
so that it is not straightforward to separate the two effects.

For MAGMA, as in \citet{hunt15},
we adopt the hypothesis that \taudep\ depends on sSFR, and assume that any deviations from the global
trend are governed by \aco\ and its metallicity dependence.
This albeit simplistic assumption was validated by \citet{accurso17} who investigated 
CO and C$^+$ emission in a sample observed by \hers, and compared it with radiative transfer models of 
photodissociation regions (PDRs).
They found that variations in \aco\ can be almost totally ascribed to metallicity variations,
with only a small dependence on sSFR.
Here we attempt to disentangle the effects of metallicity and sSFR (and their mutual correlation)
and focus primarily on metallicity and its impact on the destruction of CO through photodissociation
in metal-poor radiation fields.

\begin{figure*}[!ht]
\centering
\includegraphics[width=\textwidth]{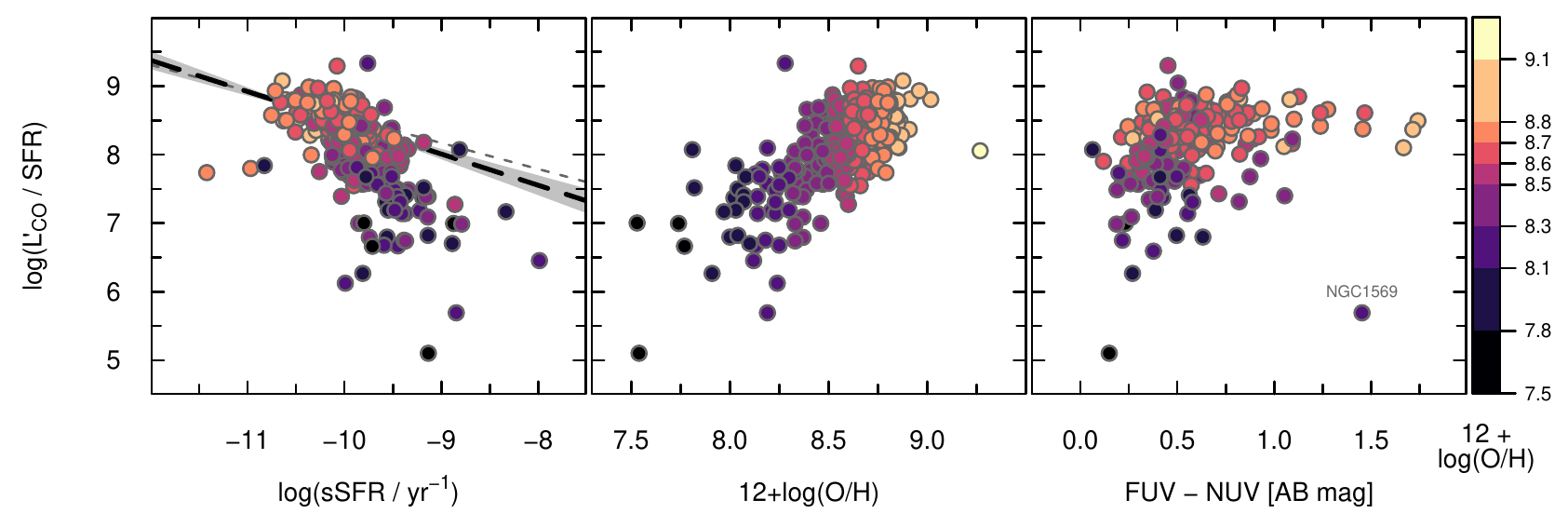}
\caption{Ratio of CO luminosity \lco\ and SFR plotted against specific SFR (left panel), \logoh\ (middle),
and GALEX \fuvnuv\ color (right).
In the left panel, the long-dashed line corresponds to the robust fit given by Eqn. (\ref{eqn:ssfr}) and the
(gray) line to the shallower slope ($-0.38$) from \citet{hunt15}.
The gray shadowed region in the leftmost panel indicates the range of uncertainty on the slope in Eqn. (\ref{eqn:ssfr}).
Galaxies are color-coded by \logoh\ as in Fig. \ref{fig:lcosfr}. 
	\label{fig:lcosfrvsssfr}
}
\end{figure*}

\begin{figure}[h!]
\centering
\includegraphics[width=0.81\linewidth]{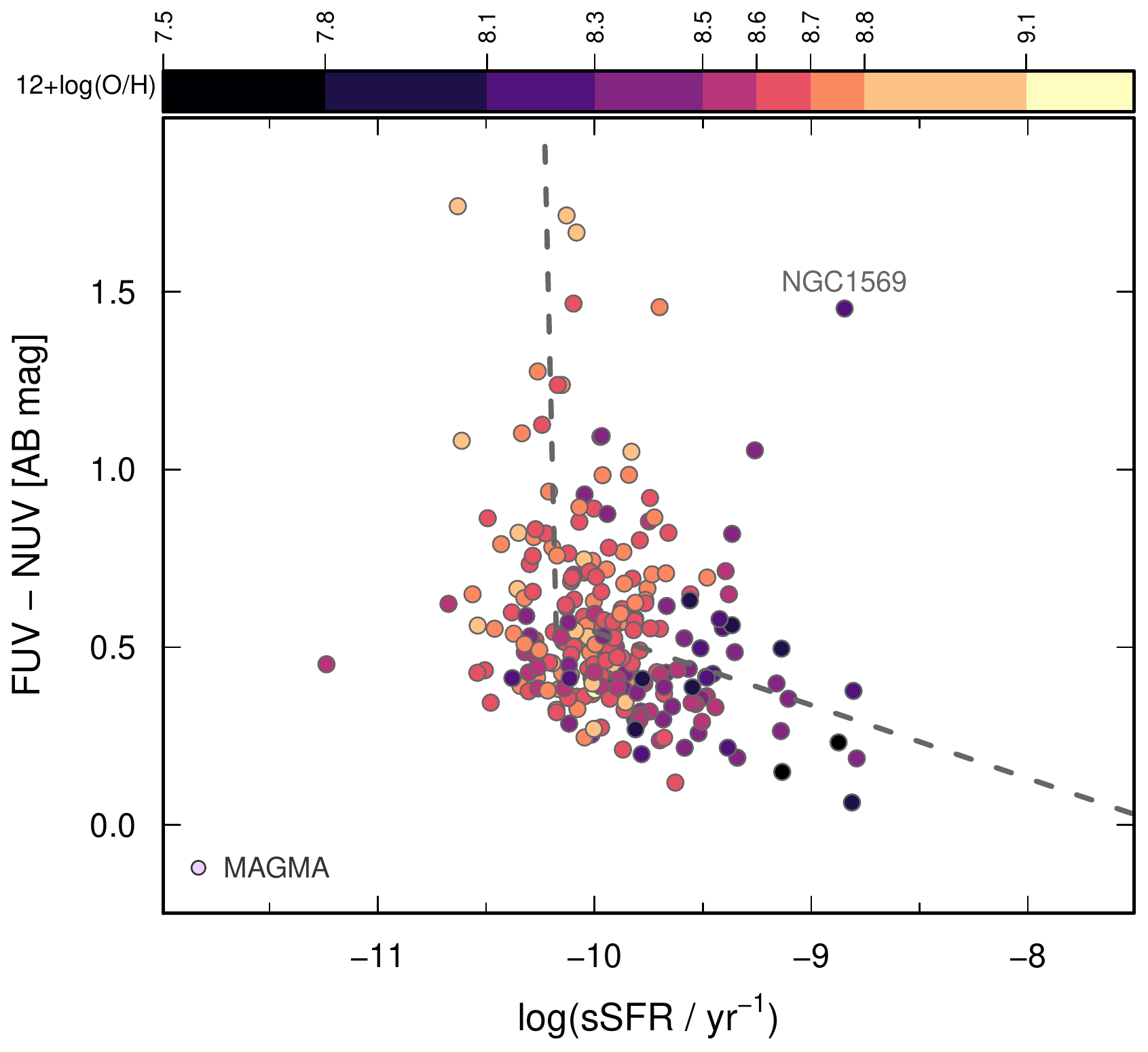} \\
\vspace{0.5\baselineskip}
\includegraphics[width=0.81\linewidth]{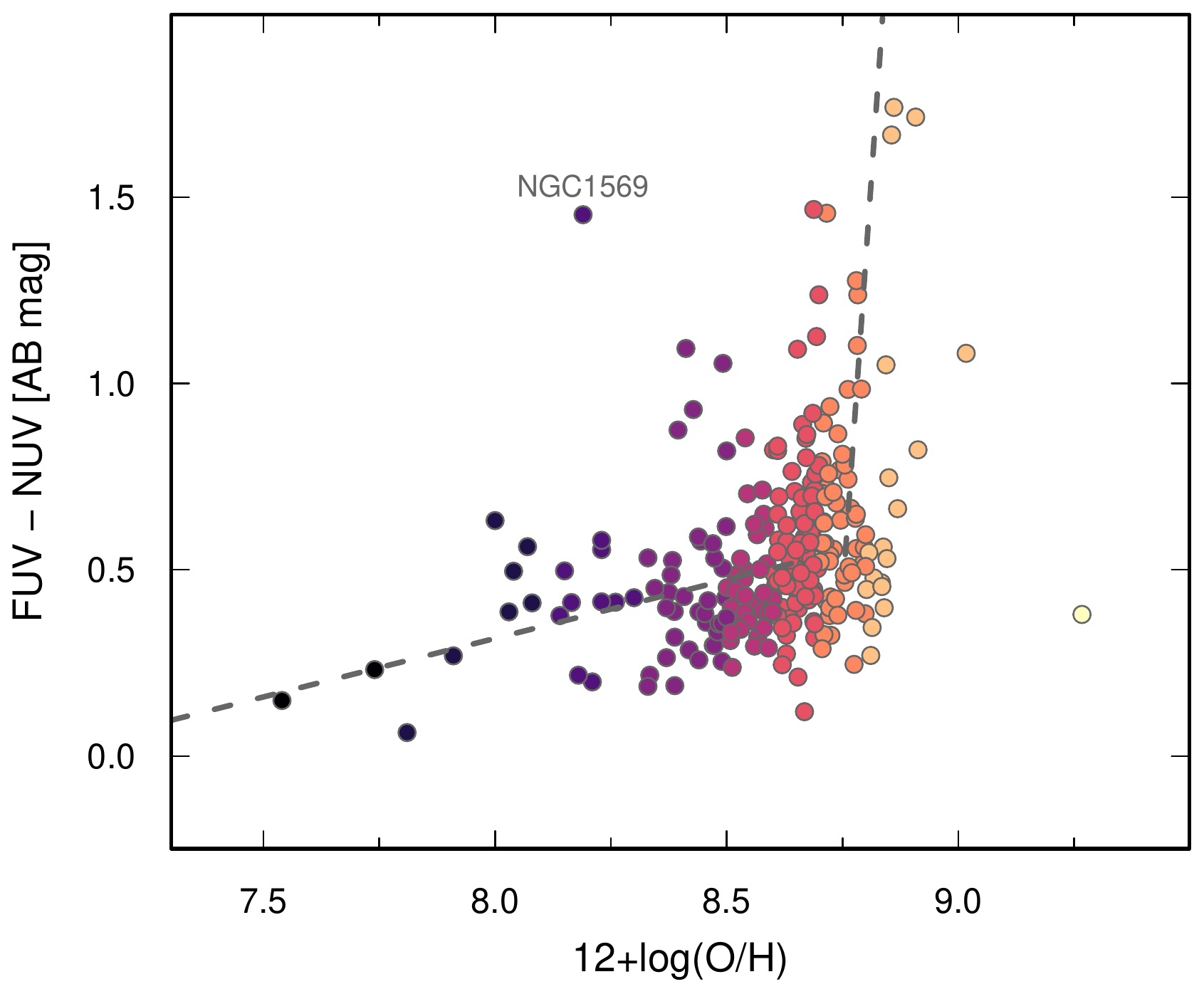} 
\caption{GALEX \fuvnuv\ color plotted against log(sSFR) (top panel), and \logoh\ (bottom).
Like in Fig. \ref{fig:lcosfr}, galaxies are color-coded by \logoh.
The long-dashed lines correspond to an approximation of the
GALEX Blue Sequence and GALEX Red Sequence as a function of O/H as described in the text.
}
\label{fig:uvsfroh}
\end{figure}

Figure \ref{fig:lcosfrvsssfr} illustrates the correlation of sSFR and the ratio
of \lco\ and SFR for MAGMA.
Using a robust least-squares fitting algorithm,
and applying it only to solar- and super-solar metallicity galaxies,
we find the best-fit regression: 
\begin{eqnarray}
\log (L^\prime_\mathrm{CO}/\mathrm{SFR}) & = & (-0.46\,\pm\,0.07)\,\log(\mathrm{sSFR}) + (3.91\,\pm\,0.69) \nonumber \\ 
&& \text{if \logoh $\geq$ (O/H)$_\odot$}\quad .
\label{eqn:ssfr}
\end{eqnarray}
The relation has a median absolute deviation (MAD) of 0.19\,dex for the 129 MAGMA galaxies at these metallicities.
The power-law slope of $-0.46$ is somewhat steeper than the value of $-0.38$ found
for the (much smaller) dataset of \citet{hunt15}, based mostly on the earlier version
of xCOLDGASS by \citet{Saintonge2011a}; both trends are shown in Fig. \ref{fig:lcosfrvsssfr}.
Low-metallicity galaxies at high sSFR deviate from the regression as indicated by the metallicity color coding in the figure. 
The correlation of \lco/SFR with O/H is plotted in 
the middle panel of Fig. \ref{fig:lcosfrvsssfr}, where the association of low \lco/SFR
with low metallicity is clearly evident.
In the right panel of Fig. \ref{fig:lcosfrvsssfr},
we have attempted to quantify radiation-field hardness by the GALEX \fuvnuv\ color, 
here corrected only for Galactic extinction as described in \pone. 
The use of \fuvnuv\ to probe radiation-field hardness is compromised
by its dependence on IMF and star-formation history (SFH),
in particular for early-type galaxies
\citep[e.g.,][]{zaritsky15,yildiz17}.

With the aim of quantifying radiation-field hardness, in Fig. \ref{fig:uvsfroh},
we explore the dependence of \fuvnuv\ on sSFR (top panel) and on O/H (bottom). 
NGC\,1569 is an outlier in both plots probably due to the difficulty in correcting \fuvnuv\ for
Galactic extinction because of its relatively low Galactic latitude: according to \citet{schlafly11}, \av\,=\,1.90\,mag, 
so it is extremely reddened by the MW.

There is no straight single power-law dependence of \fuvnuv\ on either
sSFR or O/H, but rather a linear (power-law) trend, and then a plateau.
Such behavior also emerged in the analysis of $\sim$2000 galaxies by
\citet{bouquin15} who plotted \fuvnuv\ against NUV$-$[3.6\,micron] colors, 
an approximate tracer of sSFR.
They found evidence for two trends:
a tight GALEX Blue Sequence (GBS) up to \fuvnuv$\sim$0.7, and
a broader GALEX Red Sequence (GRS) for redder colors.
By comparing the data with models, they attributed the GBS to actively
star-forming galaxies over a wide range of \mstar, while
more evolved, passive, galaxies lie on the GRS. 
At one point, however, older early-type galaxies become bluer in \fuvnuv\
because of the UV upturn as galaxies age \citep[e.g.,][]{gildepaz07,kaviraj07,rampazzo07,boissier18}.
Both the GBS (the more horizontal trends) and the GRS (the almost vertical ones) can be seen in MAGMA data in Fig. \ref{fig:uvsfroh}.
Relative to (log) sSFR and \logoh, we have performed piecewise regressions, 
corresponding roughly to the GBS and the GRS, and these are overlaid in Fig. \ref{fig:uvsfroh}.
This bimodal behavior of \fuvnuv\ makes it difficult to unambiguously trace radiation-field hardness
using this parameter.
Another difficulty is caused by the level of reddening and the shape of the dust attenuation curve,
both of which are expected to have an impact on \fuvnuv\ \citep[e.g.,][]{wyder07}.

\begin{figure*}[!t]
\centering
\includegraphics[width=0.85\textwidth]{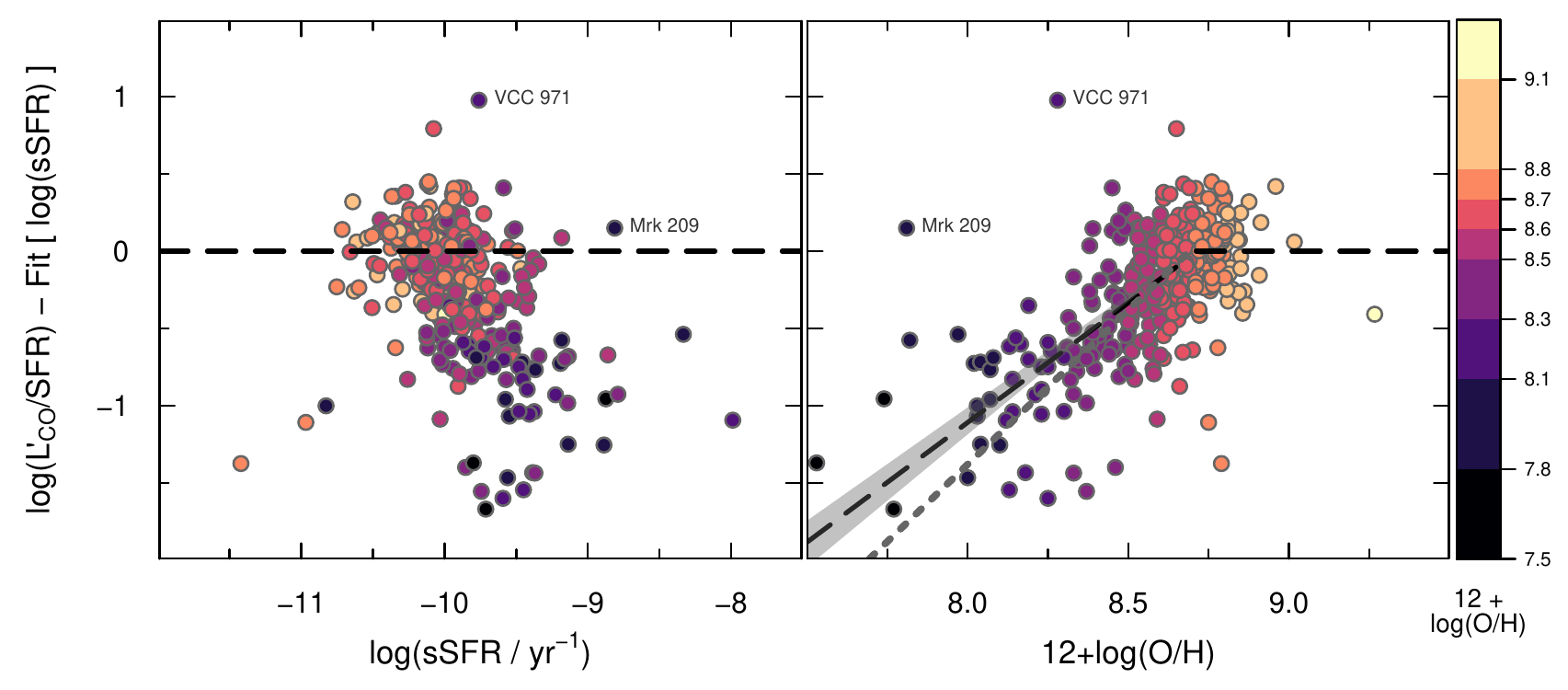} 
\caption{Residuals of the fit given by Eqn. (\ref{eqn:ssfr}) plotted against specific SFR (left panel) and \logoh\ (right).
The metal-poor portion of the piece-wise regression in the right panel is given by 
$\mathrm{\log(L^\prime_\mathrm{CO}/SFR) - fit}\,=\,(1.55\,\pm\,0.08)\,[12+\log(\mathrm{O/H})] - (13.51\,\pm\,0.71)$, 
and the ``fit" refers to Eqn. (\ref{eqn:ssfr}), both shown by long-dashed (gray) lines.
The gray shadowed region in the right panel indicates the range of uncertainty on the slope in the above equation, 
and the short-dashed (gray) line corresponds to a power-law slope of 2.0 
(compared to the best-fit value here of 1.55).
As in Fig. \ref{fig:lcosfr}, galaxies are color-coded by \logoh.
\label{fig:lcosfr_residuals}
}
\end{figure*}

The implicit assumption that we now make is that metallicity is the best
tracer of radiation-field hardness and the propensity for CO to photodissociate.
We make the further assumption that the deviations of the trends for \taudep\ and sSFR 
depend on metallicity. 
Consequently, to attempt to separate the effects on \aco\ of sSFR and metallicity,
we apply the regression of Eqn. (\ref{eqn:ssfr}) to the entire MAGMA sample,
and plot the residuals of the fit in Fig. \ref{fig:lcosfr_residuals}.
The left panel shows the residuals relative to sSFR and the right with respect to \logoh.
If we assume that \aco\ is constant at solar metallicity and above (\acosun),
we can fit the residuals at lower metallicity to a trend with \logoh\ as shown
in the right-hand panel of Fig. \ref{fig:lcosfr_residuals}.
The scatter in the fit (MAD) is 0.25\,dex for the 261 MAGMA galaxies at 
sub-solar metallicities\footnote{We have eliminated
two galaxies from this fit because of their extreme outlier nature, namely Mrk\,209 
and VCC\,971 (galaxies from DGS and HRS, respectively, with unreliable CO detections).
Despite the robust fitting algorithm, these two galaxies alone
change the power-slope from $-1.55$ (without them) to $-1.50$ (with them).}.

Assuming, as alluded to above, that the same correlation of \lco/SFR and sSFR continues to low metallicities,
and that the only deviation is due to metallicity variations,
we can convert the regression shown in the right panel of Fig. \ref{fig:lcosfr_residuals} 
to an expression for \aco:
{\small
\begin{equation}
\alpha_{\rm CO}\ = \
\begin{cases}
\alpha_{\rm CO}^{\odot}\,(Z/Z_\odot)^{-(1.55\,\pm\,0.08)}\,& \\
\quad\quad\quad\quad\quad\quad\quad\quad\quad\quad\quad \text{if 12+log(O/H) $<$ 8.69}\ ; \\
\alpha_{\rm CO}^{\odot}                              & \\
\quad\quad\quad\quad\quad\quad\quad\quad\quad\quad\quad \text{if  12+log(O/H) $\geq$ 8.69}\ ; \\
\end{cases}
\label{eqn:alphaco}
\end{equation}
}
where \zzsun\ is quantified by dex[$12+\log(\mathrm{O/H})-8.69$].
The power-law metallicity dependence of 1.55 is somewhat shallower than the quadratic dependence
found by \citet{hunt15}, but consistent with other estimates using different techniques
\citep[e.g., power-law slopes of 1.5$-$1.7]{amorin16,accurso17}.
The difference relative to our earlier, steeper, estimation of O/H dependence is 
related to the slope of the \lco/SFR ratio vs. sSFR.
Here in Eqn. (\ref{eqn:ssfr}) the regression with slope $-0.46$ is slightly steeper than the best-fit value of $\sim -0.4$
found by \citet{hunt15} and \citet{Saintonge2011b}.
Fig. \ref{fig:lcosfrvsssfr} illustrates the two regressions.
The most probable cause of this difference is the way SFRs are calculated:
\citet{Saintonge2011b} derive SFRs from fitting of the optical-UV spectral energy distributions,
while in \pone\ we use the empirical approach of \citet{Leroy2019} based on FUV, NUV, and mid-infrared
luminosities, similar to the ``ladder'' technique described in \citet{Saintonge2017}.
If we fix the slope of the regression in Eqn. (\ref{eqn:ssfr}) to $-0.4$, we find a quadratic
metallicity dependence as in \citet{hunt15}, but with a much poorer overall fit for the larger MAGMA sample. 
Such behavior exemplifies the difficulty in separating the effects of sSFR and metallicity,
showing that lingering dependencies almost certainly exist within our formulation.

The choice of a broken power-law as a function of metallicity, rather than a continuous one
\citep[e.g.,][]{amorin16,accurso17},
is motivated mainly by the data.
Rather than fitting only the lower metallicity points, we also applied
a continuous power law to fit the trend of \lco/SFR as a function of metallicity.
In this case, the slope is shallower, $\sim$1.3, rather than the best fit value of $\sim$1.55.
To better understand why,
we fit  with a power law only the metal-rich galaxies (those with \logoh\,$\geq$\,8.69), and obtained 
a negative slope of $\sim -0.3$.
This would imply that \aco\ varies with metallicity in an unphysical way, \aco\,$\propto$\,(\zzsun)$^{0.3}$,
but would explain why the slope is shallower when we fit all the galaxies, not just those with sub-Solar metallicities.

Physically, the variation of \aco\ with metallicity is motivated primarily
by considerations about dust shielding and a higher ambient UV radiation field, 
neither of which is expected to change dramatically above a limiting metallicity.
The theoretical exponential trend predicted by \citet{wolfire10} 
is roughly flat at super-Solar metallicities,
and fairly well approximated by the broken power law we use here.
\citet{glover11} carried out hydrodynamical modeling of molecular clouds coupled with a chemical network,
and also found that \aco\ could be best formulated as a broken power law.
They found that the main variable for \aco\ variation is visual extinction \av\
within the cloud, and that above \av$\,\ga\,$3, the conversion factor \aco\ is roughly constant.
Thus, our use of a broken power law, namely a variation with metallicity only below a certain 
metallicity threshold (here we take it to be $\sim$\zsun) 
can be justified also from a theoretical perspective. 

In contrast to our technique,
derivations of \aco\ based on dust-to-gas ratios (DGRs) in resolved regions within galaxies generally 
give a weaker metallicity dependence, 
possibly because of higher DGRs in the dense gas residing in molecular complexes 
\citep[e.g.,][]{leroy11,sandstrom13}.
Virial-based \htwo\ mass estimates give little or no metallicity dependence \citep[e.g.,][]{Leroy2006,leroy11},
probably because the virial measurements only reflect the CO-bright clouds \citep[e.g.,][]{bolatto08}.

In what follows, we adopt the \aco\ given by Eqn. (\ref{eqn:alphaco}),
with \acosun\,=\,3.2\,\msun\,(\lcounits)$^{-1}$ \citep[e.g.,][]{Saintonge2011b}, 
not including a factor of 1.36 to account for helium. 
When relevant, we assess the ramifications on our results of our recipe for \aco.

\section{Gas content and scaling relations with stellar mass and SFR} 
\label{sec:gasscaling}

The gas reservoir available to a galaxy regulates baryonic cycling, and
provides the fuel for star formation.
Examining the scaling relations among gas content, stellar mass, SFR, and metallicity
is crucial for understanding the physical mechanisms behind the baryonic energy exchange.
In \pone, we looked at scaling relations through several Principal Component Analyses (PCAs).
Four-dimensional PCAs with \mstar, SFR, O/H, and gas mass (either total \mgas, atomic \mhi, or molecular \mhtwo)
suggested that the parameter space for galaxies is two-dimensional (2D). 
In all the PCAs examined (including 5D, 4D, and 3D), metallicity was always the most dependent parameter 
\citep[see also][]{Hunt2012,Hunt2016a}.
Thus, it was not clear that adding information about gas content
adds another dimension to the description of a galaxy, although we re-examine this conclusion
below.
The underlying implication is that gas content, and possibly even the individual atomic and gas phases separately, can
be predicted using only \mstar\ and SFR.

\begin{figure*}[!h]
\centering
\includegraphics[width=0.48\textwidth]{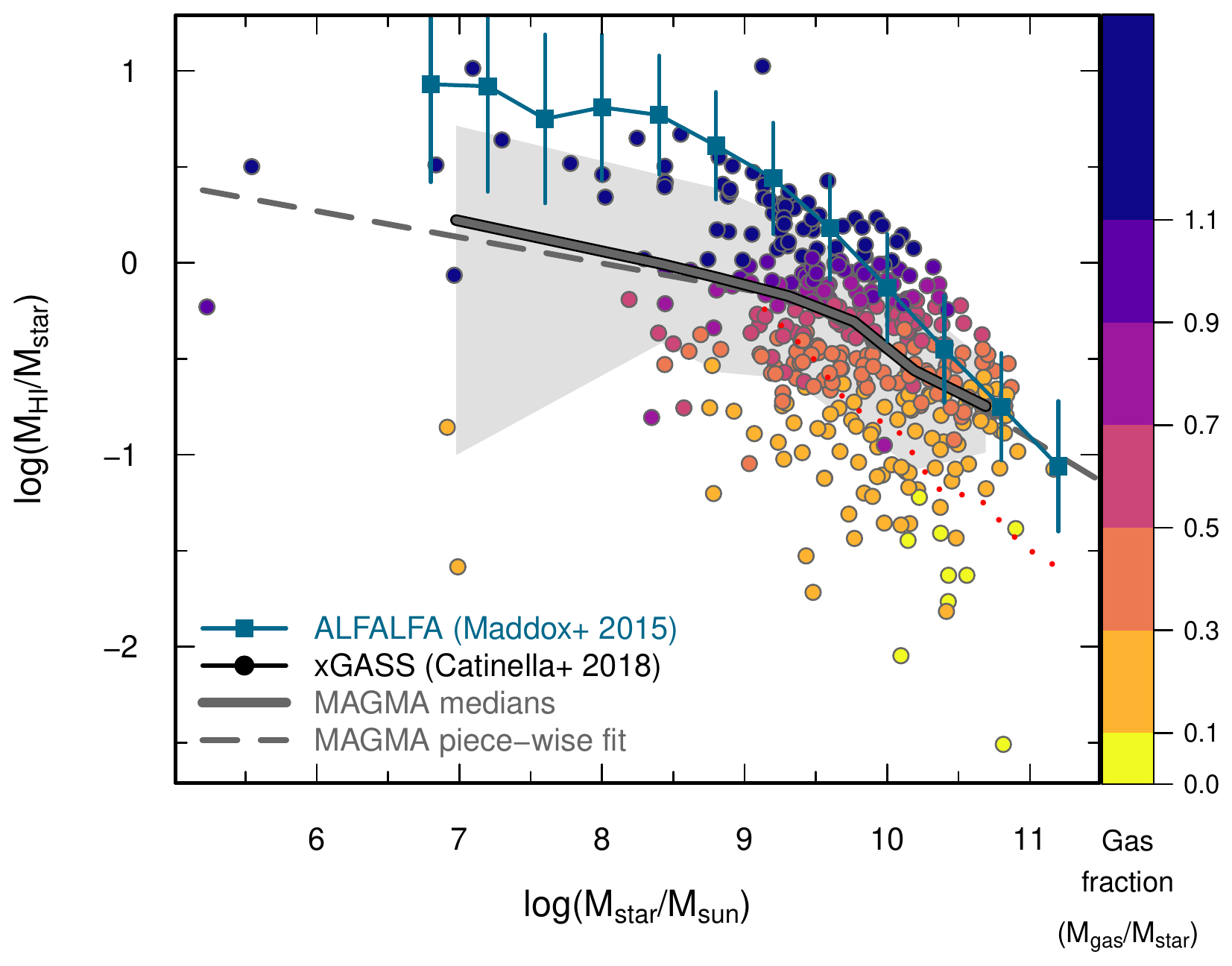} 
\hspace{0.02\textwidth}
\includegraphics[width=0.48\textwidth]{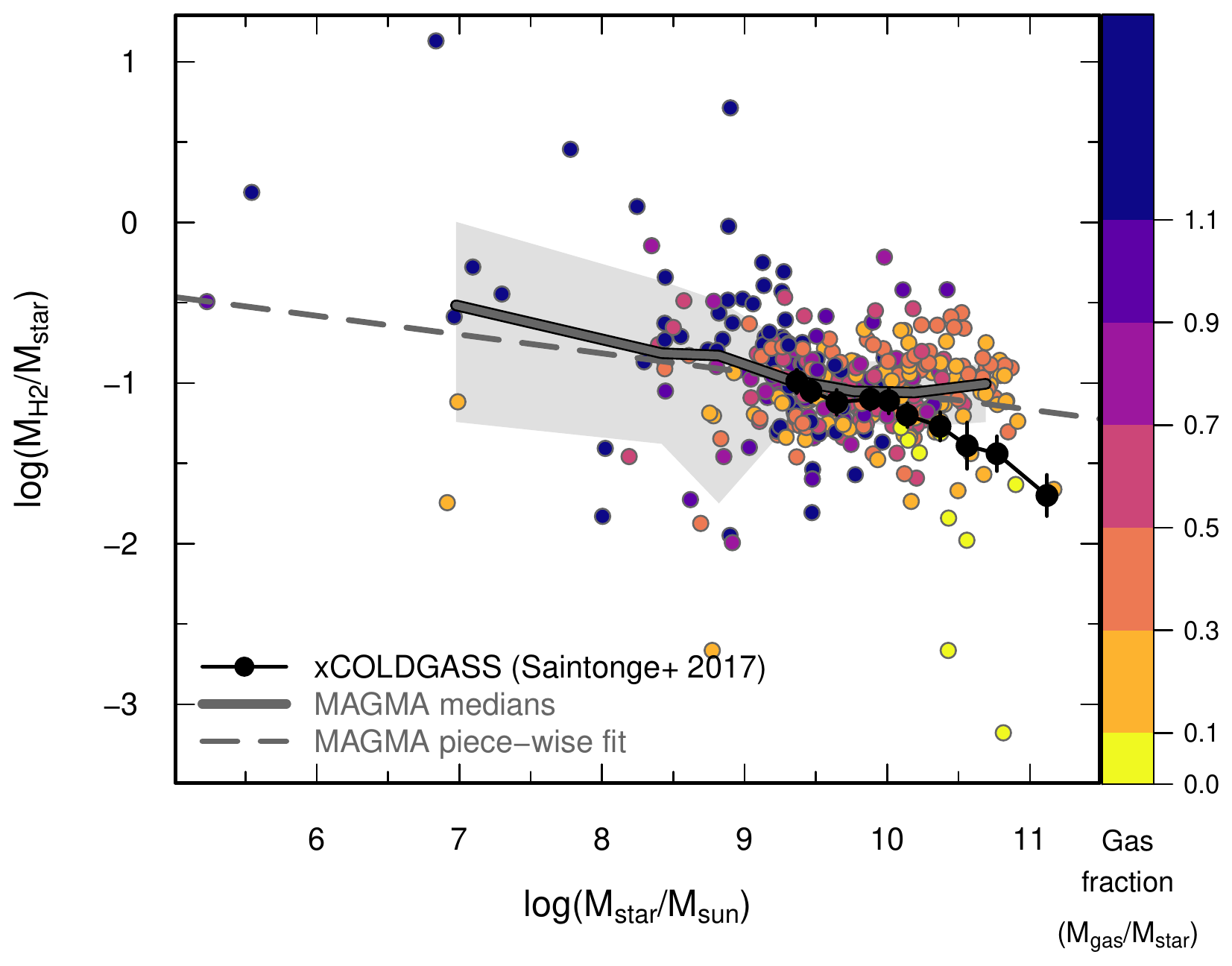} 
\caption{Left panel: \mhi/\mstar\ plotted vs. \mstar\ (logarithmic units). 
The piece-wise power-law regression for MAGMA is shown by a dashed line,
and the gray regions correspond to $\pm\,1\,\sigma$ deviations of the binned data medians.
The width of the distribution of the residuals from this fit is 0.39\,dex.
Also shown as filled blue squares are the data from \citet{maddox15} for the \hi-selected ALFALFA sample as described
in the text.
xGASS \citep{Catinella2018} data are shown as filled black circles; these include \hi\
non-detections (see their Table 1) so tend to be biased low relative to MAGMA with \hi\ detections only.
Right panel: \mhtwo/\mstar\ plotted vs. \mstar.
The filled black circles correspond to data from xCOLDGASS \citep{Saintonge2017}, and include only
main-sequence galaxies (see their Table 6). 
In both panels,
the MAGMA medians of binned data are shown by heavy gray lines, and
MAGMA galaxies are colored by gas fraction as in Fig. \ref{fig:magma_ms}.
\label{fig:hih2}
}
\end{figure*}

Observationally, molecular gas can be ``expensive'' to observe, especially in low-mass dwarf galaxies where CO and
other molecular tracers require high sensitivity, but also in
high-mass potentially quenched galaxies with low gas content.
\citet{yesuf19} use visual extinction \av, metallicity, SFR and/or the optical half-light
radius to estimate molecular and atomic gas masses to within factors of 2-3.
Here we improve on this accuracy using fewer parameters, and 
examine scaling relations with \mstar\ and SFR for total gas content in MAGMA, and also the separate gas phases, \hi\ and \htwo. 
Our study follows a long line of pioneering observational efforts 
\citep[e.g.,][]{schiminovich10,Wei2010,Haynes2011,Saintonge2011a,Cortese2011,Huang2012,Stark2013,Kannappan2013,Gavazzi2013,Bothwell2014,Boselli2014b,Saintonge2017,Cicone2017,Catinella2018}.

\begin{figure}[!h]
\centering
\includegraphics[width=\linewidth]{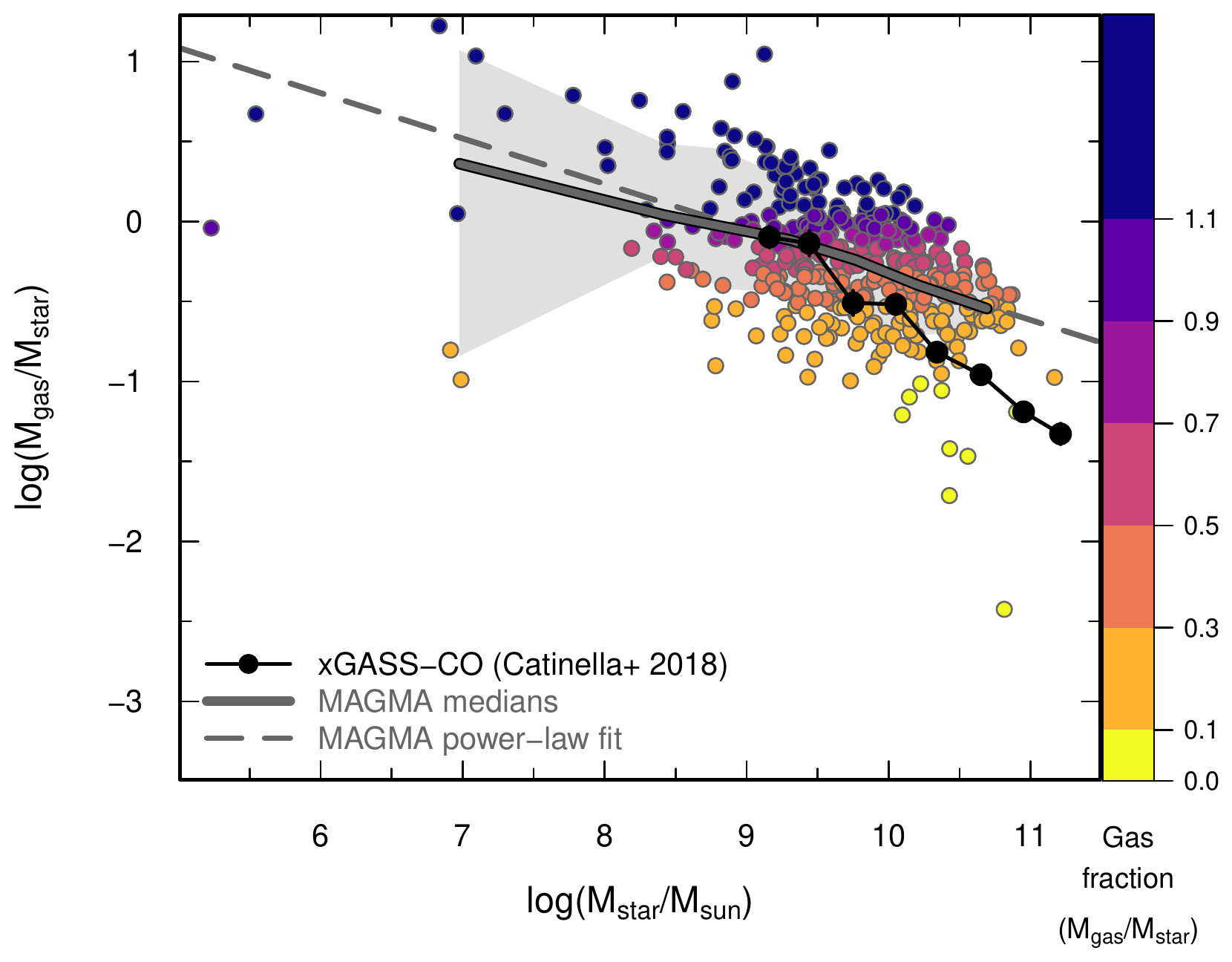} 
\caption{Total gas content ratioed by \mstar\ (\mgas/\mstar) plotted against \mstar.
The robust best fit given by Eqn. \ref{eqn:mgasmstar}
is shown by a light-gray long-dashed line.
As in Fig. \ref{fig:hih2} the means (and standard deviations) from xGASS-CO 
are shown by filled black circles, and the MAGMA medians with a heavy gray line.
MAGMA galaxies are colored by gas fraction as in previous figures
\label{fig:mgasvsmstar}
}
\end{figure}

\subsection{Trends of gas content with \mstar}
\label{sec:gasmstar}

First, we examine the relation of \mhi\ and \mhtwo\ with \mstar.
The ratio of \mhi\ to \mstar\ plotted against \mstar\
is illustrated in the left panel of Fig. \ref{fig:hih2}, and \mhtwo/\mstar\ in the right.
Like \hi-selected samples such as ALFALFA that probe low \mstar\ \citep[Arecibo Legacy Fast ALFA survey,][]{Haynes2011,Huang2012,Haynes2018},
MAGMA data show a break around \mstar=$3\times10^9$\,\msun;
below this threshold the power-law slope between \mhi/\mstar\ vs. \mstar\ is flatter than
above this limit. 
The best robust fit to the piecewise trend for MAGMA galaxies is given by:
{\small
\begin{equation}
\log(M_{\rm HI}/M_{\rm star})\ = \
\begin{cases}
(-0.14\,\pm\,0.05)\,\log(M_{\rm star}) + (1.09\,\pm\,0.49) & \\
\quad\quad\quad\quad\quad\quad\quad\quad\quad \text{log($M_{\rm star}) \leq$ 9.5}\ ; \\
(-0.45\,\pm\,0.07)\,\log(M_{\rm star}) + (4.03\,\pm\,0.68) & \\
\quad\quad\quad\quad\quad\quad\quad\quad\quad \text{log($M_{\rm star}) >$ 9.5}\ . \\
\end{cases}
\label{eqn:himstar}
\end{equation}
}

\begin{figure*}[!ht]
\begin{minipage}[c]{0.7\textwidth}
\includegraphics[width=\textwidth]{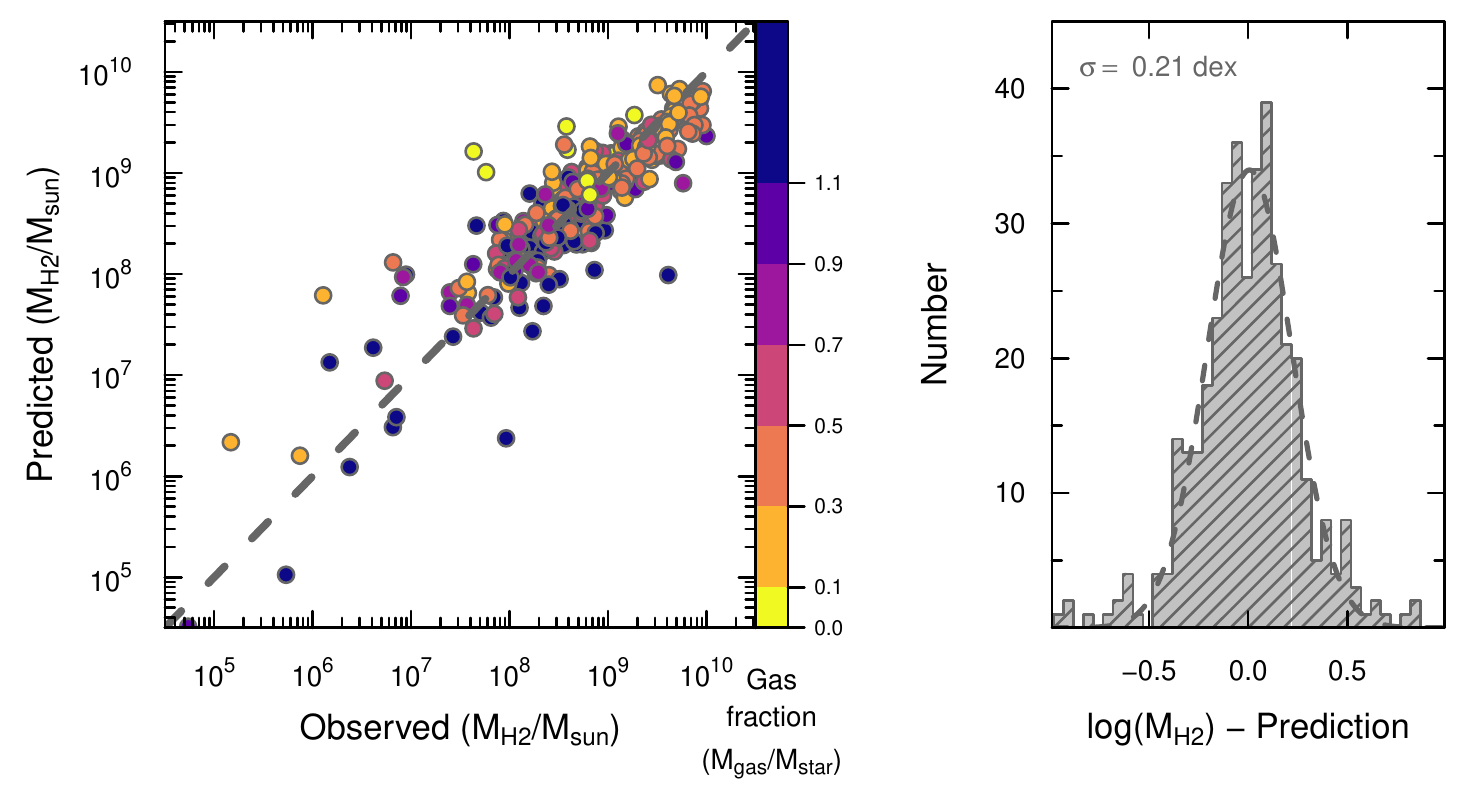} 
\end{minipage}
\hspace{0.02\textwidth}
\begin{minipage}[c]{0.26\textwidth}
\caption{Observed values of log(\mhtwo) in the MAGMA sample compared to those 
predicted by the regression as a function of \mstar\ and SFR given in Eqn. (\ref{eqn:regressionmh2}). 
The regression is illustrated by the gray-dashed line.
The right panel shows the residuals and the Gaussian fit;
the 1$\sigma$ dispersion of the Gaussian is 0.21\,dex as discussed in the text.
Galaxies are colored by gas fraction as shown in previous figures.
\label{fig:mh2vsmstarSFR}
}
\end{minipage}
\end{figure*}

There is significant scatter, 0.39\,dex (MAD) in this relation, 
illustrating the large scatter in the \mhi/\mstar\ ratio.
We experimented with different break \mstar\ around $10^9$\,\msun, and found that
log(\mstar/\msun)\,=\,9.5 provides the lowest residuals.
The spirit of this fit is similar to that reported by \citet{Huang2012}, although
the slope for the massive ALFALFA galaxies is steeper in the \mhi/\mstar\,--\,\mstar\ plane
(see their Eqn. 1).

For a given stellar mass, the gas-rich MAGMA galaxies (shown by the dark purple symbols, see color coding in 
the right-hand wedge in Fig. \ref{fig:mgasvsmstar})
resemble the ALFALFA galaxies, while the more gas-poor objects bring down the median.
This is particularly true in the mass range above the \hi\ break mass of $10^{9.5}$\,\msun\ where
the more gas-poor galaxies are fairly well represented by the xGASS trend \citep[see][]{Catinella2018}.

The right panel of Fig. \ref{fig:hih2} illustrates the \mhtwo/\mstar\ relation with \mstar.
Unlike for \hi, there is no apparent break mass and a single power-law slope is a fairly
good approximation to the data.
Here, as apparent also by eye, the MAD scatter is lower, 0.24\,dex, relative to \hi\ (0.39\,dex);
the excursions of the \mhtwo/\mstar\ ratio are significantly smaller than those of \hi.

Also shown in Fig. \ref{fig:hih2} are the xGASS and xCOLDGASS medians which, however, incorporate
non-detections \citep{Saintonge2017,Catinella2018}.
This biases \mhi\ and \mhtwo\ low and makes comparison with MAGMA difficult; however, Fig. \ref{fig:hih2} shows that the xGASS
galaxies behave similarly to the \hi-selected ALFALFA sample, although offset by $\sim -0.7$\,dex.
In any case, the agreement of MAGMA with the (binned) \htwo\ content for MS xCOLDGASS galaxies is good, at least
up to the highest-\mstar\ bins where the dearth of high-mass galaxies in MAGMA becomes apparent.
This comparison is meaningful because the formulation for \aco\ used by \citet{Saintonge2017} is
quite similar to ours \citep[see][]{accurso17}.
MAGMA considers only \htwo\ (and \hi) detections, so that the upturning of the MAGMA median trend relative
to xCOLDGASS is probably due to their inclusion of \htwo\ non-detections, similarly to the
comparison with xGASS shown in the left panel of Fig. \ref{fig:hih2}. 

Figure \ref{fig:hih2} illustrates the importance of extending statistical studies to low \mstar.
The galaxies below the \mstar\ break are gas rich (see color coding), and bridge the
gap between representative samples such as MAGMA and \hi-selected ones like ALFALFA.
The large scatter at these masses shows that more observations are needed at low \mstar, but
the requirement of CO detections for \htwo\ makes this currently very time-consuming.

The analogous relation for total gas mass \mgas\ is shown in Fig. \ref{fig:mgasvsmstar}.
The comparison with XGASS-CO is also given in Fig. \ref{fig:mgasvsmstar}, and,
as before, the similarity of the \aco\ formation used by \citet{Catinella2018} for \mgas\
gives meaning to the comparison.
The best robust fit of \mgas/\mstar\ vs. \mstar\ is given by:
\begin{equation}
\log{M_{\rm gas}/M_{\rm star}}\,=\,(-0.28\,\pm\,0.02)\ \log(M_{\rm star}) + (2.51\,\pm\,0.22)\ . 
\label{eqn:mgasmstar}
\end{equation}
The scatter in terms of MAD is 0.33\,dex, worse than
that with \mhtwo, but better than with \mhi\ alone.

\begin{figure*}[!ht]
\centering
\includegraphics[width=0.48\linewidth]{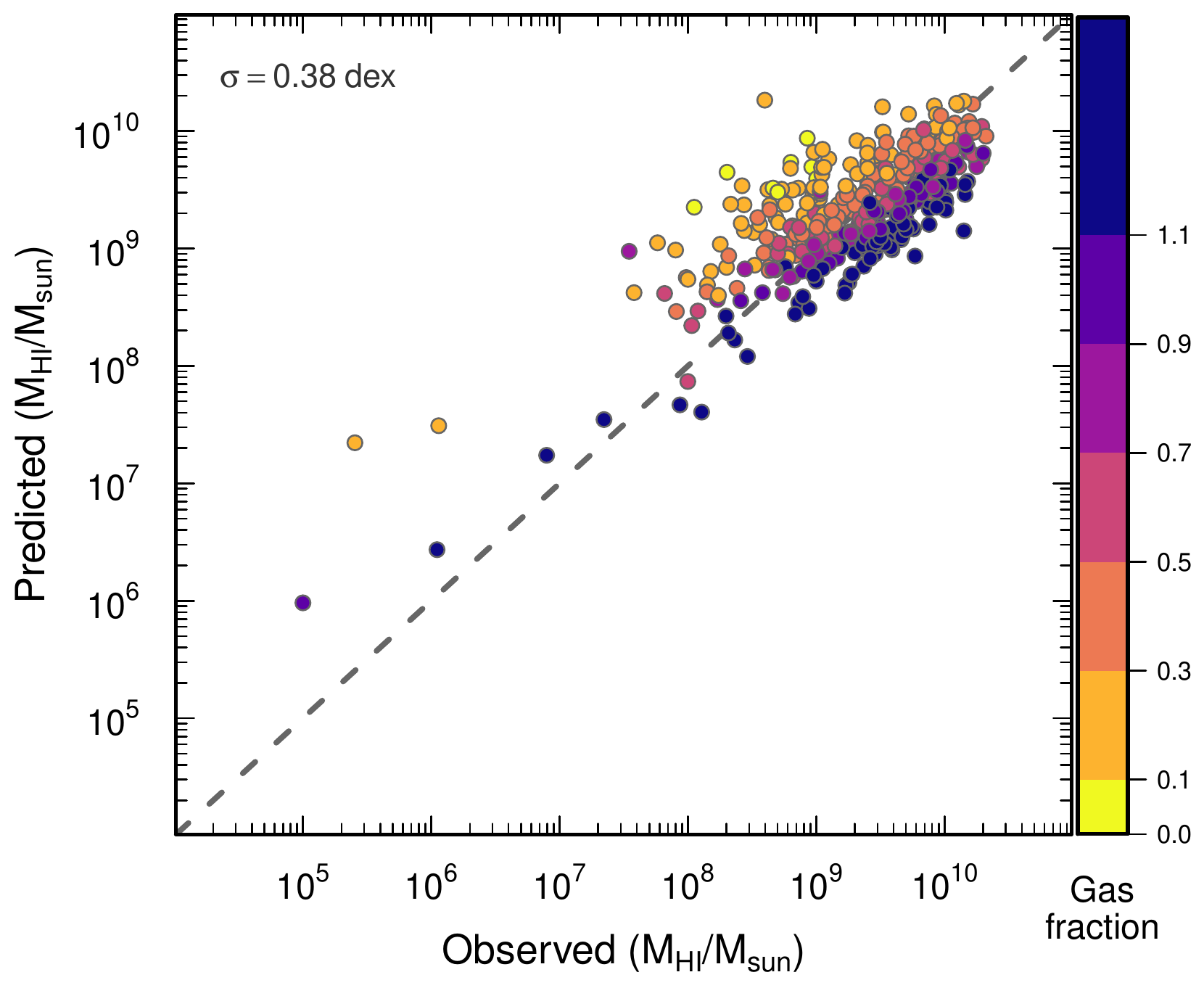} 
\includegraphics[width=0.48\linewidth]{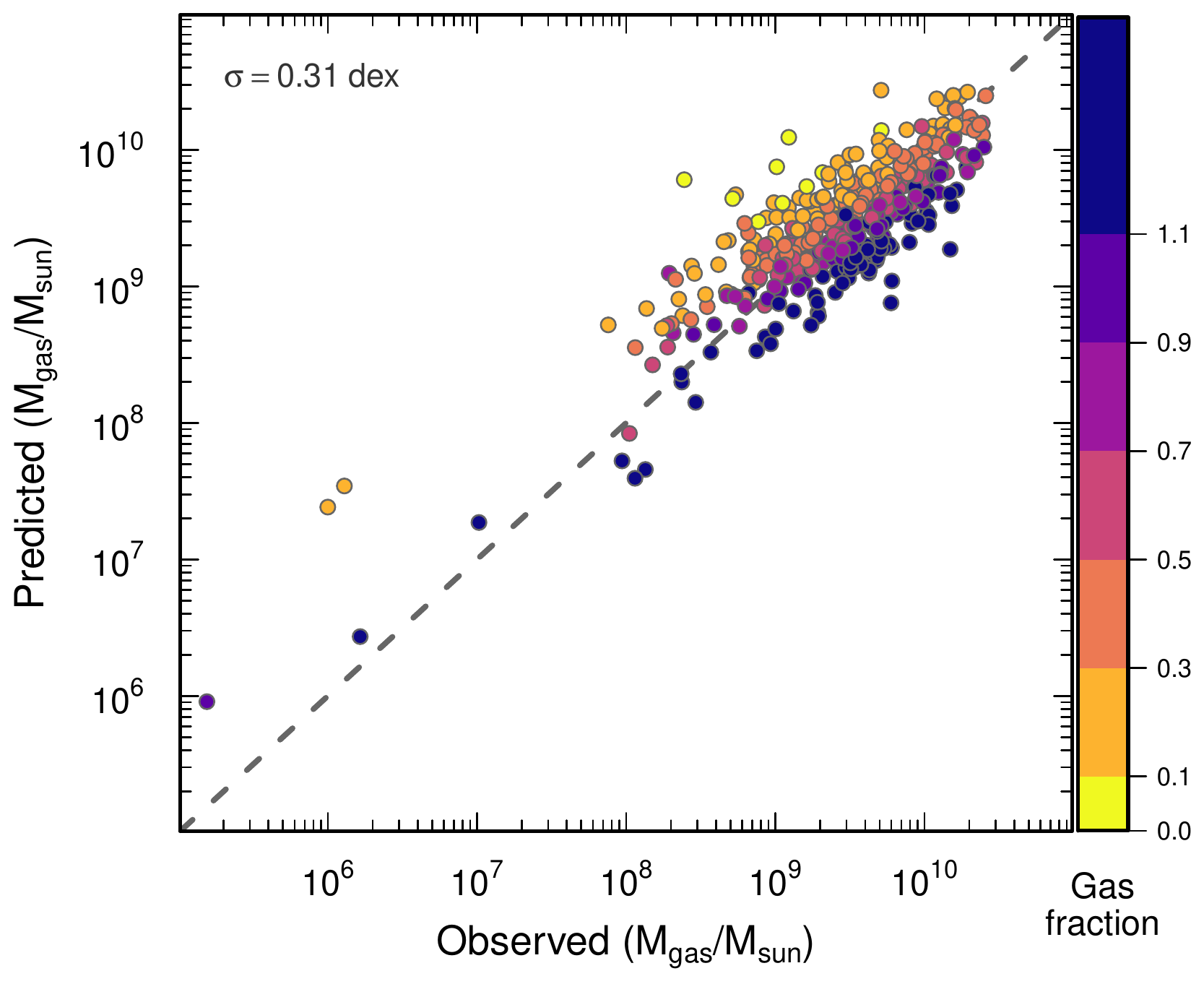} 
\caption{Left panel: Observed values of log(\mhi) in the MAGMA sample compared to those 
predicted as a function of \mstar\ and SFR given by Eqn. (\ref{eqn:regressionmhi})
and shown by a long-dashed line; 
the 1$\sigma$ dispersion of the Gaussian is 0.39\,dex as discussed in the text.
Right panel: Observed values of log(\mgas) in the MAGMA sample compared to those 
predicted by the regression as a function of \mstar\ and SFR and given in Eqn. (\ref{eqn:regressionmgas});
the gray-dashed line corresponds to this regression, and 
the 1$\sigma$ dispersion of the Gaussian is 0.31\,dex as discussed in the text.
Galaxies are colored by gas fraction as shown in previous figures.
\label{fig:mgasvsmstarSFR}
}
\end{figure*}

\subsection{Trends of gas content with \mstar\ and SFR}
\label{sec:gasmstarsfr}

We now explore the implication provided by the PCAs from \pone\ that gas content can
be well described by only \mstar\ and SFR.
As outlined in \pone, the PCAs did not lend themselves to the estimation of gas mass
(\mhtwo, \mhi\, \mgas) by inverting the PCAs.
Here we perform robust fits of gas quantities as a function of \mstar\ and SFR.
We find the following expression for \mhtwo\ as a function of \mstar\ and SFR:
{\small
\begin{eqnarray}
\log{M_{\rm H2}}& = & (0.59\,\pm\,0.04)\ \log(M_{\rm star}) \nonumber \\
	&& \ +\ (0.35\,\pm\,0.04)\ \log({\rm SFR}) + (3.02\,\pm\,0.36) \ .
\label{eqn:regressionmh2}
\end{eqnarray}
}
Figure \ref{fig:mh2vsmstarSFR} shows the comparison of predicted (using Eqn. \ref{eqn:regressionmh2}) and observed values of (log)\mhtwo, with
the right panel illustrating the deviations from the best robust-fit prediction.
The expression in Eqn. (\ref{eqn:regressionmh2})
is accurate to $\sim$0.21\,dex, assessed by fitting a Gaussian to the residuals of the fit, as shown 
in the right panel of Fig. \ref{fig:mh2vsmstarSFR} (the MAD for this fit is 0.22\,dex). 

The analogous regression of \mhi\ vs. \mstar\ and SFR is shown in the left panel of Fig. \ref{fig:mgasvsmstarSFR}, 
with the prediction for \mhi\ along the ordinate taken from the following expression:
{\small
\begin{eqnarray}
\log{M_\mathrm{HI}}& = & (0.41\,\pm\,0.06)\ \log(M_{\rm star}) \nonumber \\
	&& \ + \ (0.35\,\pm\,0.07)\ \log({\rm SFR}) + (5.43\,\pm\,0.59) \ .
\label{eqn:regressionmhi}
\end{eqnarray}
}
The scatter for \mhi\ as a function of \mstar\ and SFR is significantly
larger (0.39\,dex) than that for \mhtwo.
The scatter is the same as that obtained from the piece-wise power-law fit of \mhi\
against \mstar, showing that nothing is gained by introducing SFR as an additional parameter
(although here we find no clear evidence for a break mass).
The color coding in Fig. \ref{fig:mgasvsmstarSFR} suggests that atomic gas content is underestimated
by the regression in Eqn. (\ref{eqn:regressionmhi}) 
in the galaxies that are more gas rich (blue tones), and overestimated in the ones that are more gas poor.
The implication is that in terms of scaling relations \hi\ behaves differently than \htwo;
while \htwo\ apparently depends closely on \mstar\ and SFR,
\hi\ is the more independent variable.
This may be due to the greater variation in the availability of atomic gas in
the gas reservoir within the galaxy, and surrounding it the CGM, and from which the \htwo\ is formed
and then converted into stars. 

\begin{figure*}[!ht]
\begin{minipage}[c]{0.7\textwidth}
\includegraphics[width=\textwidth]{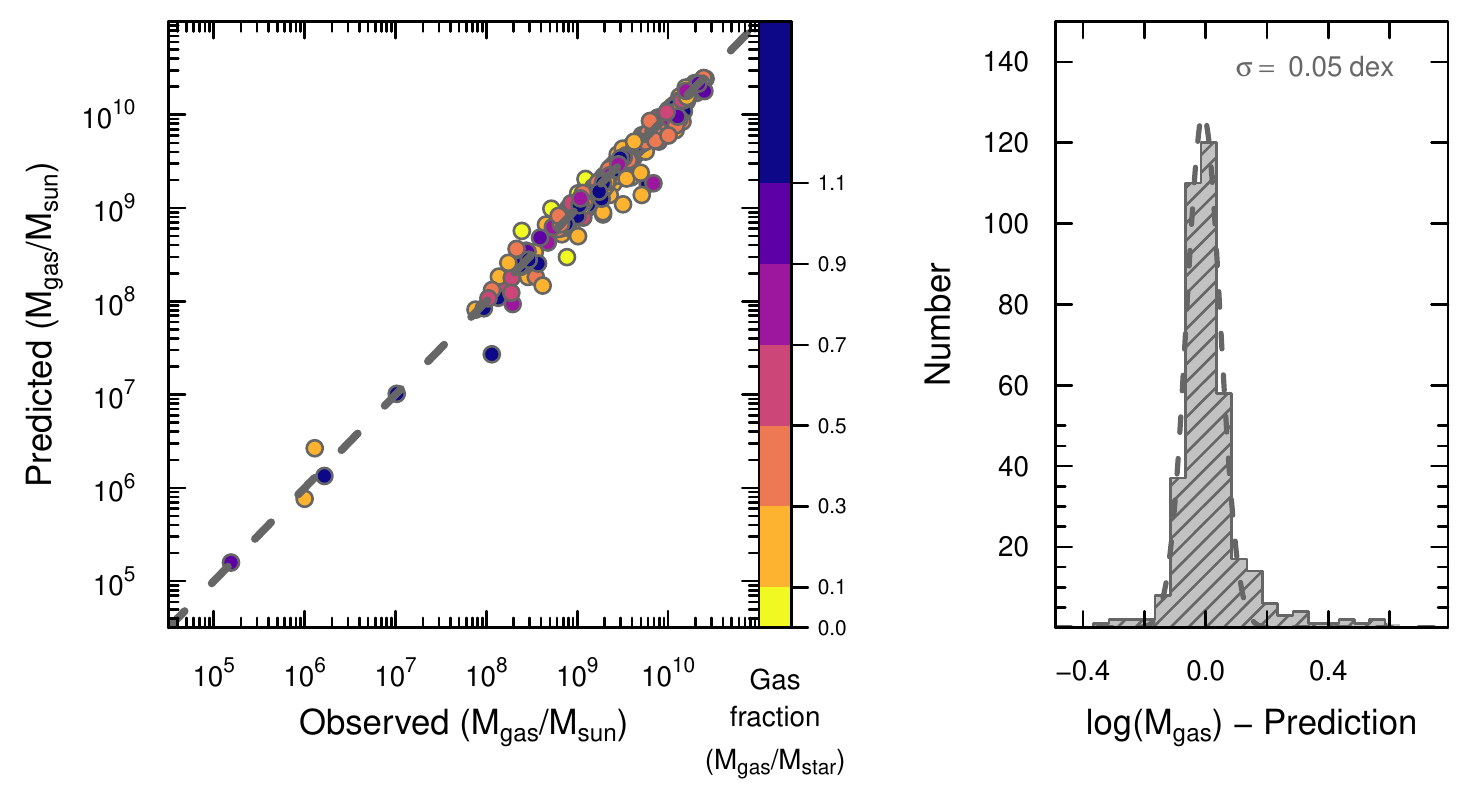}
\end{minipage}
\hspace{0.02\textwidth}
\begin{minipage}[c]{0.26\textwidth}
\caption{Observed values of log(\mgas) in the MAGMA sample compared to those 
predicted by Eqn. (\ref{eqn:regressionmgashi}) as a function of \mstar, SFR, and \mhi; 
the regression is illustrated by the gray-dashed line.
Unlike Fig. \ref{fig:mgasvsmstarSFR} and Eqn. (\ref{eqn:regressionmgas}), here the independent variables include \mhi.
The right panel shows the residuals and the Gaussian fit;
the 1$\sigma$ dispersion of the Gaussian is $\sim$0.05\,dex as discussed in the text.
Galaxies are colored by gas fraction as shown in previous figures.
\label{fig:mgasvsmstarSFRHI}
}
\end{minipage}
\end{figure*}

There have been many previous attempts to quantify \hi\ content in galaxies, with varying
degrees of scatter in the final result.
\citet{kannappan04} found that \mhi/\mstar\ correlates with optical colors with a dispersion
of $\sim$0.4\,dex, similar to the scatter in our approach described above.
Additional parameters such as surface brightness, stellar surface mass density, and hybrid UV-optical colors
improved the scatter to $\sim$0.3\,dex \citep[e.g.,][]{zhang09,li12}.
Axial ratios were also found to reduce dispersion in photometric descriptions of \hi/\mstar\
\citep{eckert15}.
More recent predictions of \hi\ content based on machine learning \citep[e.g.,][]{tei17,rafier18}
have improved the dispersion even more to $\sim$0.2\,dex;
however, these techniques require large training sets with \hi\ detections.
\citet{zu20} applies Bayesian techniques to eliminate the Malmquist bias in \hi-selected surveys,
and is thus able to overcome previous limitations for samples with low \hi\ content;
this approach is able to predict overall \mhi/\mstar\ gas fractions to within 0.27\,dex.
The salient feature of these efforts is that galaxies with blue and red colors behave very
differently in their \hi\ content, and thus to some extent thwarts attempts to derive a robust
\hi\ predictor over a large range in \mstar.
We explore this point further in Sects. \ref{sec:hi} and \ref{sec:discussion}.

For total gas content, \mhi\,$+$\,\mhtwo\,=\,\mgas, the analogous regression is only slightly noisier 
($\sim$0.3\,dex) than that for \mhtwo\ alone:
{\small
\begin{eqnarray}
 \log{M_{\rm gas}}& = & (0.42\,\pm\,0.05)\ \log(M_{\rm star}) \nonumber \\
	&& \ +\  (0.37\,\pm\,0.05)\ \log({\rm SFR}) + (5.43\,\pm\,0.47) \ .
\label{eqn:regressionmgas}
\end{eqnarray}
}

The comparison for \mgas\ is shown graphically in the right panel of Fig. \ref{fig:mgasvsmstarSFR}
with the prediction for \mgas\ given by Eqn. (\ref{eqn:regressionmgas}). 
Although similar qualitatively,
the dependencies on \mstar\ and SFR for \mgas\ are somewhat weaker than for \mhtwo, as a result of the relative
weights of \htwo\ and \hi\ in total gas content (c.f., Eqn. \ref{eqn:regressionmhi}).

Given that \hi\ seems to behave differently, more independently
(that is to say with more scatter) than \htwo\ and with less dependence on SFR (c.f., Eqns.  \ref{eqn:regressionmh2}, \ref{eqn:regressionmhi}), 
we explored the possibility of a regression for \mgas\ including \mhi\ as an
independent variable.
This is important because total baryonic mass is often inferred from \hi\ mass
and stellar mass alone, without including the effects of \htwo.
As shown in Fig. \ref{fig:mgasvsmstarSFRHI},
performing a regression gives an excellent fit, with very low dispersion: $\sigma$\,=\,0.054\,dex as fitted by
a Gaussian as in the right panel of Fig. \ref{fig:mgasvsmstarSFRHI} (MAD $\sigma$\,=\,0.063\,dex). 
The prediction for \mgas\ depicted in the vertical axis is given by the following expression: 
{\small
\begin{eqnarray}
\log{M_{\rm gas}}& = & (0.77\,\pm\,0.01)\ \log(M_\mathrm{HI}) + (0.11\,\pm\,0.01)\ \log(M_{\rm star}) + \nonumber \\
	&& (0.09\,\pm\,0.01)\ \log({\rm SFR}) + (1.16\,\pm\,0.10) \ .
\label{eqn:regressionmgashi}
\end{eqnarray}
}

Figure \ref{fig:mgasvsmstarSFRHI} shows explicitly that once
\mhi\ is included in the expression for \mgas, total gas mass can be determined
to very high accuracy ($\sim$0.05\,dex).
This implies that the three-space of \mstar, SFR, and \mhi\ is not a true
plane; the larger scatter for \mhi\ in the \mstar\,$-$\,SFR regressions
is because \mhi\ cannot be reduced to a simple \mstar\,$-$\,SFR dependence.
Molecular gas mass \mhtwo\ behaves differently because \htwo\ content depends more
strongly on \mstar\ and SFR, so that its amount can be accurately predicted
(to within $\sim$0.2\,dex) using only these variables.

For the MAGMA sample, the mean deviation from \mgas\ estimated from Eqn. (\ref{eqn:regressionmgashi})
and \mhi, typically used to infer total baryonic mass, is 0.13\,dex, $\sim$35\%.
However, the difference between \mhi\ and \mgas\ can be as large as a factor of 3 or more.
The discrepancy between \mhi\ and \mgas\ is larger for those galaxies with small gas fractions, as these are typically
\htwo-dominated.
Thus, for accurate estimates of total gas mass and baryonic content, it is important to consider total gas, rather than only \hi.
Selection effects will also play a role because if galaxies have similar specific SFRs, and a small
range in stellar mass, they will also have similar \htwo\ content (see Eqn. \ref{eqn:regressionmh2}),
and will thus give a tight, but possibly spuriously so, scatter in Tully-Fisher baryonic mass regressions.

The PCA analysis discussed in \pone\ suggests that gas mass in the Local Universe, and in particular \mhtwo, depends on only \mstar\ and SFR.
However, we have found that atomic gas is a more independent variable relative to \htwo, and the degree to which
gas mass can be well predicted using \mstar\ and SFR will depend on whether galaxy samples are
\hi-selected, selected on \mstar, or, like MAGMA, representative of isolated galaxies across a wide range of parameter
space.
In the following, we discuss the relation between the two gas phases, in the context of their relation with star
formation.

\section{The molecular and atomic gas phases across the stellar mass spectrum}
\label{sec:gasphases}

As shown above, the molecular and atomic gas components of dwarf galaxies, spiral disks, and massive
bulge-dominated galaxies differ in their behavior as a function of \mstar\ and SFR.
Here we explore the different gas phases separately,
and then compare them over different \mstar\ regimes. 
We also examine the results in the context of the feedback processes outlined in the Introduction,
namely \textit{(P1)} preventive feedback, the availability of a gas reservoir for
star formation; and \textit{(P2)} star-formation efficiency, the ability to turn
gas into stars.
The first mechanism, preventive feedback \textit{P1}, depends mainly on atomic gas, while the 
second, SFE \textit{P2} is tightly linked with the molecular component.
We first discuss the molecular gas in connection with \textit{P2}, followed by
the atomic gas for \textit{P1}, and how the two gas phases are partitioned in the ISM.

\begin{figure*}[!th]
\centering
\includegraphics[width=\textwidth]{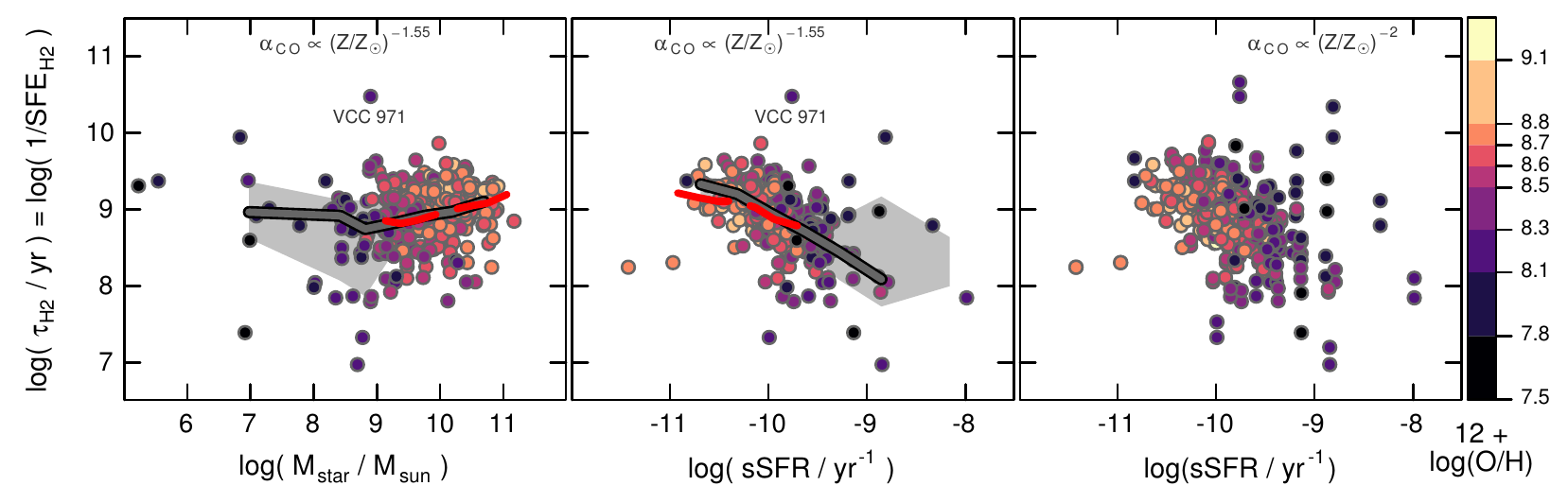}
\caption{Logarithm of \htwo\ depletion time \taudep\ plotted against \mstar\ in the left
panel, and against specific SFR in the middle and right panels.
In the middle and right panels,
the heavy dark-gray curves correspond to the binned data medians;
only bins with 5 or more galaxies are shown.
The gray shadowed regions indicate the range of $\pm\,1\sigma$ of the binned data.
The heavy red long-dashed curves give the xCOLDGASS medians \citep{Saintonge2017}.
The right panel shows the impact of using \aco$\propto (Z/Z_\odot)^{2}$, rather than the value 
we find here.
Galaxies are colored with \logoh\ is as in Fig. \ref{fig:lcosfr}. 
	\label{fig:taudep}
}
\end{figure*}

\subsection{Star-formation efficiency with molecular gas}
\label{sec:sfe}

Because molecular gas fuels the star-formation process,
molecular depletion times are a measure of the timescale of the transformation of gas into stars.
As discussed above, \taudep, the inverse of \sfehtwo, is not necessarily a measure of true efficiency, 
because of the difficulty in translating \sfehtwo\ into the amount of gas converted into stars
over a standard timescale such as free-fall time or dynamical time \citep[e.g.,][]{krumholz09}.
\sfehtwo\ is also inextricably linked to cloud lifetimes in a typical galaxy environment \citep{krumholz07},
and the global measurements for MAGMA do not constrain this quantity.
Ultimately, \taudep\ and \sfehtwo\ signify the rapidity of star formation from a given gas reservoir,
which depends on the mean lifetime of star-forming clouds and the true efficiency of star formation.
Here we explore with MAGMA global timescales for star formation and how these change with
various quantities. 

Molecular depletion times for the MAGMA sample, computed as in Eqn. (\ref{eqn:taudep}) 
with \aco\ according to Eqn. (\ref{eqn:alphaco}), are shown in Fig. \ref{fig:taudep}.
\htwo\ depletion times depend only very weakly on \mstar, as illustrated in the left panel. 
The MAGMA sample extends the trend beyond \mstar$\la 10^9$\,\msun.
Although there is a weak trend for increasing \taudep\ above this limit, in agreement with previous work
\citep[e.g.,][]{leroy13,Boselli2014b,Saintonge2017}, 
below this mass, there is no clear trend. 
There are some galaxies with very short depletion times ($\sim 10^7$\,yr), but the scatter is large.
Further molecular-gas observations of very low-mass galaxies are needed to confirm the trend with \mstar.

Unlike the weak trend of \taudep\ with \mstar, \taudep\ is strongly correlated with sSFR, 
as shown in the middle panel of Fig. \ref{fig:taudep}. 
For sSFR$\,\la\,10^{-11}$\,yr$^{-1}$ and sSFR$\,\ga\,10^{-9}$\,yr$^{-1}$, there are too few MAGMA galaxies 
to constrain \taudep.
However, within these ranges, over two orders of magnitude in sSFR, shorter molecular depletion times \taudep\
are associated with increasing sSFR (see also Sect. \ref{sec:alphaco}). 
There is also relatively small scatter associated with this fit, $\sim$0.2\,dex.
For MAGMA galaxies, \taudep\ ranges from $\sim 10^8$\,yr for sSFR $\sim 10^{-8}$\,yr$^{-1}$ 
to $\sim 3\times10^9$\,yr for sSFR $\sim 10^{-11}$\,yr$^{-1}$.
The agreement with xCOLDGASS \citep{Saintonge2017} is excellent, as shown by the heavy
red long-dashed lines in Fig. \ref{fig:taudep};
the extension of MAGMA to almost two orders of magnitude higher sSFR is also apparent in the comparison.
Such a general trend is in agreement with other previous work 
\citep[e.g.,][]{leroy13,Bothwell2014,Boselli2014b,Huang2014,huang15,genzel15,Catinella2018},
although the specific formulation depends on the form of \aco, and also on the methods for calculating \mstar\ and SFR.
This agreement is perhaps not surprising given that MAGMA comprises samples from these studies
(ALLSMOG, HRS, xCOLDGASS).
The middle and right panels in Fig. \ref{fig:taudep} compare the depletion times
(and implicitly the \htwo\ mass) calculated with the two power-law slopes for the \aco\ metallicity dependence.
The steeper slope shown in the right panel results in a noisier data distribution, with many more
outliers relative to the middle panel.

MAGMA extends the previous trends of \taudep\ and sSFR to higher sSFR and lower \mstar\
than previous samples. 
However, the MAGMA selection criteria require detections in CO (and \hi), thus systems with relatively high SFR. 
This consideration is particularly relevant for low \mstar\ and may influence the trends of \taudep.
Nevertheless, the extension toward higher sSFR seems to be roughly a continuation of the lower sSFR regime 
\citep[e.g.,][]{Saintonge2011b,Saintonge2017}.
Although caution is warranted, and new observations at low \mstar\
are needed for confirmation, \taudep\ is fairly tightly correlated with sSFR in galaxies, 
at least over the mass regime probed by MAGMA.

\begin{figure*}[!ht]
\centering
\includegraphics[width=0.85\textwidth]{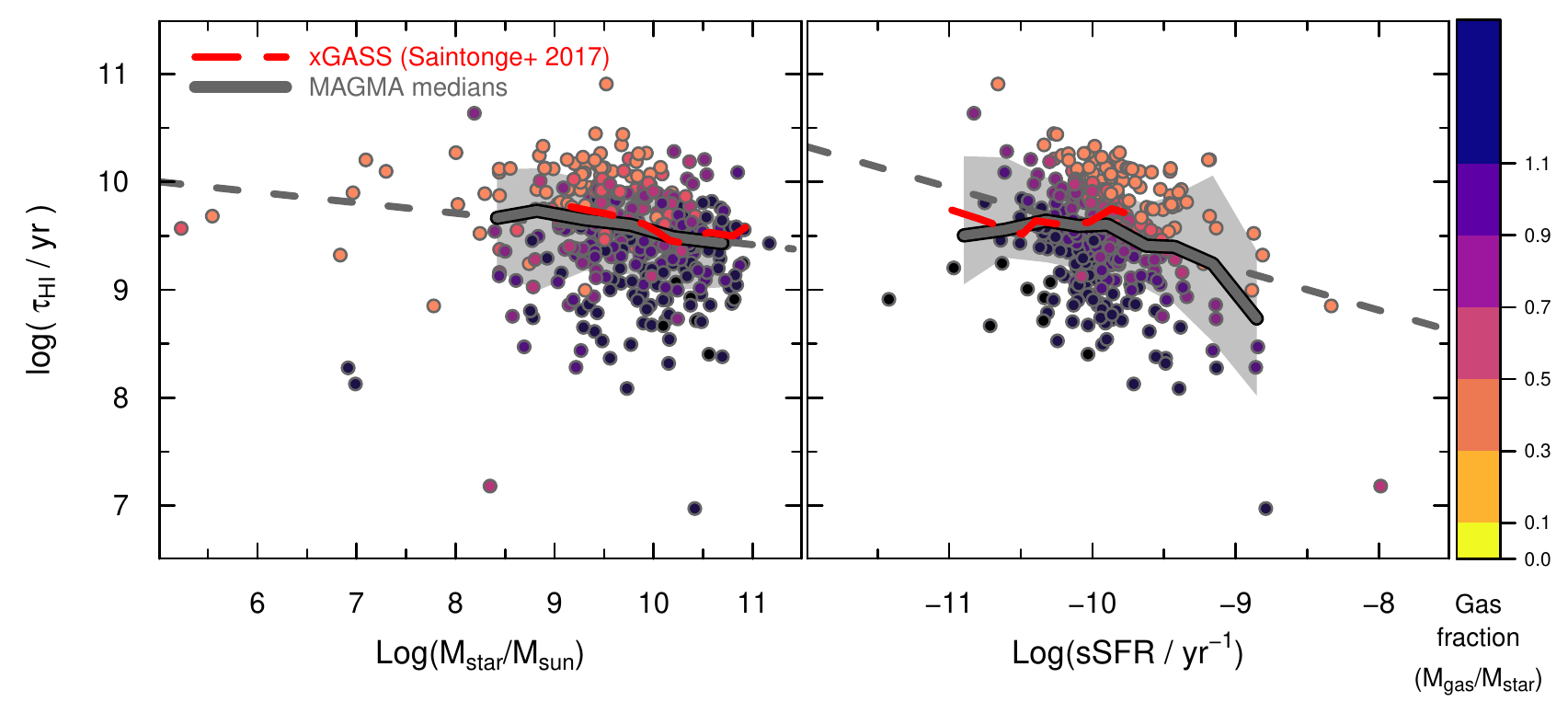} 
\caption{Atomic gas depletion time, \taudephi, plotted against \mstar\ in the
left panel, and against sSFR in the right (both in logarithmic units).
The best-fit power-law regressions are shown by short-dashed gray lines in both panels,
and the MAGMA medians by a heavy curve;
the gray regions correspond to $\pm\,1\sigma$ of the data.
The heavy red long-dashed curves give the xCOLDGASS medians \citep{Saintonge2017}.
As in Fig. \ref{fig:magma_ms}, galaxies are colored by gas fraction.
\label{fig:taudephi}
}
\end{figure*}

If the trend of \taudep\ with sSFR is taken at face value, we 
would conclude that star formation is more ``efficient'' at high sSFR, and thus also
at low metallicity (notice color coding, see also Fig. \ref{fig:magma_ms}). 
Thus if we examine the feedback process (\textit{P2}),
namely the efficiency with which gas is converted to stars,
there is an apparent contradiction with the low values of \sfedmh\ at low halo (and stellar) mass:
low-mass metal-poor galaxies with larger sSFR than in more metal-rich and massive galaxies,
are more, not less, efficient at forming stars. 
This illustrates the conundrum of associating low \sfedmh\ with a low \sfehtwo;
they are not directly connected, on the contrary.
The first is a process of hierarchical growth, and the second is a baryonic process, the
timescales of which are almost certainly different.
Hierarchical timescales depend on the availability of baryons and the accretion of (\hi) gas from the CGM,
while SFR timescales are related to the conversion of \htwo\ to stars.
The missing link here is the atomic gas content and the relation between \hi\ and \htwo,
and we examine both in the next two sections.

\subsection{Depletion times of atomic gas}
\label{sec:hi}

The availability of gas is the basic requirement for star formation,
and is directly related to the preventive feedback mechanism \textit{(P1)}
alluded to in the Introduction.
Although atomic gas does not contribute directly to star formation 
\citep[but see][]{krumholz12}, \hi\ plays a role in
regulating star formation through total gas column density.
Several theoretical arguments suggest that the conversion of \hi\ to \htwo\
is driven by total gas surface density because of the need for self-shielding \citep[e.g.,][]{wolfire95,krumholz09}, 
or the requirement of thermal/dynamical  or hydrostatic equilibrium
\citep{ostriker10,krumholz13b}. 
Thus, \hi\ content can be considered as fundamental for setting the stage for
star formation in a galaxy. 
Moreover, as shown in the previous Section, \hi\
behaves differently from \htwo, seemingly a more independent parameter in the
description of a galaxy.

Atomic gas depletion times \taudephi\ (\taudephi\,$\equiv$\,\mhi/SFR)
are illustrated in Fig. \ref{fig:taudephi}.
The trend of \taudephi\ with \mstar\ is shown in the left panel, and agrees with
much previous work showing little, if any, dependence on \mstar\
\citep[e.g.,][]{schiminovich10,Boselli2014b,Saintonge2017}.
The mean \taudephi\ of 3.2\,Gyr for MAGMA galaxies is also in excellent agreement 
with these previous estimates.
Interestingly, \hi-selected samples such as ALFALFA have a mean \taudephi\
almost three times higher, $\sim$9\,Gyr \citep{Huang2012}.
Although the dependence of both \taudep\ and \taudephi\ on \mstar\ is weak,
comparing Figs. \ref{fig:taudep} and \ref{fig:taudephi} suggest opposite
trends: for \htwo, depletion times are increasing with \mstar,
while for \hi, they are (slightly) decreasing.


In apparent contrast with previous studies,
here we find a clear dependence of \taudephi\ on sSFR, as illustrated 
in the right panel of Fig. \ref{fig:taudephi}. 
This again is due to the extension of MAGMA to lower \mstar\ and thus higher sSFR.
For MAGMA galaxies, \taudephi\ is $\sim 3\,\times\,10^8$\,yr for sSFR $\sim 10^{-9}$\,yr$^{-1}$ 
increasing to $\sim 10^{10}$\,yr for sSFR $\sim 10^{-11}$\,yr$^{-1}$.
The shortest \hi\ depletion times at large sSFR are comparable to \taudep,
but at small sSFR, the longest \taudephi\ are more than 10 times larger.
The scatter for the robust fit of \taudephi\ vs. sSFR is also twice as large (0.44\,dex) 
than that for \taudep\ (0.22\,dex).
This large degree of scatter suggests a weaker trend for \taudephi\ than for \taudep\,
and the need for more data in this high sSFR, low-\mstar\ regime.

\citet{Saintonge2017} found that \hi\ depletion times were roughly constant with both
\mstar\ and sSFR, while here, with the extension to higher sSFRs provided by MAGMA,
we find a significant trend.
The agreement is good for the range of sSFR in common, but the higher sSFRs probed
by MAGMA suggest a significant decrease down to sSFR $\la\,10^{-9}$\,yr$^{-1}$.
At a given sSFR (or \mstar), 
the most gas-rich galaxies (colored by dark blue) tend to have shorter \taudephi\
than the gas-poor ones.
Nevertheless, the trend is only hinted at by the relatively few galaxies in this extreme 
high sSFR, low-mass regime; more data are needed to confirm it.

\begin{figure*}[!ht]
\centering
\includegraphics[width=0.85\linewidth]{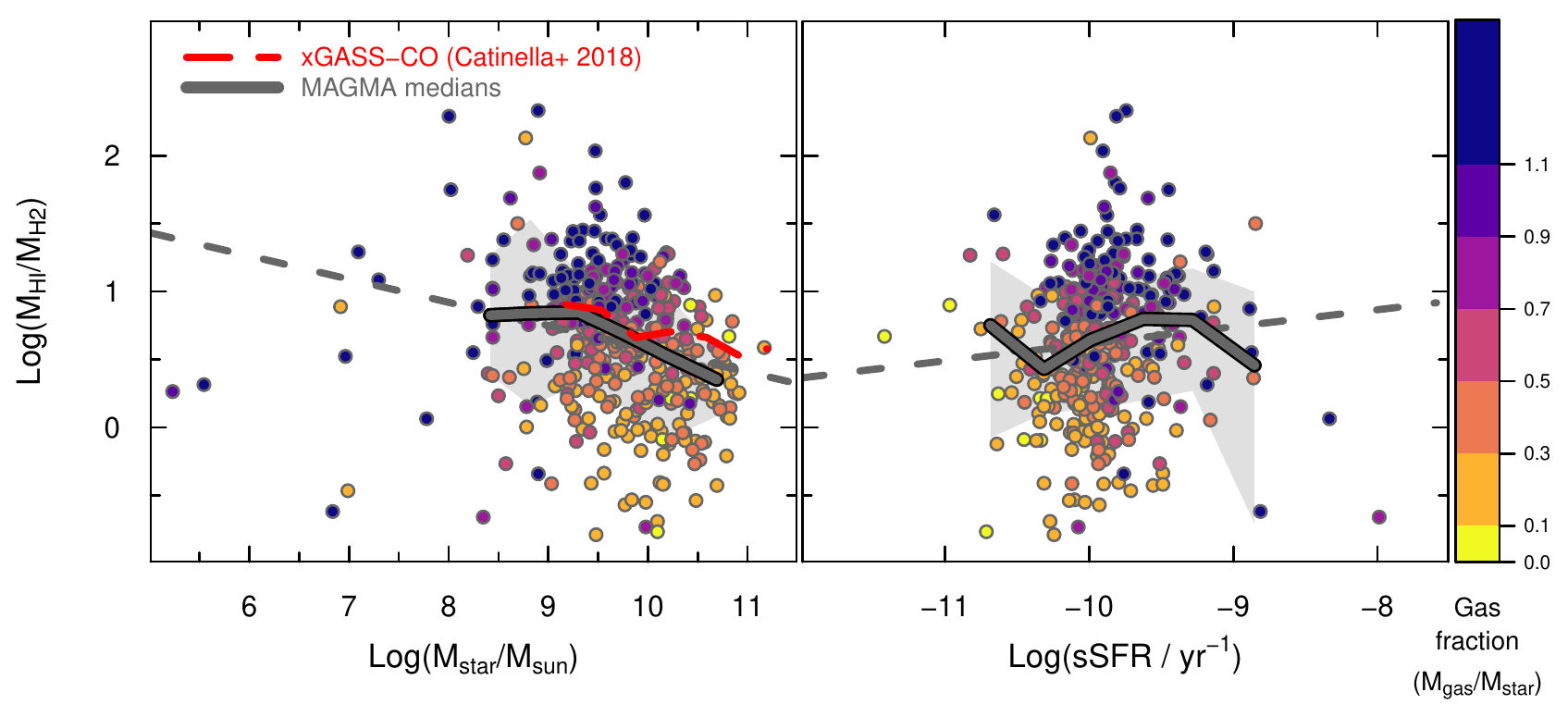}
\caption{Mass ratio of atomic-to-molecular gas plotted vs. \mstar\ in the left panel,
and vs. sSFR in the right (both in logarithmic units).
As in Fig. \ref{fig:taudephi}, the long-dashed lines correspond to the best-fit
power-law regressions, and the heavy curve to the MAGMA medians.
The gray regions show the $\pm\,1\sigma$ excursions of the data.
The heavy red long-dashed curve corresponds to the binned data from xGASS-CO
\citep[see][]{Catinella2018}.
Galaxies are colored by gas fraction as in Fig. \ref{fig:magma_ms}. 
\label{fig:h2hi_mstarssfr}
}
\end{figure*}

\subsection{The relation between molecular and atomic gas}
\label{sec:h2hi}

Understanding how the total gas is partitioned into molecular and atomic phases, and the efficiency
of the conversion of \hi\ to \htwo,
requires measurements of the \hi/\htwo\ ratio over a broad range in \mstar.
Such measurements are crucial to quantify the physical mechanisms behind star formation and baryonic cycling in galaxies. 
The results presented above suggest that both \hi\ and \htwo\ are involved,
albeit to different degrees, in the regulation of star formation.
However, previous work \citep[e.g.,][]{Bothwell2014,Boselli2014b,Catinella2018}, 
including studies of resolved galaxies \citep[e.g.,][]{Bigiel2008,leroy13}
and theoretical models \citep[e.g.,][]{krumholz11}, 
has established that molecular gas is more closely associated than the atomic component with the star formation process
\citep[although see][]{wang20}.
To understand this distinction, here, with MAGMA, 
we examine the \hi/\htwo\ ratio and the scaling across different regimes in stellar mass and SFR.

The ratios of \mhi\ and \mhtwo\ are plotted in Fig. \ref{fig:h2hi_mstarssfr}; the left panel
shows the comparison with \mstar\ and the right panel with sSFR.
In agreement with previous work \citep[e.g.,][]{Bothwell2014,Boselli2014b,Catinella2018},
\mhi/\mhtwo\ decreases with increasing \mstar, and shows little, if any, variation with sSFR.
The best robust fit for \mhi/\mhtwo\ against \mstar\ shown in Fig. \ref{fig:h2hi_mstarssfr} is given by:
{\small
\begin{equation}
\log(M_{\rm HI}/M_{\rm H2})\,=\,(-0.17\,\pm\,0.03)\,\log(M_{\rm star}/10^9) + (0.75\,\pm\,0.03) \ .
\label{eqn:hih2}
\end{equation}
}
Although the parameters of the fit are determined with fairly high accuracy,
there is much scatter, with a best-fit MAD of 0.5\,dex.
The galaxies with the lowest gas fractions \fgas\ ($\equiv$\,\mgas/\mstar, see Fig. \ref{fig:magma_ms}) can have 
\mhi/\mhtwo\ less than unity, but otherwise atomic gas is always the dominant component.
The median \mhi/\mhtwo\ ratio exceeds unity over the entire \mstar\ regime probed by MAGMA 
even for the most massive galaxies; for \mstar$\,$\ga$\,5\times10^{10}$\,\msun,
on average, \mhi/\mhtwo\ $\sim$ 2.
For galaxies 20 times less massive,  \mstar\,$\sim\,3\times10^{9}$\,\msun,
the average ratio is more than 3 times higher, \mhi/\mhtwo\ $\sim$ 7.
In the most gas-rich galaxies,
\mhi/\mhtwo\ can be as high as 10 or more, and can exceed values of 30 in some cases. 

\begin{figure*}[!ht]
\includegraphics[width=0.48\textwidth]{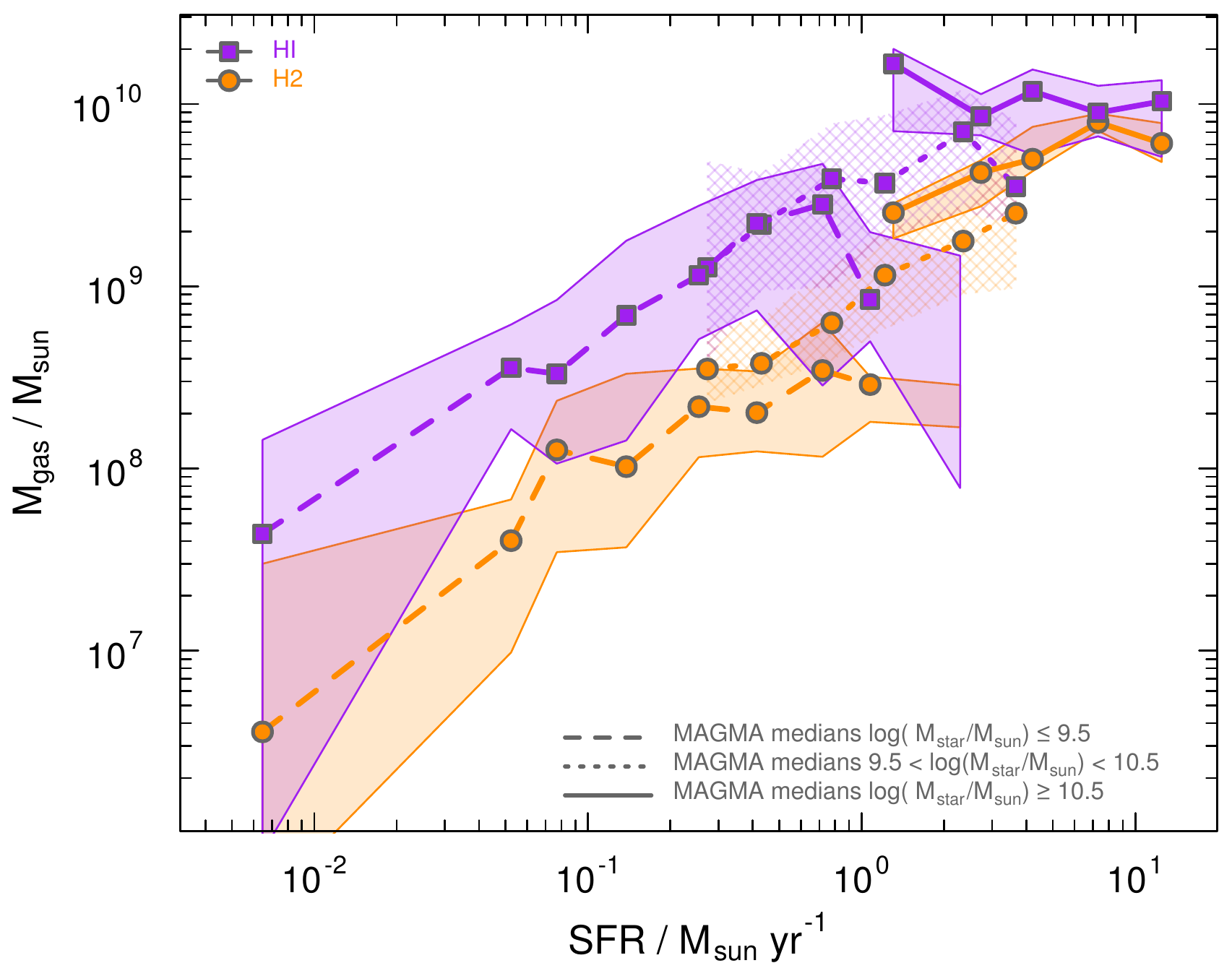} 
\hspace{0.02\textwidth}
\includegraphics[width=0.48\textwidth]{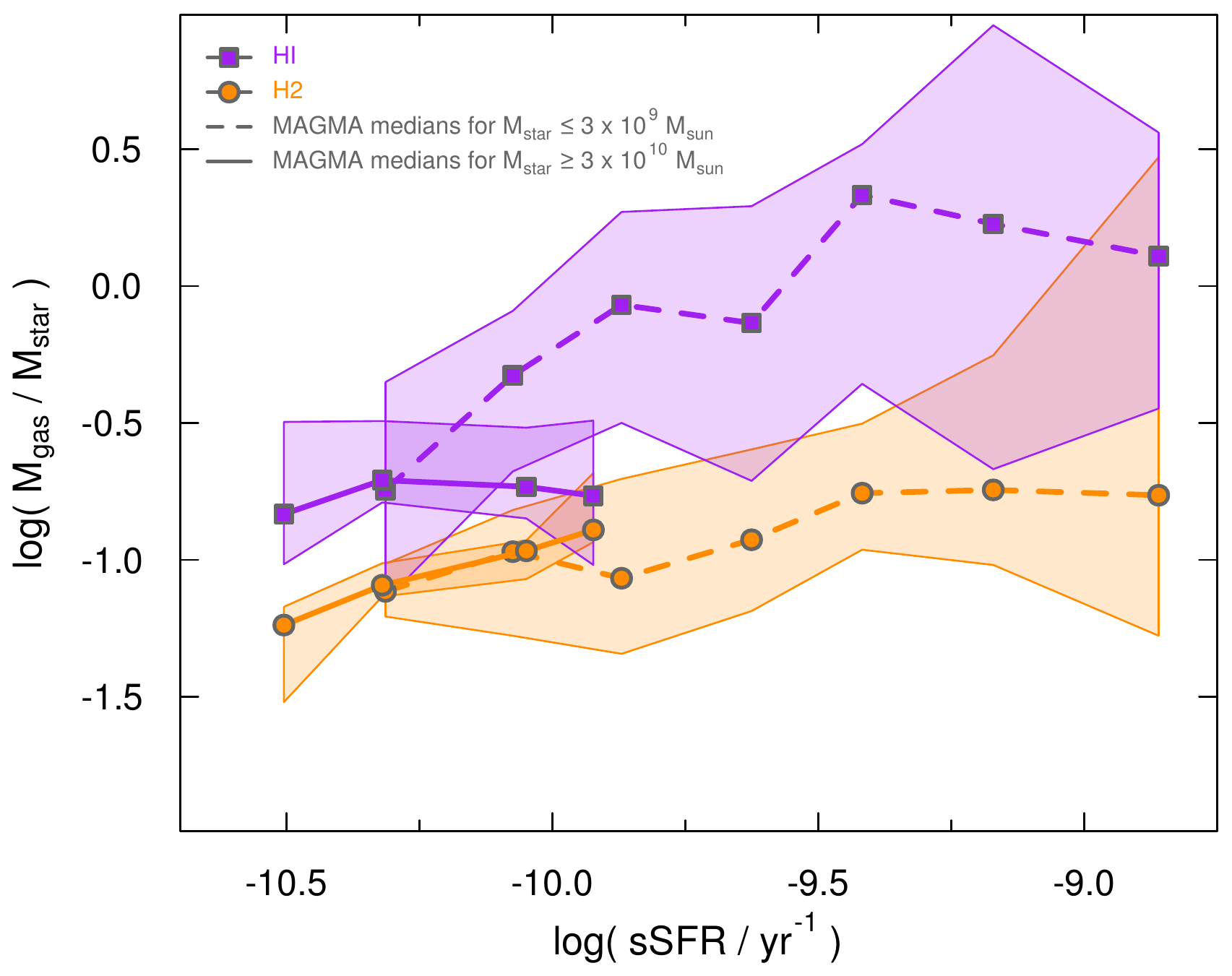} 
\caption{Gas mass plotted against SFR (left) and ratio of gas mass and stellar mass plotted against sSFR (right, in logarithmic units).
In both panels,
the MAGMA sample has been divided into three \mstar\ bins 
(\mstar\,$\leq 3\times10^{9}$\,\msun, 143 galaxies; $3\times10^9$\,\msun\,$<$\,\mstar\,$<\,3\times10^{10}$\,\msun, 210 galaxies;
\mstar\,$\geq 3\times10^{10}$\,\msun, 39 galaxies).
The binned medians for \mhtwo\ are shown by orange circles and \mhi\ by purple squares;
only bins with number of galaxies $\geq$4 are plotted.
The low-\mstar\ bins are connected with dashed lines, the high-mass bins with solid ones,
and the intermediate bin with dot-dashed;
the hatched area shows the spreads of the intermediate \mstar\ regime.
The shaded regions correspond to $\pm\,1\sigma$ spreads of the binned parameters.
In the right panel, only the highest and lowest \mstar\ bins are shown.
\label{fig:h2hibin_sfr}
}
\end{figure*}

That even galaxies above the characteristic Schechter mass \mstar\,$\sim 3-4\times10^{10}$\,\msun\
\citep[e.g.,][]{schechter76,baldry12}
have a significant atomic gas reservoir is not a new result.
Recently \citet{zhang19} claimed that even ``quiescent'' massive disk galaxies
have an \hi\ content comparable to their star-forming counterparts.
However, \citet{cortese20} countered this claim by analyzing the SFRs used by \citet{zhang19}.
They found that the SFRs used by \citet{zhang19} were systematically low,
thus the galaxies they considered quiescent were in fact star forming.
Either way, the massive galaxies in MAGMA \citep[and previously xGASS, xCOLDGASS:][]{Catinella2018,Saintonge2017} 
clearly host significant amounts of gas, both \hi\ and \htwo.
The question is: how does the partition of \hi\ and \htwo\ change with SFR?

To answer this, we have divided MAGMA into three stellar mass bins:
one corresponding to the galaxies having \mstar\ below the \hi\ break of \mstar\,$\sim 3\times10^{9}$\,\msun\ shown in Fig. \ref{fig:hih2},
one with \mstar\ above the characteristic Schechter mass of \mstar\,$\sim 3-4\times10^{10}$\,\msun,
and one between these two limits.
The \mstar\ value of $3-4\times10^{10}$\,\msun\ is significant not only for the local stellar-mass function
\citep[usually fit by a Schechter function,][]{schechter76,baldry12}, but also for many other
physical processes which we discuss in Sect. \ref{sec:discussion}.

Figure \ref{fig:h2hibin_sfr} shows gas mass plotted against SFR for the three mass bins (left panel),
and for the two extreme bins, \mgas/\mstar\ vs. sSFR (right).
The high-mass regime clearly differs from the low-mass one.
The upper portion of the left panel illustrates the \hi\ and \htwo\ masses for the massive galaxies;
\mhi\ does not vary with SFR over this \mstar\ bin, while \mhtwo\ increases with increasing SFR.
This behavior at the high-mass end is in excellent agreement with the trends shown by \citet{zhang19}, and
suggests that the \hi\ in these massive systems does not participate directly in the
star-formation process.

In contrast, \mhi\ and \mhtwo\ in the low-mass bin behave similarly to one another; both show a marked increase
in gas content with increasing SFR although, at relatively high SFRs (SFR\,$\ga$\,0.6\,\msunyr), 
both gas components in the low-mass bin show a downturn.
Instead, at similar SFRs, in the intermediate-\mstar\ bin both the atomic and molecular phases are monotonically increasing with SFR. 
The implication is that at intermediate masses, both gas phases are involved in ongoing episodes of star formation, 
while in low-mass dwarf galaxies \hi\ the star formation is unable to keep up with the amount of gas.

The right panel of Fig. \ref{fig:h2hibin_sfr} instead shows \mgas/\mstar\ plotted against sSFR
(here in logarithmic scale).
Again, the high-\mstar\ bin shows that \hi\ is disconnected from the star-formation process,
as there is little variation in \mhi/\mstar\ over a factor of 3 change in sSFR.
However, the molecular component, relative to stellar mass, increases systematically with sSFR,
again suggesting that at these masses \htwo\ is directly fueling SFR.
For low-mass galaxies, the ratio of \mhtwo\ and \mstar\ is relatively constant with sSFR, while
the analogous ratio for \hi\ shows a fairly constant rise, up to the highest sSFRs probed by MAGMA.
This is yet another indication that at low \mstar, atomic gas is 
necessary ingredient for star formation to proceed and 
participating (at least indirectly) in the process of star formation.


\section{Three mass regimes in galaxy evolution}
\label{sec:discussion}

Based on the NFGS, \citet{Kannappan2013} identified two transitions in
galaxy morphology, gas fractions, and fueling regimes.
As seen above, the MAGMA sample also shows evidence for transitions at low 
and high \mstar\ (see e.g., Fig. \ref{fig:h2hibin_sfr}),
and here we explore the resulting ramifications for galaxy evolution.
The \htwo\ and \hi\ detections for MAGMA enable us to probe the total gas content, and its relation
with star-formation properties; \citet{Kannappan2013} had only few galaxies with \htwo\ (CO) measurements, so were
impeded by the lack of data for \mhtwo\ and thus total \mgas. 
In addition, there were very few galaxies in \citet{Kannappan2013} with \mstar$\la 3\times10^9$\,\msun.
Thus, MAGMA allows a more detailed study of the behavior of low-mass galaxies.

Before discussing in detail changes in gas properties
over the different mass regimes, we need to consider whether or not the
\hi\ and \htwo\ contents in MAGMA galaxies could be subject to a systematic bias.
Our \hi\ (and \htwo) measurements are mainly global measurements \citep[although see][]{hunt15},
so it is virtually impossible to identify the location of the gas components within the
galaxy.
A bias could occur if the \hi\ in low-\mstar\ galaxies were to lie predominantly
outside the galaxy, while in more massive galaxies to be more closely confined to
the stellar star-forming disk, similarly to \htwo. 
To test this, we have adopted the approach of \citet{wang20} who deduced the
amount of \hi\ within the disks of xGASS galaxies and compared it to \htwo\ content.
Our analysis is described in detail in Appendix \ref{app:hiradius}.
The results suggest that although much of the \hi\ gas falls outside the stellar disk,
as expected and in agreement with \citet{wang20},
there is no systematic variation with stellar mass.
Thus, the \hi\ gas in low-mass galaxies is spatially configured in a similar way as the \hi\
in high-mass systems, and no bias should be introduced in our analysis of \htwo\
and \hi\ gas properties for this reason.

\subsection{Accretion-dominated, low-mass galaxies}
\label{sec:lowmass}

The first transition noted by \citet{Kannappan2013} was considered a ``gas-richness'' threshold;
galaxies below a transition stellar mass of log(\mstar/\msun)$\sim$9.5 were found to be ``accretion dominated'',
characterized by overwhelming gas accretion.
In the MAGMA sample, this mass scale emerges as a break in the trend of \mhi\ and \mstar\
(see Sect. \ref{sec:gasscaling} and Fig. \ref{fig:hih2}).
This same low-mass break also emerges in the MAGMA galaxies as a gas-accretion threshold signature for effective 
metallicity yields \citep[e.g.,][]{garnett02,dalcanton07}. 
We explore the relation
between gas content and metallicity in a companion paper (Tortora et al. 2020, in prep.)

Fig. \ref{fig:h2hibin_sfr} shows that for low-mass MAGMA galaxies, 
\mhtwo/\mstar\ remains relatively constant over a factor of 10 increase in sSFR.
In contrast, \mhi/\mstar\ increases with sSFR$\sim 1.5\times10^{-10}$\,yr$^{-1}$,
remains flat until sSFR$\sim 3\times10^{-10}$\,yr$^{-1}$, then increases
with sSFR up to the highest sSFRs ($\sim 2\times10^{-8}$\,yr$^{-1}$) probed by MAGMA.
We interpret this as a confirmation of the scenario proposed by \citet{Kannappan2013},
namely that these low-mass galaxies with high sSFR are accreting
gas faster than they can process it.
High sSFR means that these systems are forming stars rapidly,
but does not signify much in terms of their star-forming \text{efficiency}.
This is particularly true for the long timescales, and thus low ``efficiency'', implied by the relatively large \hi\ depletion times 
(see Fig. \ref{fig:taudephi}).
Indeed, as stated by \citet{Kannappan2013}:
``if the gas is pouring in faster than even the highest-sSFR galaxies can possibly consume it, 
such galaxies might be better termed `overwhelmed' rather than `inefficient'".

Central to the notion of ``overwhelmed'' is the idea that there may be a limit to how fast, 
and how efficiently, gas can be converted into stars.
Limitations governed by the interplay between pressure and turbulence in galaxy disks
may be impossible to overcome.
In a volume-limited sample of galaxies, \citet{karach13} noticed that there is a maximum sSFR, 
log(sSFR/yr$^{-1}$)\,$\sim\,-9.4$, above which there are very few galaxies.
A similar sSFR limit was found by \citet{schiminovich10} in a sample of \hi-selected galaxies from ALFALFA.
Because of the presence of low-mass starbursts, MAGMA extends
the range of maximum sSFR values to somewhat higher values (see Fig. \ref{fig:magma_ms}),
but the general trend toward lower \mstar\ (and higher sSFR) is very shallow.
Although definitely not a typical stellar disk,
even the prototypical ultraluminous infrared galaxy, Arp\,220, 
has an sSFR comparable to the highest values in MAGMA with log(sSFR/yr$^{-1}$)\,$\sim\,-8.6$ 
\citep{u13}\footnote{Arp\,220 has a stellar mass of $\sim 8\times10^{10}$\,\msun\ \citep{u13}.}.

A possible explanation for such a limit is furnished by theoretical arguments
that propose that star formation in disk galaxies is regulated through stellar feedback and turbulence 
\citep[e.g.,][]{ostriker10,ostriker11,hayward17}.
Although ``maximum intensity starbursts'' are usually defined in terms of surface brightness 
\citep[e.g.,][]{meurer97,hathi08,hopkins10}, sSFR is a closely-related proxy. 
Thus, if star formation is truly regulated by internal feedback and turbulence as proposed theoretically,
given copious gas accretion, it may not be possible for galaxies to convert the accreted gas
into stars more rapidly than these limitations allow.

\subsection{Gas-poor, high-mass galaxies}
\label{sec:highmass}

The second transition identified by \citet{Kannappan2013} corresponds to a bimodality threshold,
roughly indicating a ``quenched'' regime at log(\mstar/\msun)$\,\ga\,$10.5, populated in their sample by 
spheroid-dominated, gas-poor galaxies with low rates of star formation over the last Gyr.
This \mstar\ threshold also signifies several transitions in galaxy demographics:
a signature of bimodality in galaxy color
\citep[the ``blue cloud'' to the ``red sequence'', e.g.,][]{kauffmann03,nelson18};
a transition in morphology from disk-dominated, low surface brightness systems
to centrally concentrated, high S\'ersic index, and high surface brightness 
\citep[e.g.,][]{bundy05,driver06};
a break in the galaxy stellar-mass function at $z \sim 0$ \citep[e.g.,][]{baldry12}; 
an inflection in the galaxy SFR main sequence at 
low redshift \citep[e.g.,][]{whitaker12,whitaker14,lee15,Saintonge2016}\footnote{The break mass 
in the SFMS is not always found strictly to be the characteristic log(\mstar/\msun)\,=\,10.5, but this
depends on methods for definition of \mstar\ and SFR.};
and a break in the mass-metallicity relation \citep[e.g.,][]{tremonti04}.
A minimum in the dark matter fraction and the steepest galaxy mass profiles 
and population gradients are also found at this \mstar\ threshold
\citep[e.g.,][]{tortora10,tortora19}.

Interestingly, this \mstar\ transition also corresponds to changes in gas-accretion modes with ``cold accretion'' thought
to dominate in those galaxies with \mstar\ below this threshold.
Beyond this, a ``hot mode'' is expected to take over, induced by shock heating of infalling gas to the virial temperature in massive haloes 
\citep[e.g.,][]{birnboim03,keres05,dekelbirnboim06,keres09}.
Such a coincidence in transition masses for this wide variety of physical processes
and observational trends is telling us something about changes in the
way that galaxy evolution is regulated above this mass scale,
relative to less-massive systems.

For galaxies above the high-\mstar\ threshold,
in MAGMA we find that
\mhtwo/\mstar\ increases with both SFR and sSFR, while \mhi/\mstar\ remains constant
over a factor of 10 in SFR and a factor of $\sim$3 in sSFR.
This implies that \hi\ is not participating in the current star-formation process at all, but
rather is a potential reservoir for future conversion into stars, or a past relic of gas accretion
that will remain ``quiescent'' over the galaxy's lifetime.

\subsection{Gas-equilibrium, intermediate-mass galaxies}
\label{sec:intermediatemass}

Galaxies in the intermediate mass range with \mstar\ between $3\times10^9$ and $3\times10^{10}$\,\msun\
were considered by \citet{Kannappan2013} to be in a ``processing-dominated'' regime.
These galaxies are able to consume gas through star formation as fast as it is accreted,
and in this \mstar\ range gas accretion and star formation proceed together \citep{fraternali12}.
Metal-enriched gas outflows driven by feedback from SNe and massive stellar winds tend
to mix with the surrounding hot gas in galaxy coronae, thus enabling cooling
and subsequent cold-mode accretion \citep{marinacci10,armillotta16}. 

In the intermediate-mass or ``gas-equilibrium'' regime, 
Fig. \ref{fig:h2hibin_sfr} for MAGMA implies that the (\hi) accretion rate has slowed somewhat,
because the divarication in \mhi/\mstar\ and \mhtwo/\mstar\ is narrowing as SFR increases.
Thus, it seems that the galaxies are able to form stars in pace with the arrival of incoming gas,
in agreement with the results of \citet{fraternali12} and \citet{tacchella16}.
Such behavior in this mass regime would be consistent with the gas-regulator models proposed by \citet{Lilly2013}
and others.
Gas that comes in as accretion is processed by star formation and goes out as stellar winds,
maintaining an ``equilibrium'' configuration expected to endure over most of cosmic time \citep[e.g.,][]{Dave2012}.

\begin{figure}[!ht]
\centering
\includegraphics[width=0.92\linewidth]{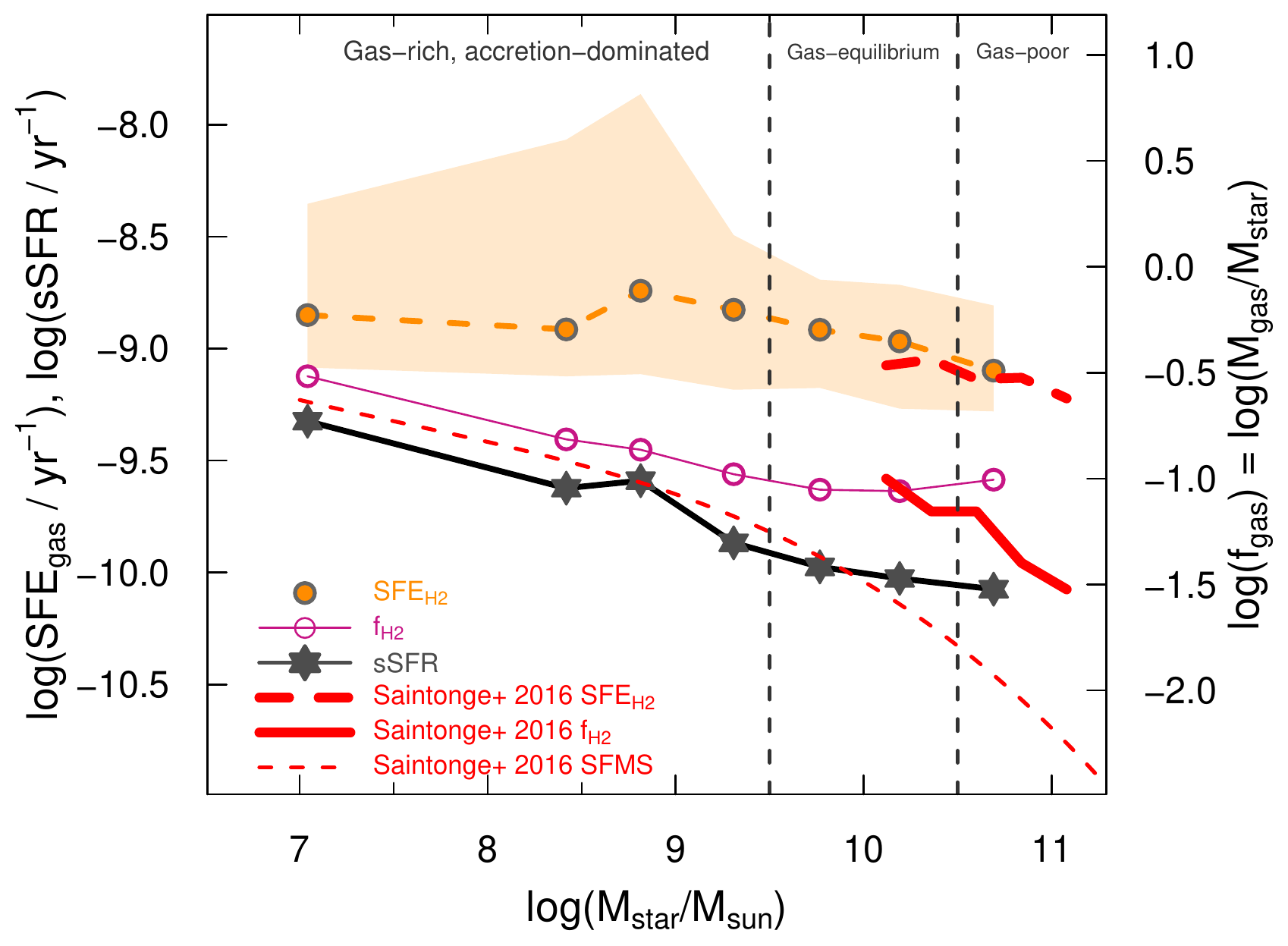} \\
\includegraphics[width=0.92\linewidth]{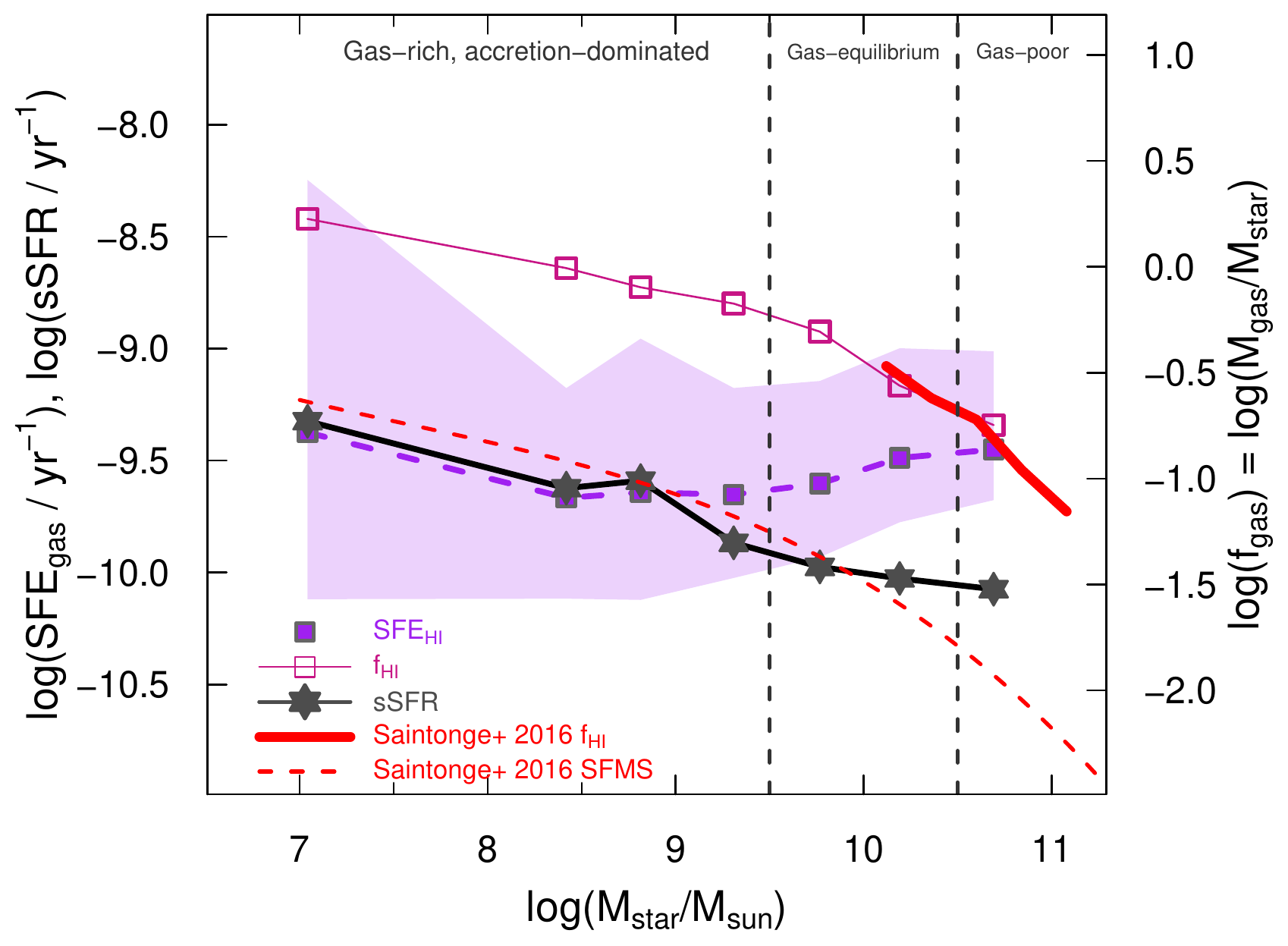} \\
\includegraphics[width=0.92\linewidth]{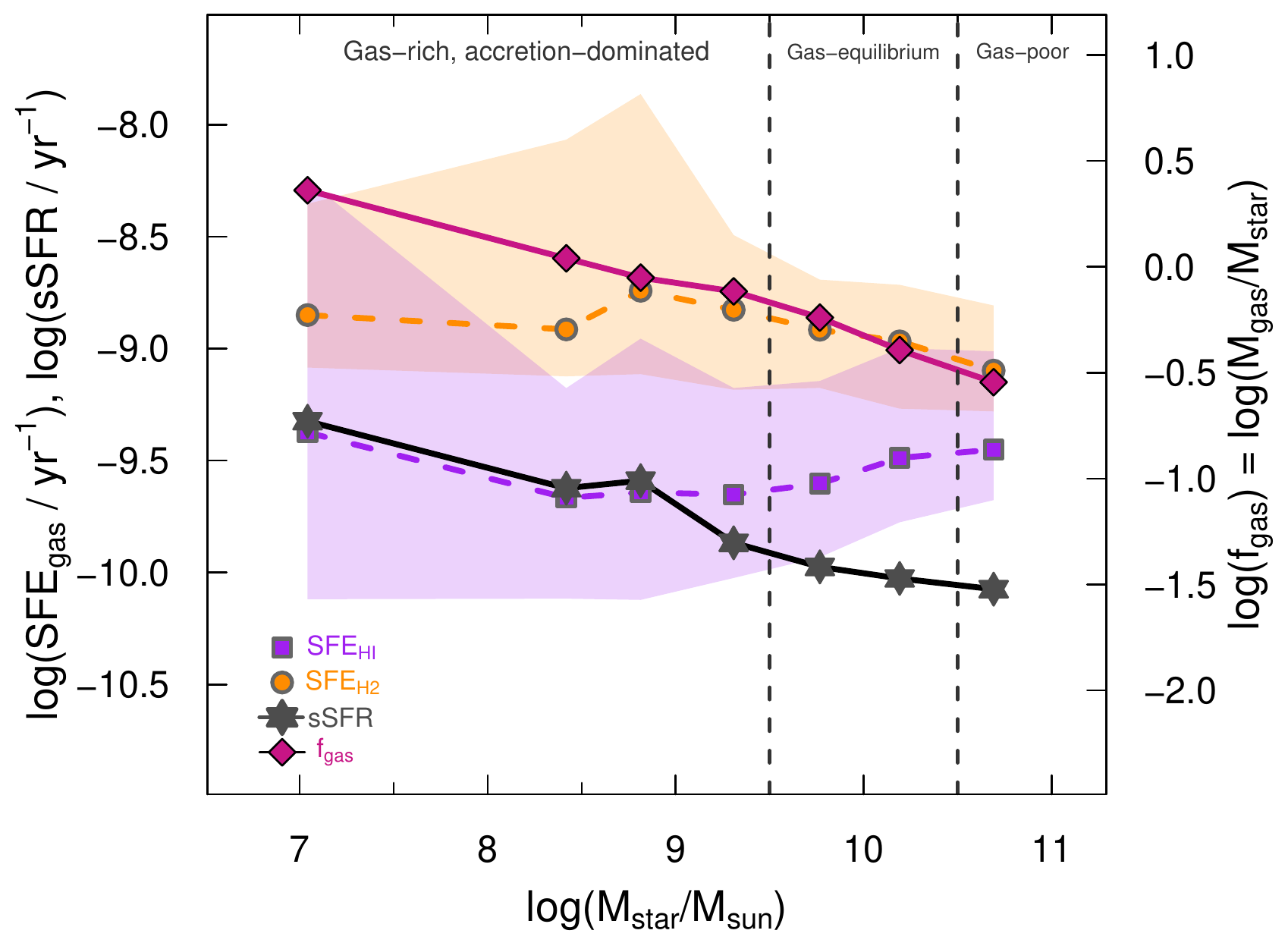} \\
\caption{Star-formation efficiency with \htwo\ (\sfehtwo) and \htwo\ gas fraction \fhtwo\ plotted against stellar mass
(top panel); 
the analogous quantities but for \hi\ (middle);
and with total gas fraction \fgas\ (bottom). 
In all panels, the left-hand axis gives the SFE and sSFR units, 
while the right-hand one corresponds to gas fractions, both logarithmic scale. 
The binned medians for \taudep\ are shown by orange circles and \fhtwo\ by open violet circles;
the same quantities for \hi\ are indicated by purple squares and open violet squares.
Only bins with $\geq$ 5 data points are shown.
Also plotted is the (binned) SFMS reported in Fig. \ref{fig:magma_ms}, here shown by six-sided stars.
Trends from \citet{Saintonge2016} are also shown by red curves as indicated in the legend in the lower left-hand corner.
The shaded region correspond to $\pm\,1\sigma$ spreads of the binned parameters.
The three \mstar\ bins discussed in the text are delineated by vertical dashed lines.
\label{fig:3bin}
}
\end{figure}

\subsection{Gas regimes in MAGMA}
\label{sec:gasregimes}

The results for the MAGMA sample confirm and extend the findings of \citet{Kannappan2013}.
For galaxies in the low \mstar\ bin, \mstar\,$\la\,3\times10^{9}$\,\msun,
the gas-rich, accretion-dominated, regime,
\hi\ is a necessary, perhaps even fundamental, ingredient for star formation.
This could be because of the need for self-shielding to foster the formation of \htwo\ \citep[e.g,][]{wolfire95},
possibly linked with high column density \citep[e.g.,][]{krumholz09,krumholz11},
thermal/dynamical or hydrostatic equilibrium \citep{ostriker10,krumholz13b}, 
or hydrostatic pressure in the disk \citep[e.g.,][]{blitz06}.
The preponderance of ``excess gas'' in the form of \hi\ at these low masses could
also be connected with the efficiency of cold-accretion modes \citep{fraternali17}.

For galaxies in the intermediate \mstar\ ``gas-equilibrium'' bin,
\hi\ and \htwo\ content are more tightly related with each other and with SFR.
Thus, stars can form roughly as fast as gas is accreted from the CGM and beyond
\citep{fraternali12},
although \hi\ content seems to dominate over \htwo\ by a factor of 3 or so
(see Fig. \ref{fig:h2hi_mstarssfr}).

In the high-mass ``bimodality'' regime, with \mstar$\ga 3\times 10^{10}$, the atomic
gas seems not to be directly linked to SFR.
Even at these masses, \hi\ content still dominates over \htwo, but
overall gas content is low; on average, \mhi/\mstar$\sim$0.15 while \mhtwo/\mstar$\sim$0.07.
As shown in Fig. \ref{fig:h2hibin_sfr}, \mhi/\mstar\ is roughly constant
over a factor of 10 in SFR, while \mhtwo/\mstar\ and SFR are fairly well correlated.
It is true that the high-mass bin in MAGMA is underpopulated ($\sim$\,10\% of the sample), but 
the emerging trends are fairly clear. 
However, through the requirement of \hi\ and CO (\htwo) detections,
the high-\mstar\ MAGMA galaxies tend not to be quenched systems.
Thus, our conclusions for this bin may be slightly biased by the possibly enhanced star-forming nature of the sample
at high \mstar\footnote{In fact, evidence points to a dichotomy in gas and star-formation properties
at these high masses; all high-\mstar\ galaxies are not quenched, ``red and dead'', but rather
there is also a class that contains gas and is currently forming stars.
We plan to explore this distinction in future work.}.

A primary question posed by these results is the physics that regulates star formation 
over different mass regimes:
is it the timescales (depletion times \taudep, or their inverse, the ``efficiency'' \sfehtwo),
or is it the available gas reservoir (\fgas)?
We examine this by noting \citep[see also][]{brownson20}:
\begin{eqnarray}
\log(\mathrm{sSFR/yr}^{-1})&\,=\,&\log(\mathrm{SFR}/M_\mathrm{{H2}}) + \log(M_\mathrm{H2}/M_\mathrm{star}) \nonumber \\
&\,=\,&\log(\mathrm{SFE}_\mathrm{{H2}}) + \log(f_\mathrm{H2}) \ ,
\end{eqnarray}
\noindent
where \fhtwo\ is defined in the same way as \fgas, namely \fhtwo$\,\equiv\,M_\mathrm{H2}/M_\mathrm{star}$.
These two ingredients for sSFR are shown in the 
top panel of Fig. \ref{fig:3bin}, together
with sSFR as a function of \mstar.
The middle panel of Fig. \ref{fig:3bin} reports analogous quantities for \hi,
and total gas fraction \fgas\ is shown in the lower one. 
The different quantities, sSFR, SFE and \fgas, are shown to scale as indicated in
the left-hand axis label (for sSFR and SFE) and the right-hand one (for \fgas).

Figure \ref{fig:3bin} (top) 
illustrates that in the accretion-dominated gas-rich regime at low \mstar, 
\sfehtwo\ (or equivalently \taudep, see Fig. \ref{fig:taudep}), 
is relatively constant, although in the lowest-\mstar\ bins, it falls by a factor of $\sim$3. 
In the ``gas-equilibrium'' and ``bimodality'' regimes, at higher \mstar, 
\sfehtwo\ is diminishing with increasing \mstar.
Conversely, \fhtwo\ is relatively constant at higher \mstar, while it tends to
grow toward lower \mstar, roughly in tandem with sSFR.
In the highest-\mstar\ bin, at \mstar$\sim 5\times10^{10}$\,\msun, growing \fhtwo\ is reflected in the upward trend of sSFR.
As discussed above, the increases in \fhtwo\ and sSFR in the highest \mstar\ bin are probably a selection effect,
but they are internally consistent within MAGMA.

The implication is that in the gas-equilibrium and bimodality \mstar\ regimes, while \sfehtwo\ is falling, \fhtwo\ is rising, so
that both timescales/efficiency and available \htwo\ contribute to the regulation of star formation.
Instead, in the low-mass bin, although the available \htwo\ is increasing,
\sfehtwo\ is constant, even falling toward the lowest \mstar.
We interpret this as a signature of the accretion-dominated, ``overwhelmed'' regime in which the galaxies are unable to
process all of the available gas.

The shape of the trend of \fhi\ with \mstar\ in the middle panel of Fig. \ref{fig:3bin} is very similar
to the form of the SFMS in terms of sSFR.
Atomic gas content is quite high toward low-\mstar\ (see also Fig. \ref{fig:h2hibin_sfr}), but falls precipitously at higher masses,
similar to the SFMS.
On the other hand, \sfehi\ is rising for \mstar\,$\ga 10^9$\,\msun,
changing direction
roughly at the break mass of the \mhi\ scaling with \mstar\ shown in Fig. \ref{fig:hih2}.
Over the same range in \mstar, \sfehtwo\ is falling as is the overall gas fraction shown in the lower 
panel of Fig. \ref{fig:3bin}.
The similarity of the \fgas\ curve (bottom panel) 
with the \fhi\ trend (middle) is due to the dominance
of atomic gas in the total gas budget, and is reflected in the low scatter from the fit in Eqn. (\ref{eqn:regressionmgashi}).
Beyond \mstar\,$\sim 10^9$\,\msun, the shape of the \fgas\ curve also mirrors \sfehtwo.
This is yet another indication that \hi\ and the total gas reservoir, not just \htwo, are participating,
at least indirectly, in the star-formation process.
A similar conclusion was reached by \citet{wang20} who found that \hi\ within the massive disks of star-forming galaxies plays an
important role as an intermediate step in fueling star formation.

Finally, another piece of the fueling puzzle can be found in 
the relation between gas fractions and star formation over longer timescales. 
\citet{Kannappan2013} formulated a quantity defined as the ratio of the mass of stars formed over the last Gyr relative
to the pre-existing stellar mass.
They found that this long-term fractional stellar mass growth rate correlates
quite well with the $U-J$ color \citep[see also][]{kannappan04}.
We explore the correlation of gas fraction and hybrid NUV$-$NIR color for MAGMA
in Appendix \ref{app:colors}, with an analysis of the hybrid color 
GALEX NUV and W1 (3.4\,\micron), two of the measurements used in \pone\ to estimate SFR
and \mstar.
We confirm and extend the trends of gas fraction with hybrid NUV$-$NIR color (in our case NUV$-$W1) 
found by \citet{Kannappan2013}.
These relations provide further evidence that atomic gas content is related with long-term SFR;
\hi\ mass is linked with the potential for star formation over the lifetime of the galaxy.



\section{Conclusions}
\label{sec:conclusions}

We have analyzed in terms of gas content and SFE the MAGMA sample of 392 galaxies with \hi\ and CO detections (for \htwo)
presented in \pone. 
The broad \mstar\ range of MAGMA (from $\sim\,10^7$\,\msun $\ -\ 10^{11}$\,\msun),
with $\sim$36\% of the sample having \mstar $\la\,3\times10^9$\,\msun,
enables for the first time an in-depth comparison of the gas properties of dwarf galaxies
with more massive systems.
Our main focus is on the availability of a cold gas reservoir for star formation,
gas depletion times as the inverse of SFE, and the partition of molecular and atomic gas
over a range in \mstar, as described below:

\begin{itemize}[listparindent=0pt,parsep=0.5\baselineskip,font=\itshape,style=sameline,labelsep=1pt,itemindent=0pt]
\item
{\itemfont\selectfont The conversion factor \aco\ as a function of metallicity.}
We have used MAGMA galaxies to revisit the dependence of \aco\ as a function of gas metallicity, $Z$,
under the assumption that \htwo\ depletion time \taudep\ depends on sSFR, and any deviations
are due to metallicity as a proxy of hardness of the radiation field.
Quantifying the hardness of the interstellar radiation field through FUV and NUV colors
was unsuccessful because of a dichotomy in \fuvnuv\ (see Sect. \ref{sec:alphaco}).
Results show that \aco$\,\propto\,(Z/Z_\odot)^{-1.55}$, less steep than a (previous) quadratic dependence,
but consistent with other techniques.
\item
{\itemfont\selectfont Simple relations for \mhtwo, \mhi, and total gas content \mgas\ as functions of \mstar\ and SFR.}
Molecular gas mass \mhtwo\ is strongly correlated with \mstar\ and SFR, enabling an expression for \mhtwo\
based on these two parameters that is accurate to within $\sim$0.2\,dex.
The relation of the atomic gas mass \mhi\ with stellar mass
shows a break at \mstar$\sim 3\times 10^9$\,\msun,
similar to that found by \citet{Huang2012} for an \hi-selected sample.
It is possible to express \mhi\ as a function of \mstar\ and SFR only to within $\sim$0.4\,dex,
while the analogous expression for total gas \mgas\ is accurate to within $\sim$0.3\,dex.
The behavior of \hi\ in these scaling relations suggests that it is an additional independent parameter,
and when it is included in the expression for \mgas, we find a relation good to within $\sim$0.05\,dex
[see Fig. \ref{fig:mgasvsmstarSFRHI} and Eqn. (\ref{eqn:regressionmgashi})].
\item
{\itemfont\selectfont SFE, gas depletion times, and and their relation with \mstar\ and sSFR.}
MAGMA galaxies span more than two orders of magnitude in \taudep, ranging from $\la 10^{-11}$\,yr$^{-1}$ to
$\ga 10^{-9}$\,yr$^{-1}$.
In agreement with previous work, we find \taudep\ to be correlated with sSFR, but only weakly with \mstar.
We also find that \taudephi\ is weakly correlated with \mstar, but well correlated with sSFR;
this result is new, stemming from the extension of MAGMA to a significant number of low-mass dwarf galaxies.
The short \taudep\ found for low-mass galaxies would imply that they are more efficient
at forming stars than more massive ones, in contrast with the low efficiencies predicted
by DMH abundance matching.
However, it is potentially spurious to assume that short or long depletion times (and their inverse
SFE) are manifestations of high- or low-efficiency star formation; rather they should be interpreted as 
simply timescales.
Inefficient star formation inferred from DMH assessments is related to the availability of baryons
and hierarchical growth, distinct from the conversion of molecular gas to stars.
``Efficient'' star formation inferred from \taudep\ is just more rapid star formation; 
whether or not it is truly efficient, that is to say able to convert more \htwo\ into stars over the mean
lifetime of a molecular cloud, is not straightforward to determine with the observables available for MAGMA.
\item
{\itemfont\selectfont The partition of cold gas into its molecular and atomic components.}
The ratio of atomic to molecular gas, \mhi/\mhtwo, only weakly depends on either \mstar\ or sSFR,
with large scatter.
Even galaxies more massive than $\sim 5\times10^{10}$\,\mstar\ tend to be dominated by \hi,
with an average \mhi/\mhtwo\,$\approx$\,2;
galaxies 20 times less massive have more than 3 times more \hi, and the most gas-rich galaxies
can have \mhi/\mhtwo\ as high as 10 or more.
In the lowest \mstar\ bin with \mstar$\,\leq 3\times10^9$\,\mstar, 
\hi\ and \htwo\ are both related to SFR, however with an inversion in the trend at relatively
high SFR, SFR\,$\sim$\,0.6\,\msunyr.
In the highest \mstar\ bin, with \mstar$\,\geq 3\times10^{10}$\,\mstar,
\hi\ content is roughly constant over a factor of 10 range in SFR, while in the same range of SFR \mhtwo\ is increasing.
\item
{\itemfont\selectfont Evidence for a differentiation of gas properties over three \mstar\ regimes.}
Following \citet{Kannappan2013}, we have divided the MAGMA sample into three \mstar\ bins:
\mstar\,$\leq 3\times10^9$\,\msun;
$3\times10^9$\,\msun\,$<$\mstar\,$< 3\times10^{10}$\,\msun;
\mstar\,$\geq 3\times10^{10}$\,\msun.
These intervals in \mstar\ are associated with a broad variety of transitions of physical processes in galaxy evolution
(see Sect. \ref{sec:discussion}), and also manifest different behavior of the molecular and atomic gas phases.
Our results extend those of \citet{Kannappan2013} because of the inclusion in MAGMA of many more dwarf galaxies,
and the requirement of \htwo\ (CO) detections. 
The properties of the gas in the lowest-mass ``gas-rich, accretion-dominated'' MAGMA bin suggest that there is simply too
much gas to process, even with the large sSFR and short \taudep\ common to these galaxies; they are ``overwhelmed''
by gas in the form of \hi. 
In the intermediate \mstar\ bin, ``gas equilibrium'' galaxies apparently form stars apace with their gas content,
able to keep up with the supply of gas.
For the galaxies in the most massive ``gas-poor, bimodality'' regime,
gas content is on average almost 10 times lower than in low-mass dwarfs.
While the \htwo\ content depends on SFR (or physically vice-versa),
the atomic gas seems not to participate (at least directly) in the current star formation episodes.
The implication is that in these high-mass galaxies \hi\ constitutes either a reservoir for future
star formation, or a past remnant of gas accretion that will remain inert over the galaxy's lifetime.
For galaxies in gas equilibrium and above the bimodality threshold, both the available \htwo\ reservoir and timescale/efficiency considerations
regulate star formation; in contrast, for dwarf galaxies \hi\ plays the main role. 
\end{itemize}

Although MAGMA substantially extends the parameter space for gas, stars, and metallicity to dwarf galaxies,
the coverage at low \mstar\ is still relatively sparse. 
We aim at quantifying gas content in a statistically significant number of low-mass galaxies with
future observations.
In addition, in a future paper,
we will analyze the effect of outflows on metallicity and gas properties,
and further confirm the existence of three different mass regimes.

Ideally, 
the study of gas content and SFE would be coupled with an
assessment of dark-matter halo mass.
This could be accomplished through \hi\ velocity profiles and/or rotation curves,
which we intend to incorporate in future analysis.

\section*{Acknowledgements}

We are grateful to the anonymous referee whose careful and constructive comments 
made this a better paper.
We also thank Sheila Kannappan for insightful discussions and comments on the manuscript.
LKH, CT, and MG acknowledge funding from the INAF PRIN-SKA 2017 program 1.05.01.88.04.

\bibliographystyle{aa}
\bibliography{biblio_baryonic,biblio_ERCplus} 

\begin{thebibliography}{156}
\expandafter\ifx\csname natexlab\endcsname\relax\def\natexlab#1{#1}\fi

\bibitem[{{Accurso} {et~al.}(2017){Accurso}, {Saintonge}, {Catinella},
  {Cortese}, {Dav{\'e}}, {Dunsheath}, {Genzel}, {Gracia-Carpio}, {Heckman},
  {Jimmy}, {Kramer}, {Li}, {Lutz}, {Schiminovich}, {Schuster}, {Sternberg},
  {Sturm}, {Tacconi}, {Tran}, \& {Wang}}]{accurso17}
{Accurso}, G., {Saintonge}, A., {Catinella}, B., {et~al.} 2017, \mnras, 470,
  4750

\bibitem[{{Amor{\'{\i}}n} {et~al.}(2016){Amor{\'{\i}}n},
  {Mu{\~n}oz-Tu{\~n}{\'o}n}, {Aguerri}, \& {Planesas}}]{amorin16}
{Amor{\'{\i}}n}, R., {Mu{\~n}oz-Tu{\~n}{\'o}n}, C., {Aguerri}, J.~A.~L., \&
  {Planesas}, P. 2016, \aap, 588, A23

\bibitem[{{Aniano} {et~al.}(2020){Aniano}, {Draine}, {Hunt}, {Sandstrom},
  {Calzetti}, {Kennicutt}, {Dale}, {Galametz}, {Gordon}, {Leroy}, {Smith},
  {Roussel}, {Sauvage}, {Walter}, {Armus}, {Bolatto}, {Boquien}, {Crocker}, {De
  Looze}, {Donovan Meyer}, {Helou}, {Hinz}, {Johnson}, {Koda}, {Miller},
  {Montiel}, {Murphy}, {Rela{\~n}o}, {Rix}, {Schinnerer}, {Skibba}, {Wolfire},
  \& {Engelbracht}}]{Aniano2020}
{Aniano}, G., {Draine}, B.~T., {Hunt}, L.~K., {et~al.} 2020, \apj, 889, 150

\bibitem[{{Armillotta} {et~al.}(2016){Armillotta}, {Fraternali}, \&
  {Marinacci}}]{armillotta16}
{Armillotta}, L., {Fraternali}, F., \& {Marinacci}, F. 2016, \mnras, 462, 4157

\bibitem[{{Asplund} {et~al.}(2009){Asplund}, {Grevesse}, {Sauval}, \&
  {Scott}}]{Asplund2009}
{Asplund}, M., {Grevesse}, N., {Sauval}, A.~J., \& {Scott}, P. 2009, \araa, 47,
  481

\bibitem[{{Baldry} {et~al.}(2012){Baldry}, {Driver}, {Loveday}, {Taylor},
  {Kelvin}, {Liske}, {Norberg}, {Robotham}, {Brough}, {Hopkins}, {Bamford},
  {Peacock}, {Bland-Hawthorn}, {Conselice}, {Croom}, {Jones}, {Parkinson},
  {Popescu}, {Prescott}, {Sharp}, \& {Tuffs}}]{baldry12}
{Baldry}, I.~K., {Driver}, S.~P., {Loveday}, J., {et~al.} 2012, \mnras, 421,
  621

\bibitem[{{Begum} {et~al.}(2005){Begum}, {Chengalur}, \&
  {Karachentsev}}]{begum05}
{Begum}, A., {Chengalur}, J.~N., \& {Karachentsev}, I.~D. 2005, \aap, 433, L1

\bibitem[{{Behroozi} {et~al.}(2019){Behroozi}, {Wechsler}, {Hearin}, \&
  {Conroy}}]{behroozi19}
{Behroozi}, P., {Wechsler}, R.~H., {Hearin}, A.~P., \& {Conroy}, C. 2019,
  \mnras, 488, 3143

\bibitem[{{Behroozi} {et~al.}(2013){Behroozi}, {Wechsler}, \&
  {Conroy}}]{behroozi13}
{Behroozi}, P.~S., {Wechsler}, R.~H., \& {Conroy}, C. 2013, \apj, 770, 57

\bibitem[{{Bigiel} {et~al.}(2008){Bigiel}, {Leroy}, {Walter}, {Brinks}, {de
  Blok}, {Madore}, \& {Thornley}}]{Bigiel2008}
{Bigiel}, F., {Leroy}, A., {Walter}, F., {et~al.} 2008, \aj, 136, 2846

\bibitem[{{Birnboim} \& {Dekel}(2003)}]{birnboim03}
{Birnboim}, Y. \& {Dekel}, A. 2003, \mnras, 345, 349

\bibitem[{{Blitz} \& {Rosolowsky}(2006)}]{blitz06}
{Blitz}, L. \& {Rosolowsky}, E. 2006, \apj, 650, 933

\bibitem[{{Boissier} {et~al.}(2018){Boissier}, {Cucciati}, {Boselli}, {Mei}, \&
  {Ferrarese}}]{boissier18}
{Boissier}, S., {Cucciati}, O., {Boselli}, A., {Mei}, S., \& {Ferrarese}, L.
  2018, \aap, 611, A42

\bibitem[{{Bolatto} {et~al.}(2008){Bolatto}, {Leroy}, {Rosolowsky}, {Walter},
  \& {Blitz}}]{bolatto08}
{Bolatto}, A.~D., {Leroy}, A.~K., {Rosolowsky}, E., {Walter}, F., \& {Blitz},
  L. 2008, \apj, 686, 948

\bibitem[{{Bolatto} {et~al.}(2013){Bolatto}, {Wolfire}, \& {Leroy}}]{bolatto13}
{Bolatto}, A.~D., {Wolfire}, M., \& {Leroy}, A.~K. 2013, \araa, 51, 207

\bibitem[{{Boquien} {et~al.}(2019){Boquien}, {Burgarella}, {Roehlly}, {Buat},
  {Ciesla}, {Corre}, {Inoue}, \& {Salas}}]{Boquien2019}
{Boquien}, M., {Burgarella}, D., {Roehlly}, Y., {et~al.} 2019, \aap, 622, A103

\bibitem[{{Boselli} {et~al.}(2014a){Boselli}, {Cortese}, \&
  {Boquien}}]{Boselli2014a}
{Boselli}, A., {Cortese}, L., \& {Boquien}, M. 2014a, \aap, 564, A65

\bibitem[{{Boselli} {et~al.}(2014b){Boselli}, {Cortese}, {Boquien}, {Boissier},
  {Catinella}, {Lagos}, \& {Saintonge}}]{Boselli2014b}
{Boselli}, A., {Cortese}, L., {Boquien}, M., {et~al.} 2014b, \aap, 564, A66

\bibitem[{{Boselli} {et~al.}(2010){Boselli}, {Eales}, {Cortese}, {Bendo},
  {Chanial}, {Buat}, {Davies}, {Auld}, {Rigby}, {Baes}, {Barlow}, {Bock},
  {Bradford}, {Castro-Rodriguez}, {Charlot}, {Clements}, {Cormier}, {Dwek},
  {Elbaz}, {Galametz}, {Galliano}, {Gear}, {Glenn}, {Gomez}, {Griffin}, {Hony},
  {Isaak}, {Levenson}, {Lu}, {Madden}, {O'Halloran}, {Okamura}, {Oliver},
  {Page}, {Panuzzo}, {Papageorgiou}, {Parkin}, {Perez-Fournon}, {Pohlen},
  {Rangwala}, {Roussel}, {Rykala}, {Sacchi}, {Sauvage}, {Schulz}, {Schirm},
  {Smith}, {Spinoglio}, {Stevens}, {Symeonidis}, {Vaccari}, {Vigroux},
  {Wilson}, {Wozniak}, {Wright}, \& {Zeilinger}}]{Boselli2010}
{Boselli}, A., {Eales}, S., {Cortese}, L., {et~al.} 2010, \pasp, 122, 261

\bibitem[{{Boselli} {et~al.}(2015){Boselli}, {Fossati}, {Gavazzi}, {Ciesla},
  {Buat}, {Boissier}, \& {Hughes}}]{Boselli2015}
{Boselli}, A., {Fossati}, M., {Gavazzi}, G., {et~al.} 2015, \aap, 579, A102

\bibitem[{{Bothwell} {et~al.}(2014){Bothwell}, {Wagg}, {Cicone}, {Maiolino},
  {M{\o}ller}, {Aravena}, {De Breuck}, {Peng}, {Espada}, {Hodge},
  {Impellizzeri}, {Mart{\'{\i}}n}, {Riechers}, \& {Walter}}]{Bothwell2014}
{Bothwell}, M.~S., {Wagg}, J., {Cicone}, C., {et~al.} 2014, \mnras, 445, 2599

\bibitem[{{Bouquin} {et~al.}(2015){Bouquin}, {Gil de Paz}, {Boissier},
  {Mu{\~n}oz-Mateos}, {Sheth}, {Zaritsky}, {Laine}, {Gallego}, {Peletier},
  {R{\"o}ck}, \& {Knapen}}]{bouquin15}
{Bouquin}, A. Y.~K., {Gil de Paz}, A., {Boissier}, S., {et~al.} 2015, \apjl,
  800, L19

\bibitem[{{Broeils} \& {Rhee}(1997)}]{broeils97}
{Broeils}, A.~H. \& {Rhee}, M.~H. 1997, \aap, 324, 877

\bibitem[{{Brownson} {et~al.}(2020){Brownson}, {Belfiore}, {Maiolino}, {Lin},
  \& {Carniani}}]{brownson20}
{Brownson}, S., {Belfiore}, F., {Maiolino}, R., {Lin}, L., \& {Carniani}, S.
  2020, arXiv e-prints, arXiv:2007.02976

\bibitem[{{Bundy} {et~al.}(2005){Bundy}, {Ellis}, \& {Conselice}}]{bundy05}
{Bundy}, K., {Ellis}, R.~S., \& {Conselice}, C.~J. 2005, \apj, 625, 621

\bibitem[{{Calzetti} {et~al.}(2005){Calzetti}, {Kennicutt}, {Bianchi},
  {Thilker}, {Dale}, {Engelbracht}, {Leitherer}, {Meyer}, {Sosey}, {Mutchler},
  {Regan}, {Thornley}, {Armus}, {Bendo}, {Boissier}, {Boselli}, {Draine},
  {Gordon}, {Helou}, {Hollenbach}, {Kewley}, {Madore}, {Martin}, {Murphy},
  {Rieke}, {Rieke}, {Roussel}, {Sheth}, {Smith}, {Walter}, {White}, {Yi},
  {Scoville}, {Polletta}, \& {Lindler}}]{calzetti05}
{Calzetti}, D., {Kennicutt}, R.~C., J., {Bianchi}, L., {et~al.} 2005, \apj,
  633, 871

\bibitem[{{Catinella} {et~al.}(2018){Catinella}, {Saintonge}, {Janowiecki},
  {Cortese}, {Dav{\'e}}, {Lemonias}, {Cooper}, {Schiminovich}, {Hummels},
  {Fabello}, {Ger{\'e}b}, {Kilborn}, \& {Wang}}]{Catinella2018}
{Catinella}, B., {Saintonge}, A., {Janowiecki}, S., {et~al.} 2018, \mnras, 476,
  875

\bibitem[{{Chabrier}(2003)}]{Chabrier2003}
{Chabrier}, G. 2003, Publications of the Astronomical Society of the Pacific,
  115, 763

\bibitem[{{Cicone} {et~al.}(2017){Cicone}, {Bothwell}, {Wagg}, {M{\o}ller}, {De
  Breuck}, {Zhang}, {Mart{\'\i}n}, {Maiolino}, {Severgnini}, {Aravena},
  {Belfiore}, {Espada}, {Fl{\"u}tsch}, {Impellizzeri}, {Peng}, {Raj}, {Ram{\'\i
  }rez-Olivencia}, {Riechers}, \& {Schawinski}}]{Cicone2017}
{Cicone}, C., {Bothwell}, M., {Wagg}, J., {et~al.} 2017, \aap, 604, A53

\bibitem[{{Cormier} {et~al.}(2014){Cormier}, {Madden}, {Lebouteiller}, {Hony},
  {Aalto}, {Costagliola}, {Hughes}, {R{\'e}my-Ruyer}, {Abel}, {Bayet},
  {Bigiel}, {Cannon}, {Cumming}, {Galametz}, {Galliano}, {Viti}, \&
  {Wu}}]{Cormier2014}
{Cormier}, D., {Madden}, S.~C., {Lebouteiller}, V., {et~al.} 2014, \aap, 564,
  A121

\bibitem[{{Cortese} {et~al.}(2011){Cortese}, {Catinella}, {Boissier},
  {Boselli}, \& {Heinis}}]{Cortese2011}
{Cortese}, L., {Catinella}, B., {Boissier}, S., {Boselli}, A., \& {Heinis}, S.
  2011, \mnras, 415, 1797

\bibitem[{{Cortese} {et~al.}(2020){Cortese}, {Catinella}, {Cook}, \&
  {Janowiecki}}]{cortese20}
{Cortese}, L., {Catinella}, B., {Cook}, R.~H.~W., \& {Janowiecki}, S. 2020,
  \mnras, 494, L42

\bibitem[{{Dalcanton}(2007)}]{dalcanton07}
{Dalcanton}, J.~J. 2007, \apj, 658, 941

\bibitem[{{Dale} {et~al.}(2009){Dale}, {Cohen}, {Johnson}, {Schuster},
  {Calzetti}, {Engelbracht}, {Gil de Paz}, {Kennicutt}, {Lee}, {Begum},
  {Block}, {Dalcanton}, {Funes}, {Gordon}, {Johnson}, {Marble}, {Sakai},
  {Skillman}, {van Zee}, {Walter}, {Weisz}, {Williams}, {Wu}, \&
  {Wu}}]{Dale2009}
{Dale}, D.~A., {Cohen}, S.~A., {Johnson}, L.~C., {et~al.} 2009, \apj, 703, 517

\bibitem[{{Dav{\'e}} {et~al.}(2012){Dav{\'e}}, {Finlator}, \&
  {Oppenheimer}}]{Dave2012}
{Dav{\'e}}, R., {Finlator}, K., \& {Oppenheimer}, B.~D. 2012, \mnras, 421, 98

\bibitem[{{Dekel} \& {Birnboim}(2006)}]{dekelbirnboim06}
{Dekel}, A. \& {Birnboim}, Y. 2006, \mnras, 368, 2

\bibitem[{{Driver} {et~al.}(2006){Driver}, {Allen}, {Graham}, {Cameron},
  {Liske}, {Ellis}, {Cross}, {De Propris}, {Phillipps}, \& {Couch}}]{driver06}
{Driver}, S.~P., {Allen}, P.~D., {Graham}, A.~W., {et~al.} 2006, \mnras, 368,
  414

\bibitem[{{Eckert} {et~al.}(2015){Eckert}, {Kannappan}, {Stark}, {Moffett},
  {Norris}, {Snyder}, \& {Hoversten}}]{eckert15}
{Eckert}, K.~D., {Kannappan}, S.~J., {Stark}, D.~V., {et~al.} 2015, \apj, 810,
  166

\bibitem[{{Elbaz} {et~al.}(2007){Elbaz}, {Daddi}, {Le Borgne}, {Dickinson},
  {Alexander}, {Chary}, {Starck}, {Brand t}, {Kitzbichler}, {MacDonald},
  {Nonino}, {Popesso}, {Stern}, \& {Vanzella}}]{elbaz07}
{Elbaz}, D., {Daddi}, E., {Le Borgne}, D., {et~al.} 2007, \aap, 468, 33

\bibitem[{{Fraternali}(2017)}]{fraternali17}
{Fraternali}, F. 2017, Astrophysics and Space Science Library, Vol. 430, {Gas
  Accretion via Condensation and Fountains}, ed. A.~{Fox} \& R.~{Dav{\'e}}, 323

\bibitem[{{Fraternali} \& {Tomassetti}(2012)}]{fraternali12}
{Fraternali}, F. \& {Tomassetti}, M. 2012, \mnras, 426, 2166

\bibitem[{{Gallart} {et~al.}(2015){Gallart}, {Monelli}, {Mayer}, {Aparicio},
  {Battaglia}, {Bernard}, {Cassisi}, {Cole}, {Dolphin}, {Drozdovsky},
  {Hidalgo}, {Navarro}, {Salvadori}, {Skillman}, {Stetson}, \&
  {Weisz}}]{gallart15}
{Gallart}, C., {Monelli}, M., {Mayer}, L., {et~al.} 2015, \apjl, 811, L18

\bibitem[{{Gao} \& {Solomon}(2004)}]{Gao2004}
{Gao}, Y. \& {Solomon}, P.~M. 2004, \apj, 606, 271

\bibitem[{{Garnett}(2002)}]{garnett02}
{Garnett}, D.~R. 2002, \apj, 581, 1019

\bibitem[{{Gavazzi} {et~al.}(2013){Gavazzi}, {Fumagalli}, {Fossati}, {Galardo},
  {Grossetti}, {Boselli}, {Giovanelli}, \& {Haynes}}]{Gavazzi2013}
{Gavazzi}, G., {Fumagalli}, M., {Fossati}, M., {et~al.} 2013, \aap, 553, A89

\bibitem[{{Genzel} {et~al.}(2012){Genzel}, {Tacconi}, {Combes}, {Bolatto},
  {Neri}, {Sternberg}, {Cooper}, {Bouch{\'e}}, {Bournaud}, {Burkert},
  {Comerford}, {Cox}, {Davis}, {F{\"o}rster Schreiber}, {Garcia-Burillo},
  {Gracia-Carpio}, {Lutz}, {Naab}, {Newman}, {Saintonge}, {Shapiro}, {Shapley},
  \& {Weiner}}]{genzel12}
{Genzel}, R., {Tacconi}, L.~J., {Combes}, F., {et~al.} 2012, \apj, 746, 69

\bibitem[{{Genzel} {et~al.}(2015){Genzel}, {Tacconi}, {Lutz}, {Saintonge},
  {Berta}, {Magnelli}, {Combes}, {Garc{\'\i}a-Burillo}, {Neri}, {Bolatto},
  {Contini}, {Lilly}, {Boissier}, {Boone}, {Bouch{\'e}}, {Bournaud}, {Burkert},
  {Carollo}, {Colina}, {Cooper}, {Cox}, {Feruglio}, {F{\"o}rster Schreiber},
  {Freundlich}, {Gracia-Carpio}, {Juneau}, {Kovac}, {Lippa}, {Naab}, {Salome},
  {Renzini}, {Sternberg}, {Walter}, {Weiner}, {Weiss}, \& {Wuyts}}]{genzel15}
{Genzel}, R., {Tacconi}, L.~J., {Lutz}, D., {et~al.} 2015, \apj, 800, 20

\bibitem[{{Gil de Paz} {et~al.}(2007){Gil de Paz}, {Boissier}, {Madore},
  {Seibert}, {Joe}, {Boselli}, {Wyder}, {Thilker}, {Bianchi}, {Rey}, {Rich},
  {Barlow}, {Conrow}, {Forster}, {Friedman}, {Martin}, {Morrissey}, {Neff},
  {Schiminovich}, {Small}, {Donas}, {Heckman}, {Lee}, {Milliard}, {Szalay}, \&
  {Yi}}]{gildepaz07}
{Gil de Paz}, A., {Boissier}, S., {Madore}, B.~F., {et~al.} 2007, \apjs, 173,
  185

\bibitem[{{Ginolfi} {et~al.}(2020){Ginolfi}, {Hunt}, {Tortora}, {Schneider}, \&
  {Cresci}}]{ginolfi20}
{Ginolfi}, M., {Hunt}, L.~K., {Tortora}, C., {Schneider}, R., \& {Cresci}, G.
  2020, \aap, 638, A4

\bibitem[{{Glover} \& {Mac Low}(2011)}]{glover11}
{Glover}, S.~C.~O. \& {Mac Low}, M.~M. 2011, \mnras, 412, 337

\bibitem[{{Gnedin} \& {Kravtsov}(2010)}]{gnedin10}
{Gnedin}, N.~Y. \& {Kravtsov}, A.~V. 2010, \apj, 714, 287

\bibitem[{{Gnedin} {et~al.}(2009){Gnedin}, {Tassis}, \& {Kravtsov}}]{gnedin09}
{Gnedin}, N.~Y., {Tassis}, K., \& {Kravtsov}, A.~V. 2009, \apj, 697, 55

\bibitem[{{Graziani} {et~al.}(2017){Graziani}, {de Bennassuti}, {Schneider},
  {Kawata}, \& {Salvadori}}]{graziani17}
{Graziani}, L., {de Bennassuti}, M., {Schneider}, R., {Kawata}, D., \&
  {Salvadori}, S. 2017, \mnras, 469, 1101

\bibitem[{{Graziani} {et~al.}(2015){Graziani}, {Salvadori}, {Schneider},
  {Kawata}, {de Bennassuti}, \& {Maselli}}]{graziani15}
{Graziani}, L., {Salvadori}, S., {Schneider}, R., {et~al.} 2015, \mnras, 449,
  3137

\bibitem[{{Graziani} {et~al.}(2020){Graziani}, {Schneider}, {Ginolfi}, {Hunt},
  {Maio}, {Glatzle}, \& {Ciardi}}]{graziani20}
{Graziani}, L., {Schneider}, R., {Ginolfi}, M., {et~al.} 2020, \mnras, 494,
  1071

\bibitem[{{Grossi} {et~al.}(2016){Grossi}, {Corbelli}, {Bizzocchi},
  {Giovanardi}, {Bomans}, {Coelho}, {De Looze}, {Gon{\c c}alves}, {Hunt},
  {Leonardo}, {Madden}, {Men{\'e}ndez-Delmestre}, {Pappalardo}, \&
  {Riguccini}}]{Grossi2016}
{Grossi}, M., {Corbelli}, E., {Bizzocchi}, L., {et~al.} 2016, \aap, 590, A27

\bibitem[{{Hathi} {et~al.}(2008){Hathi}, {Malhotra}, \& {Rhoads}}]{hathi08}
{Hathi}, N.~P., {Malhotra}, S., \& {Rhoads}, J.~E. 2008, \apj, 673, 686

\bibitem[{{Haynes} {et~al.}(2018){Haynes}, {Giovanelli}, {Kent}, {Adams},
  {Balonek}, {Craig}, {Fertig}, {Finn}, {Giovanardi}, {Hallenbeck}, {Hess},
  {Hoffman}, {Huang}, {Jones}, {Koopmann}, {Kornreich}, {Leisman}, {Miller},
  {Moorman}, {O'Connor}, {O'Donoghue}, {Papastergis}, {Troischt}, {Stark}, \&
  {Xiao}}]{Haynes2018}
{Haynes}, M.~P., {Giovanelli}, R., {Kent}, B.~R., {et~al.} 2018, \apj, 861, 49

\bibitem[{{Haynes} {et~al.}(2011){Haynes}, {Giovanelli}, {Martin}, {Hess},
  {Saintonge}, {Adams}, {Hallenbeck}, {Hoffman}, {Huang}, {Kent}, {Koopmann},
  {Papastergis}, {Stierwalt}, {Balonek}, {Craig}, {Higdon}, {Kornreich},
  {Miller}, {O'Donoghue}, {Olowin}, {Rosenberg}, {Spekkens}, {Troischt}, \&
  {Wilcots}}]{Haynes2011}
{Haynes}, M.~P., {Giovanelli}, R., {Martin}, A.~M., {et~al.} 2011, \aj, 142,
  170

\bibitem[{{Hayward} \& {Hopkins}(2017)}]{hayward17}
{Hayward}, C.~C. \& {Hopkins}, P.~F. 2017, \mnras, 465, 1682

\bibitem[{{Hopkins} {et~al.}(2014){Hopkins}, {Kere{\v s}}, {O{\~n}orbe},
  {Faucher-Gigu{\`e}re}, {Quataert}, {Murray}, \& {Bullock}}]{hopkins14}
{Hopkins}, P.~F., {Kere{\v s}}, D., {O{\~n}orbe}, J., {et~al.} 2014, \mnras,
  445, 581

\bibitem[{{Hopkins} {et~al.}(2010){Hopkins}, {Murray}, {Quataert}, \&
  {Thompson}}]{hopkins10}
{Hopkins}, P.~F., {Murray}, N., {Quataert}, E., \& {Thompson}, T.~A. 2010,
  \mnras, 401, L19

\bibitem[{{Huang} \& {Kauffmann}(2014)}]{Huang2014}
{Huang}, M.-L. \& {Kauffmann}, G. 2014, \mnras, 443, 1329

\bibitem[{{Huang} \& {Kauffmann}(2015)}]{huang15}
{Huang}, M.-L. \& {Kauffmann}, G. 2015, \mnras, 450, 1375

\bibitem[{{Huang} {et~al.}(2012){Huang}, {Haynes}, {Giovanelli}, \&
  {Brinchmann}}]{Huang2012}
{Huang}, S., {Haynes}, M.~P., {Giovanelli}, R., \& {Brinchmann}, J. 2012, \apj,
  756, 113

\bibitem[{{Hunt} {et~al.}(2016a){Hunt}, {Dayal}, {Magrini}, \&
  {Ferrara}}]{Hunt2016a}
{Hunt}, L., {Dayal}, P., {Magrini}, L., \& {Ferrara}, A. 2016a, \mnras, 463,
  2002

\bibitem[{{Hunt} {et~al.}(2012){Hunt}, {Magrini}, {Galli}, {Schneider},
  {Bianchi}, {Maiolino}, {Romano}, {Tosi}, \& {Valiante}}]{Hunt2012}
{Hunt}, L., {Magrini}, L., {Galli}, D., {et~al.} 2012, \mnras, 427, 906

\bibitem[{{Hunt} {et~al.}(2019){Hunt}, {De Looze}, {Boquien}, {Nikutta},
  {Rossi}, {Bianchi}, {Dale}, {Granato}, {Kennicutt}, {Silva}, {Ciesla},
  {Rela{\~n}o}, {Viaene}, {Brandl}, {Calzetti}, {Croxall}, {Draine},
  {Galametz}, {Gordon}, {Groves}, {Helou}, {Herrera-Camus}, {Hinz}, {Koda},
  {Salim}, {Sandstrom}, {Smith}, {Wilson}, \& {Zibetti}}]{Hunt2019}
{Hunt}, L.~K., {De Looze}, I., {Boquien}, M., {et~al.} 2019, \aap, 621, A51

\bibitem[{{Hunt} {et~al.}(2015){Hunt}, {Garc{\'{\i}}a-Burillo}, {Casasola},
  {Caselli}, {Combes}, {Henkel}, {Lundgren}, {Maiolino}, {Menten}, {Testi}, \&
  {Weiss}}]{hunt15}
{Hunt}, L.~K., {Garc{\'{\i}}a-Burillo}, S., {Casasola}, V., {et~al.} 2015,
  \aap, 583, A114

\bibitem[{{Ilbert} {et~al.}(2013){Ilbert}, {McCracken}, {Le F{\`e}vre},
  {Capak}, {Dunlop}, {Karim}, {Renzini}, {Caputi}, {Boissier}, {Arnouts},
  {Aussel}, {Comparat}, {Guo}, {Hudelot}, {Kartaltepe}, {Kneib}, {Krogager},
  {Le Floc'h}, {Lilly}, {Mellier}, {Milvang-Jensen}, {Moutard}, {Onodera},
  {Richard}, {Salvato}, {Sanders}, {Scoville}, {Silverman}, {Taniguchi},
  {Tasca}, {Thomas}, {Toft}, {Tresse}, {Vergani}, {Wolk}, \& {Zirm}}]{ilbert13}
{Ilbert}, O., {McCracken}, H.~J., {Le F{\`e}vre}, O., {et~al.} 2013, \aap, 556,
  A55

\bibitem[{{Israel}(1997)}]{israel97}
{Israel}, F.~P. 1997, \aap, 328, 471

\bibitem[{{Kannappan}(2004)}]{kannappan04}
{Kannappan}, S.~J. 2004, \apjl, 611, L89

\bibitem[{{Kannappan} {et~al.}(2013){Kannappan}, {Stark}, {Eckert}, {Moffett},
  {Wei}, {Pisano}, {Baker}, {Vogel}, {Fabricant}, {Laine}, {Norris}, {Jogee},
  {Lepore}, {Hough}, \& {Weinberg-Wolf}}]{Kannappan2013}
{Kannappan}, S.~J., {Stark}, D.~V., {Eckert}, K.~D., {et~al.} 2013, \apj, 777,
  42

\bibitem[{{Karachentsev} \& {Kaisina}(2013)}]{karach13}
{Karachentsev}, I.~D. \& {Kaisina}, E.~I. 2013, \aj, 146, 46

\bibitem[{{Kauffmann} {et~al.}(2003){Kauffmann}, {Heckman}, {White}, {Charlot},
  {Tremonti}, {Peng}, {Seibert}, {Brinkmann}, {Nichol}, {SubbaRao}, \&
  {York}}]{kauffmann03}
{Kauffmann}, G., {Heckman}, T.~M., {White}, S. D.~M., {et~al.} 2003, \mnras,
  341, 54

\bibitem[{{Kaviraj} {et~al.}(2007){Kaviraj}, {Schawinski}, {Devriendt},
  {Ferreras}, {Khochfar}, {Yoon}, {Yi}, {Deharveng}, {Boselli}, {Barlow},
  {Conrow}, {Forster}, {Friedman}, {Martin}, {Morrissey}, {Neff},
  {Schiminovich}, {Seibert}, {Small}, {Wyder}, {Bianchi}, {Donas}, {Heckman},
  {Lee}, {Madore}, {Milliard}, {Rich}, \& {Szalay}}]{kaviraj07}
{Kaviraj}, S., {Schawinski}, K., {Devriendt}, J.~E.~G., {et~al.} 2007, \apjs,
  173, 619

\bibitem[{{Kennicutt} {et~al.}(2011){Kennicutt}, {Calzetti}, {Aniano},
  {Appleton}, {Armus}, {Beir{\~a}o}, {Bolatto}, {Brandl}, {Crocker}, {Croxall},
  {Dale}, {Donovan Meyer}, {Draine}, {Engelbracht}, {Galametz}, {Gordon},
  {Groves}, {Hao}, {Helou}, {Hinz}, {Hunt}, {Johnson}, {Koda}, {Krause},
  {Leroy}, {Li}, {Meidt}, {Montiel}, {Murphy}, {Rahman}, {Rix}, {Roussel},
  {Sandstrom}, {Sauvage}, {Schinnerer}, {Skibba}, {Smith}, {Srinivasan},
  {Vigroux}, {Walter}, {Wilson}, {Wolfire}, \& {Zibetti}}]{Kennicutt2011}
{Kennicutt}, R.~C., {Calzetti}, D., {Aniano}, G., {et~al.} 2011, PASP, 123,
  1347

\bibitem[{{Kere{\v{s}}} {et~al.}(2009){Kere{\v{s}}}, {Katz}, {Fardal},
  {Dav{\'e}}, \& {Weinberg}}]{keres09}
{Kere{\v{s}}}, D., {Katz}, N., {Fardal}, M., {Dav{\'e}}, R., \& {Weinberg},
  D.~H. 2009, \mnras, 395, 160

\bibitem[{{Kere{\v{s}}} {et~al.}(2005){Kere{\v{s}}}, {Katz}, {Weinberg}, \&
  {Dav{\'e}}}]{keres05}
{Kere{\v{s}}}, D., {Katz}, N., {Weinberg}, D.~H., \& {Dav{\'e}}, R. 2005,
  \mnras, 363, 2

\bibitem[{{Kewley} \& {Ellison}(2008)}]{Kewley2008}
{Kewley}, L.~J. \& {Ellison}, S.~L. 2008, \apj, 681, 1183

\bibitem[{{Kroupa}(2001)}]{Kroupa2001}
{Kroupa}, P. 2001, \mnras, 322, 231

\bibitem[{{Krumholz}(2012)}]{krumholz12}
{Krumholz}, M.~R. 2012, \apj, 759, 9

\bibitem[{{Krumholz}(2013b)}]{krumholz13b}
{Krumholz}, M.~R. 2013b, \mnras, 436, 2747

\bibitem[{{Krumholz} {et~al.}(2011){Krumholz}, {Leroy}, \&
  {McKee}}]{krumholz11}
{Krumholz}, M.~R., {Leroy}, A.~K., \& {McKee}, C.~F. 2011, \apj, 731, 25

\bibitem[{{Krumholz} {et~al.}(2009){Krumholz}, {McKee}, \&
  {Tumlinson}}]{krumholz09}
{Krumholz}, M.~R., {McKee}, C.~F., \& {Tumlinson}, J. 2009, \apj, 699, 850

\bibitem[{{Krumholz} \& {Tan}(2007)}]{krumholz07}
{Krumholz}, M.~R. \& {Tan}, J.~C. 2007, \apj, 654, 304

\bibitem[{{Lee} {et~al.}(2015){Lee}, {Sanders}, {Casey}, {Toft}, {Scoville},
  {Hung}, {Le Floc'h}, {Ilbert}, {Zahid}, {Aussel}, {Capak}, {Kartaltepe},
  {Kewley}, {Li}, {Schawinski}, {Sheth}, \& {Xiao}}]{lee15}
{Lee}, N., {Sanders}, D.~B., {Casey}, C.~M., {et~al.} 2015, \apj, 801, 80

\bibitem[{{Leroy} {et~al.}(2006){Leroy}, {Bolatto}, {Walter}, \&
  {Blitz}}]{Leroy2006}
{Leroy}, A., {Bolatto}, A., {Walter}, F., \& {Blitz}, L. 2006, \apj, 643, 825

\bibitem[{{Leroy} {et~al.}(2009){Leroy}, {Bolatto}, {Bot}, {Engelbracht},
  {Gordon}, {Israel}, {Rubio}, {Sandstrom}, \& {Stanimirovi{\'c}}}]{leroy09}
{Leroy}, A.~K., {Bolatto}, A., {Bot}, C., {et~al.} 2009, \apj, 702, 352

\bibitem[{{Leroy} {et~al.}(2011){Leroy}, {Bolatto}, {Gordon}, {Sandstrom},
  {Gratier}, {Rosolowsky}, {Engelbracht}, {Mizuno}, {Corbelli}, {Fukui}, \&
  {Kawamura}}]{leroy11}
{Leroy}, A.~K., {Bolatto}, A., {Gordon}, K., {et~al.} 2011, \apj, 737, 12

\bibitem[{{Leroy} {et~al.}(2019){Leroy}, {Sandstrom}, {Lang}, {Lewis}, {Salim},
  {Behrens}, {Chastenet}, {Chiang}, {Gallagher}, {Kessler}, \&
  {Utomo}}]{Leroy2019}
{Leroy}, A.~K., {Sandstrom}, K.~M., {Lang}, D., {et~al.} 2019, \apjs, 244, 24

\bibitem[{{Leroy} {et~al.}(2013){Leroy}, {Walter}, {Sandstrom}, {Schruba},
  {Munoz-Mateos}, {Bigiel}, {Bolatto}, {Brinks}, {de Blok}, {Meidt}, {Rix},
  {Rosolowsky}, {Schinnerer}, {Schuster}, \& {Usero}}]{leroy13}
{Leroy}, A.~K., {Walter}, F., {Sandstrom}, K., {et~al.} 2013, \aj, 146, 19

\bibitem[{{Li} {et~al.}(2012){Li}, {Kauffmann}, {Fu}, {Wang}, {Catinella},
  {Fabello}, {Schiminovich}, \& {Zhang}}]{li12}
{Li}, C., {Kauffmann}, G., {Fu}, J., {et~al.} 2012, \mnras, 424, 1471

\bibitem[{{Lilly} {et~al.}(2013){Lilly}, {Carollo}, {Pipino}, {Renzini}, \&
  {Peng}}]{Lilly2013}
{Lilly}, S.~J., {Carollo}, C.~M., {Pipino}, A., {Renzini}, A., \& {Peng}, Y.
  2013, \apj, 772, 119

\bibitem[{{Madau} \& {Dickinson}(2014)}]{madau14}
{Madau}, P. \& {Dickinson}, M. 2014, \araa, 52, 415

\bibitem[{{Madau} {et~al.}(2014){Madau}, {Weisz}, \&
  {Conroy}}]{madau14reversal}
{Madau}, P., {Weisz}, D.~R., \& {Conroy}, C. 2014, \apjl, 790, L17

\bibitem[{{Maddox} {et~al.}(2015){Maddox}, {Hess}, {Obreschkow}, {Jarvis}, \&
  {Blyth}}]{maddox15}
{Maddox}, N., {Hess}, K.~M., {Obreschkow}, D., {Jarvis}, M.~J., \& {Blyth},
  S.~L. 2015, \mnras, 447, 1610

\bibitem[{{Marinacci} {et~al.}(2010){Marinacci}, {Binney}, {Fraternali},
  {Nipoti}, {Ciotti}, \& {Londrillo}}]{marinacci10}
{Marinacci}, F., {Binney}, J., {Fraternali}, F., {et~al.} 2010, \mnras, 404,
  1464

\bibitem[{{Martinsson} {et~al.}(2016){Martinsson}, {Verheijen}, {Bershady},
  {Westfall}, {Andersen}, \& {Swaters}}]{martinsson16}
{Martinsson}, T. P.~K., {Verheijen}, M. A.~W., {Bershady}, M.~A., {et~al.}
  2016, \aap, 585, A99

\bibitem[{{McGaugh}(2012)}]{mcgaugh12}
{McGaugh}, S.~S. 2012, \aj, 143, 40

\bibitem[{{McGaugh} {et~al.}(2000){McGaugh}, {Schombert}, {Bothun}, \& {de
  Blok}}]{mcgaugh00}
{McGaugh}, S.~S., {Schombert}, J.~M., {Bothun}, G.~D., \& {de Blok}, W.~J.~G.
  2000, \apjl, 533, L99

\bibitem[{{McQuinn} {et~al.}(2010){McQuinn}, {Skillman}, {Cannon}, {Dalcanton},
  {Dolphin}, {Hidalgo-Rodr{\'{\i}}guez}, {Holtzman}, {Stark}, {Weisz}, \&
  {Williams}}]{mcquinn10}
{McQuinn}, K.~B.~W., {Skillman}, E.~D., {Cannon}, J.~M., {et~al.} 2010, \apj,
  721, 297

\bibitem[{{Meurer} {et~al.}(1997){Meurer}, {Heckman}, {Lehnert}, {Leitherer},
  \& {Lowenthal}}]{meurer97}
{Meurer}, G.~R., {Heckman}, T.~M., {Lehnert}, M.~D., {Leitherer}, C., \&
  {Lowenthal}, J. 1997, \aj, 114, 54

\bibitem[{{Morrissey} {et~al.}(2007){Morrissey}, {Conrow}, {Barlow}, {Small},
  {Seibert}, {Wyder}, {Budav{\'a}ri}, {Arnouts}, {Friedman}, {Forster},
  {Martin}, {Neff}, {Schiminovich}, {Bianchi}, {Donas}, {Heckman}, {Lee},
  {Madore}, {Milliard}, {Rich}, {Szalay}, {Welsh}, \& {Yi}}]{Morrissey2007}
{Morrissey}, P., {Conrow}, T., {Barlow}, T.~A., {et~al.} 2007, \apjs, 173, 682

\bibitem[{{Moster} {et~al.}(2013){Moster}, {Naab}, \& {White}}]{moster13}
{Moster}, B.~P., {Naab}, T., \& {White}, S.~D.~M. 2013, \mnras, 428, 3121

\bibitem[{{Nelson} {et~al.}(2018){Nelson}, {Pillepich}, {Springel},
  {Weinberger}, {Hernquist}, {Pakmor}, {Genel}, {Torrey}, {Vogelsberger},
  {Kauffmann}, {Marinacci}, \& {Naiman}}]{nelson18}
{Nelson}, D., {Pillepich}, A., {Springel}, V., {et~al.} 2018, \mnras, 475, 624

\bibitem[{{Ostriker} {et~al.}(2010){Ostriker}, {McKee}, \&
  {Leroy}}]{ostriker10}
{Ostriker}, E.~C., {McKee}, C.~F., \& {Leroy}, A.~K. 2010, \apj, 721, 975

\bibitem[{{Ostriker} \& {Shetty}(2011)}]{ostriker11}
{Ostriker}, E.~C. \& {Shetty}, R. 2011, \apj, 731, 41

\bibitem[{{Pettini} \& {Pagel}(2004)}]{Pettini2004}
{Pettini}, M. \& {Pagel}, B.~E.~J. 2004, \mnras, 348, L59

\bibitem[{{Rafieferantsoa} {et~al.}(2018){Rafieferantsoa}, {Andrianomena}, \&
  {Dav{\'e}}}]{rafier18}
{Rafieferantsoa}, M., {Andrianomena}, S., \& {Dav{\'e}}, R. 2018, \mnras, 479,
  4509

\bibitem[{{Rampazzo} {et~al.}(2007){Rampazzo}, {Marino}, {Tantalo}, {Bettoni},
  {Buson}, {Chiosi}, {Galletta}, {Gr{\"u}tzbauch}, \& {Rich}}]{rampazzo07}
{Rampazzo}, R., {Marino}, A., {Tantalo}, R., {et~al.} 2007, \mnras, 381, 245

\bibitem[{{Saintonge} {et~al.}(2016){Saintonge}, {Catinella}, {Cortese},
  {Genzel}, {Giovanelli}, {Haynes}, {Janowiecki}, {Kramer}, {Lutz},
  {Schiminovich}, {Tacconi}, {Wuyts}, \& {Accurso}}]{Saintonge2016}
{Saintonge}, A., {Catinella}, B., {Cortese}, L., {et~al.} 2016, \mnras, 462,
  1749

\bibitem[{{Saintonge} {et~al.}(2017){Saintonge}, {Catinella}, {Tacconi},
  {Kauffmann}, {Genzel}, {Cortese}, {Dav{\'e}}, {Fletcher},
  {Graci{\'a}-Carpio}, {Kramer}, {Heckman}, {Janowiecki}, {Lutz}, {Rosario},
  {Schiminovich}, {Schuster}, {Wang}, {Wuyts}, {Borthakur}, {Lamperti}, \&
  {Roberts-Borsani}}]{Saintonge2017}
{Saintonge}, A., {Catinella}, B., {Tacconi}, L.~J., {et~al.} 2017, The
  Astrophysical Journal Supplement Series, 233, 22

\bibitem[{{Saintonge} {et~al.}(2011a){Saintonge}, {Kauffmann}, {Kramer},
  {Tacconi}, {Buchbender}, {Catinella}, {Fabello}, {Graci{\'a}-Carpio}, {Wang},
  {Cortese}, {Fu}, {Genzel}, {Giovanelli}, {Guo}, {Haynes}, {Heckman},
  {Krumholz}, {Lemonias}, {Li}, {Moran}, {Rodriguez-Fernandez}, {Schiminovich},
  {Schuster}, \& {Sievers}}]{Saintonge2011a}
{Saintonge}, A., {Kauffmann}, G., {Kramer}, C., {et~al.} 2011a, \mnras, 415, 32

\bibitem[{{Saintonge} {et~al.}(2011b){Saintonge}, {Kauffmann}, {Wang},
  {Kramer}, {Tacconi}, {Buchbender}, {Catinella}, {Graci{\'a}-Carpio},
  {Cortese}, {Fabello}, {Fu}, {Genzel}, {Giovanelli}, {Guo}, {Haynes},
  {Heckman}, {Krumholz}, {Lemonias}, {Li}, {Moran}, {Rodriguez-Fernand ez},
  {Schiminovich}, {Schuster}, \& {Sievers}}]{Saintonge2011b}
{Saintonge}, A., {Kauffmann}, G., {Wang}, J., {et~al.} 2011b, \mnras, 415, 61

\bibitem[{{Salim}(2014)}]{salim14alone}
{Salim}, S. 2014, Serbian Astronomical Journal, 189, 1

\bibitem[{{Salim} {et~al.}(2018){Salim}, {Boquien}, \& {Lee}}]{Salim2018}
{Salim}, S., {Boquien}, M., \& {Lee}, J.~C. 2018, \apj, 859, 11

\bibitem[{{Salim} {et~al.}(2016){Salim}, {Lee}, {Janowiecki}, {da Cunha},
  {Dickinson}, {Boquien}, {Burgarella}, {Salzer}, \& {Charlot}}]{Salim2016}
{Salim}, S., {Lee}, J.~C., {Janowiecki}, S., {et~al.} 2016, \apjs, 227, 2

\bibitem[{{Salim} {et~al.}(2014){Salim}, {Lee}, {Ly}, {Brinchmann}, {Dav{\'e}},
  {Dickinson}, {Salzer}, \& {Charlot}}]{salim14}
{Salim}, S., {Lee}, J.~C., {Ly}, C., {et~al.} 2014, \apj, 797, 126

\bibitem[{{Sandstrom} {et~al.}(2013){Sandstrom}, {Leroy}, {Walter}, {Bolatto},
  {Croxall}, {Draine}, {Wilson}, {Wolfire}, {Calzetti}, {Kennicutt}, {Aniano},
  {Donovan Meyer}, {Usero}, {Bigiel}, {Brinks}, {de Blok}, {Crocker}, {Dale},
  {Engelbracht}, {Galametz}, {Groves}, {Hunt}, {Koda}, {Kreckel}, {Linz},
  {Meidt}, {Pellegrini}, {Rix}, {Roussel}, {Schinnerer}, {Schruba}, {Schuster},
  {Skibba}, {van der Laan}, {Appleton}, {Armus}, {Brandl}, {Gordon}, {Hinz},
  {Krause}, {Montiel}, {Sauvage}, {Schmiedeke}, {Smith}, \&
  {Vigroux}}]{sandstrom13}
{Sandstrom}, K.~M., {Leroy}, A.~K., {Walter}, F., {et~al.} 2013, \apj, 777, 5

\bibitem[{{Schawinski} {et~al.}(2014){Schawinski}, {Urry}, {Simmons},
  {Fortson}, {Kaviraj}, {Keel}, {Lintott}, {Masters}, {Nichol}, {Sarzi},
  {Skibba}, {Treister}, {Willett}, {Wong}, \& {Yi}}]{schawinski14}
{Schawinski}, K., {Urry}, C.~M., {Simmons}, B.~D., {et~al.} 2014, \mnras, 440,
  889

\bibitem[{{Schechter}(1976)}]{schechter76}
{Schechter}, P. 1976, \apj, 203, 297

\bibitem[{{Schiminovich} {et~al.}(2010){Schiminovich}, {Catinella},
  {Kauffmann}, {Fabello}, {Wang}, {Hummels}, {Lemonias}, {Moran}, {Wu},
  {Giovanelli}, {Haynes}, {Heckman}, {Basu-Zych}, {Blanton}, {Brinchmann},
  {Budav{\'a}ri}, {Gon{\c{c}}alves}, {Johnson}, {Kennicutt}, {Madore},
  {Martin}, {Rich}, {Tacconi}, {Thilker}, {Wild}, \& {Wyder}}]{schiminovich10}
{Schiminovich}, D., {Catinella}, B., {Kauffmann}, G., {et~al.} 2010, \mnras,
  408, 919

\bibitem[{{Schlafly} \& {Finkbeiner}(2011)}]{schlafly11}
{Schlafly}, E.~F. \& {Finkbeiner}, D.~P. 2011, \apj, 737, 103

\bibitem[{{Schruba} {et~al.}(2012){Schruba}, {Leroy}, {Walter}, {Bigiel},
  {Brinks}, {de Blok}, {Kramer}, {Rosolowsky}, {Sandstrom}, {Schuster},
  {Usero}, {Weiss}, \& {Wiesemeyer}}]{schruba12}
{Schruba}, A., {Leroy}, A.~K., {Walter}, F., {et~al.} 2012, \aj, 143, 138

\bibitem[{{Stark} {et~al.}(2013){Stark}, {Kannappan}, {Wei}, {Baker}, {Leroy},
  {Eckert}, \& {Vogel}}]{Stark2013}
{Stark}, D.~V., {Kannappan}, S.~J., {Wei}, L.~H., {et~al.} 2013, \apj, 769, 82

\bibitem[{{Swaters} {et~al.}(2002){Swaters}, {van Albada}, {van der Hulst}, \&
  {Sancisi}}]{swaters02}
{Swaters}, R.~A., {van Albada}, T.~S., {van der Hulst}, J.~M., \& {Sancisi}, R.
  2002, \aap, 390, 829

\bibitem[{{Tacchella} {et~al.}(2016){Tacchella}, {Dekel}, {Carollo},
  {Ceverino}, {DeGraf}, {Lapiner}, {Mand elker}, \& {Primack
  Joel}}]{tacchella16}
{Tacchella}, S., {Dekel}, A., {Carollo}, C.~M., {et~al.} 2016, \mnras, 457,
  2790

\bibitem[{{Teimoorinia} {et~al.}(2017){Teimoorinia}, {Ellison}, \&
  {Patton}}]{tei17}
{Teimoorinia}, H., {Ellison}, S.~L., \& {Patton}, D.~R. 2017, \mnras, 464, 3796

\bibitem[{{Tolstoy} {et~al.}(2009){Tolstoy}, {Hill}, \& {Tosi}}]{tolstoy09}
{Tolstoy}, E., {Hill}, V., \& {Tosi}, M. 2009, \araa, 47, 371

\bibitem[{{Tortora} {et~al.}(2010){Tortora}, {Napolitano}, {Cardone},
  {Capaccioli}, {Jetzer}, \& {Molinaro}}]{tortora10}
{Tortora}, C., {Napolitano}, N.~R., {Cardone}, V.~F., {et~al.} 2010, \mnras,
  407, 144

\bibitem[{{Tortora} {et~al.}(2019){Tortora}, {Posti}, {Koopmans}, \&
  {Napolitano}}]{tortora19}
{Tortora}, C., {Posti}, L., {Koopmans}, L.~V.~E., \& {Napolitano}, N.~R. 2019,
  \mnras, 489, 5483

\bibitem[{{Tremonti} {et~al.}(2004){Tremonti}, {Heckman}, {Kauffmann},
  {Brinchmann}, {Charlot}, {White}, {Seibert}, {Peng}, {Schlegel}, {Uomoto},
  {Fukugita}, \& {Brinkmann}}]{tremonti04}
{Tremonti}, C.~A., {Heckman}, T.~M., {Kauffmann}, G., {et~al.} 2004, \apj, 613,
  898

\bibitem[{{Tumlinson} {et~al.}(2017){Tumlinson}, {Peeples}, \&
  {Werk}}]{tumlinson17}
{Tumlinson}, J., {Peeples}, M.~S., \& {Werk}, J.~K. 2017, \araa, 55, 389

\bibitem[{{U} {et~al.}(2012){U}, {Sanders}, {Mazzarella}, {Evans}, {Howell},
  {Surace}, {Armus}, {Iwasawa}, {Kim}, {Casey}, {Vavilkin}, {Dufault},
  {Larson}, {Barnes}, {Chan}, {Frayer}, {Haan}, {Inami}, {Ishida},
  {Kartaltepe}, {Melbourne}, \& {Petric}}]{u13}
{U}, V., {Sanders}, D.~B., {Mazzarella}, J.~M., {et~al.} 2012, \apjs, 203, 9

\bibitem[{{van Dishoeck} \& {Black}(1988)}]{vandishoeck88}
{van Dishoeck}, E.~F. \& {Black}, J.~H. 1988, \apj, 334, 771

\bibitem[{{van Zee} {et~al.}(1998){van Zee}, {Skillman}, \&
  {Salzer}}]{vanzee98}
{van Zee}, L., {Skillman}, E.~D., \& {Salzer}, J.~J. 1998, \aj, 116, 1186

\bibitem[{{Wang} {et~al.}(2020){Wang}, {Catinella}, {Saintonge}, {Pan},
  {Serra}, \& {Shao}}]{wang20}
{Wang}, J., {Catinella}, B., {Saintonge}, A., {et~al.} 2020, \apj, 890, 63

\bibitem[{{Wang} {et~al.}(2014){Wang}, {Fu}, {Aumer}, {Kauffmann}, {J{\'o}zsa},
  {Serra}, {Huang}, {Brinchmann}, {van der Hulst}, \& {Bigiel}}]{wang14}
{Wang}, J., {Fu}, J., {Aumer}, M., {et~al.} 2014, \mnras, 441, 2159

\bibitem[{{Wang} {et~al.}(2013){Wang}, {Kauffmann}, {J{\'o}zsa}, {Serra}, {van
  der Hulst}, {Bigiel}, {Brinchmann}, {Verheijen}, {Oosterloo}, {Wang}, {Li},
  {den Heijer}, \& {Kerp}}]{wang13}
{Wang}, J., {Kauffmann}, G., {J{\'o}zsa}, G. I.~G., {et~al.} 2013, \mnras, 433,
  270

\bibitem[{{Wang} {et~al.}(2016){Wang}, {Koribalski}, {Serra}, {van der Hulst},
  {Roychowdhury}, {Kamphuis}, \& {Chengalur}}]{wang16}
{Wang}, J., {Koribalski}, B.~S., {Serra}, P., {et~al.} 2016, \mnras, 460, 2143

\bibitem[{{Wei} {et~al.}(2010){Wei}, {Kannappan}, {Vogel}, \&
  {Baker}}]{Wei2010}
{Wei}, L.~H., {Kannappan}, S.~J., {Vogel}, S.~N., \& {Baker}, A.~J. 2010, \apj,
  708, 841

\bibitem[{{Weisz} {et~al.}(2011){Weisz}, {Dalcanton}, {Williams}, {Gilbert},
  {Skillman}, {Seth}, {Dolphin}, {McQuinn}, {Gogarten}, {Holtzman}, {Rosema},
  {Cole}, {Karachentsev}, \& {Zaritsky}}]{weisz11}
{Weisz}, D.~R., {Dalcanton}, J.~J., {Williams}, B.~F., {et~al.} 2011, \apj,
  739, 5

\bibitem[{{Whitaker} {et~al.}(2014){Whitaker}, {Franx}, {Leja}, {van Dokkum},
  {Henry}, {Skelton}, {Fumagalli}, {Momcheva}, {Brammer}, {Labb{\'e}},
  {Nelson}, \& {Rigby}}]{whitaker14}
{Whitaker}, K.~E., {Franx}, M., {Leja}, J., {et~al.} 2014, \apj, 795, 104

\bibitem[{{Whitaker} {et~al.}(2012){Whitaker}, {van Dokkum}, {Brammer}, \&
  {Franx}}]{whitaker12}
{Whitaker}, K.~E., {van Dokkum}, P.~G., {Brammer}, G., \& {Franx}, M. 2012,
  \apjl, 754, L29

\bibitem[{{Wilson}(1995)}]{wilson95}
{Wilson}, C.~D. 1995, \apjl, 448, L97

\bibitem[{{Wolfire} {et~al.}(2010){Wolfire}, {Hollenbach}, \&
  {McKee}}]{wolfire10}
{Wolfire}, M.~G., {Hollenbach}, D., \& {McKee}, C.~F. 2010, \apj, 716, 1191

\bibitem[{{Wolfire} {et~al.}(1995){Wolfire}, {Hollenbach}, {McKee}, {Tielens},
  \& {Bakes}}]{wolfire95}
{Wolfire}, M.~G., {Hollenbach}, D., {McKee}, C.~F., {Tielens}, A.~G.~G.~M., \&
  {Bakes}, E.~L.~O. 1995, \apj, 443, 152

\bibitem[{{Wright} {et~al.}(2010){Wright}, {Eisenhardt}, {Mainzer}, {Ressler},
  {Cutri}, {Jarrett}, {Kirkpatrick}, {Padgett}, {McMillan}, {Skrutskie},
  {Stanford}, {Cohen}, {Walker}, {Mather}, {Leisawitz}, {Gautier}, {McLean},
  {Benford}, {Lonsdale}, {Blain}, {Mendez}, {Irace}, {Duval}, {Liu}, {Royer},
  {Heinrichsen}, {Howard}, {Shannon}, {Kendall}, {Walsh}, {Larsen}, {Cardon},
  {Schick}, {Schwalm}, {Abid}, {Fabinsky}, {Naes}, \& {Tsai}}]{Wright2010}
{Wright}, E.~L., {Eisenhardt}, P. R.~M., {Mainzer}, A.~K., {et~al.} 2010, \aj,
  140, 1868

\bibitem[{{Wyder} {et~al.}(2007){Wyder}, {Martin}, {Schiminovich}, {Seibert},
  {Budav{\'a}ri}, {Treyer}, {Barlow}, {Forster}, {Friedman}, {Morrissey},
  {Neff}, {Small}, {Bianchi}, {Donas}, {Heckman}, {Lee}, {Madore}, {Milliard},
  {Rich}, {Szalay}, {Welsh}, \& {Yi}}]{wyder07}
{Wyder}, T.~K., {Martin}, D.~C., {Schiminovich}, D., {et~al.} 2007, \apjs, 173,
  293

\bibitem[{{Yesuf} \& {Ho}(2019)}]{yesuf19}
{Yesuf}, H.~M. \& {Ho}, L.~C. 2019, \apj, 884, 177

\bibitem[{{Y{\i}ld{\i}z} {et~al.}(2017){Y{\i}ld{\i}z}, {Serra}, {Peletier},
  {Oosterloo}, \& {Duc}}]{yildiz17}
{Y{\i}ld{\i}z}, M.~K., {Serra}, P., {Peletier}, R.~F., {Oosterloo}, T.~A., \&
  {Duc}, P.-A. 2017, \mnras, 464, 329

\bibitem[{{Zaritsky} {et~al.}(2015){Zaritsky}, {Gil de Paz}, \&
  {Bouquin}}]{zaritsky15}
{Zaritsky}, D., {Gil de Paz}, A., \& {Bouquin}, A. r. Y.~K. 2015, \mnras, 446,
  2030

\bibitem[{{Zhang} {et~al.}(2019){Zhang}, {Peng}, {Ho}, {Maiolino}, {Dekel},
  {Guo}, {Mannucci}, {Li}, {Yuan}, {Renzini}, {Dou}, {Guo}, {Man}, \&
  {Li}}]{zhang19}
{Zhang}, C., {Peng}, Y., {Ho}, L.~C., {et~al.} 2019, \apjl, 884, L52

\bibitem[{{Zhang} {et~al.}(2009){Zhang}, {Li}, {Kauffmann}, {Zou}, {Catinella},
  {Shen}, {Guo}, \& {Chang}}]{zhang09}
{Zhang}, W., {Li}, C., {Kauffmann}, G., {et~al.} 2009, \mnras, 397, 1243

\bibitem[{{Zu}(2020)}]{zu20}
{Zu}, Y. 2020, \mnras

\end{thebibliography}

\clearpage

\appendix

\section{The location of atomic gas in MAGMA galaxies \label{app:hiradius}}
\hi\ gas is known to be generally extended beyond the galaxy stellar
disk \citep[e.g.,][]{swaters02,wang13,martinsson16}, an effect that is more pronounced in low-mass
dwarf galaxies \citep[e.g.,][]{vanzee98,begum05}. 
Here our goal is to investigate the fraction of \hi\
gas that lies outside the stellar confines of the galaxy, and assess the dependence
of this fraction on \mstar.

\begin{figure}[!ht]
\includegraphics[width=0.5\textwidth]{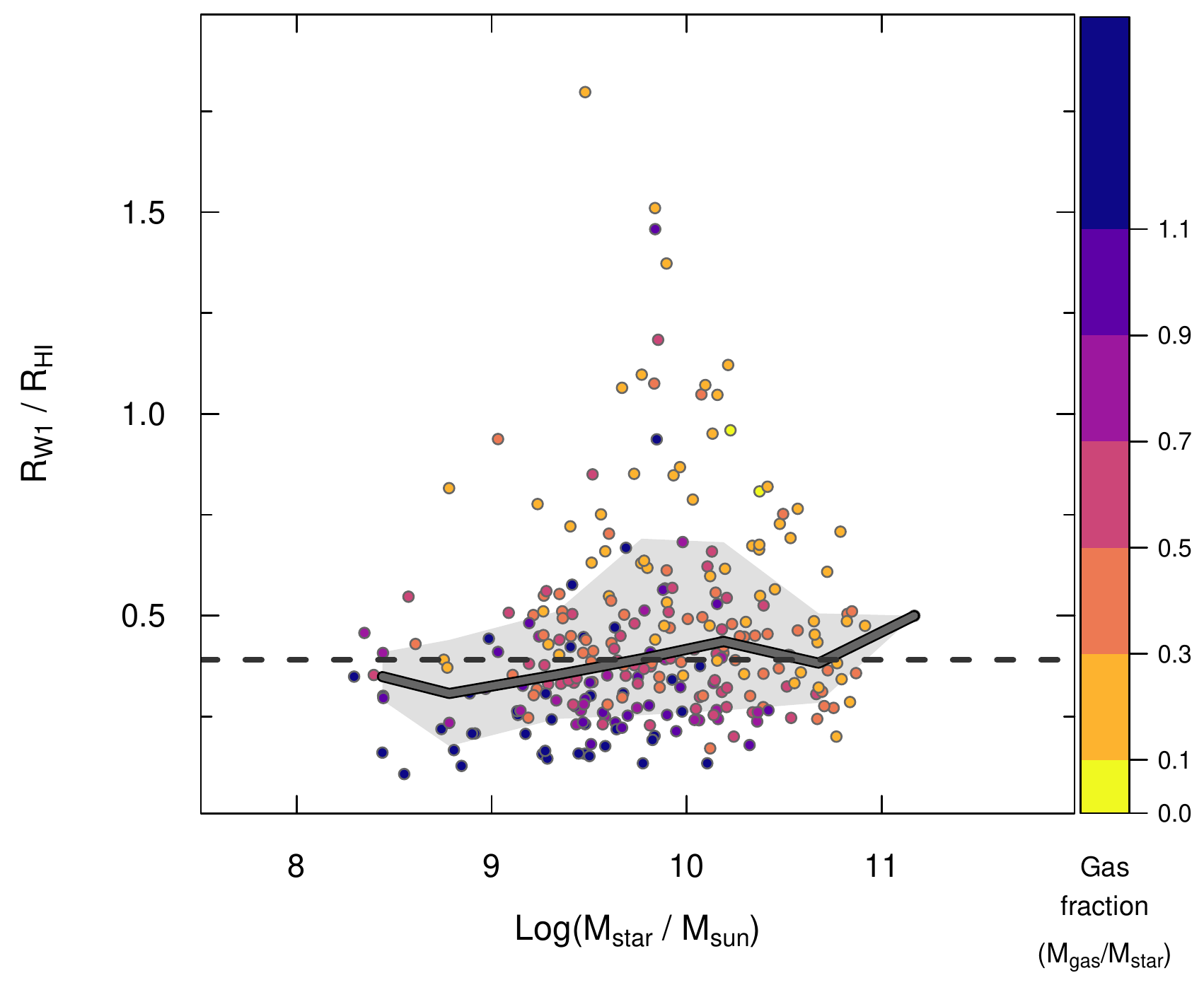}
\includegraphics[width=0.5\textwidth]{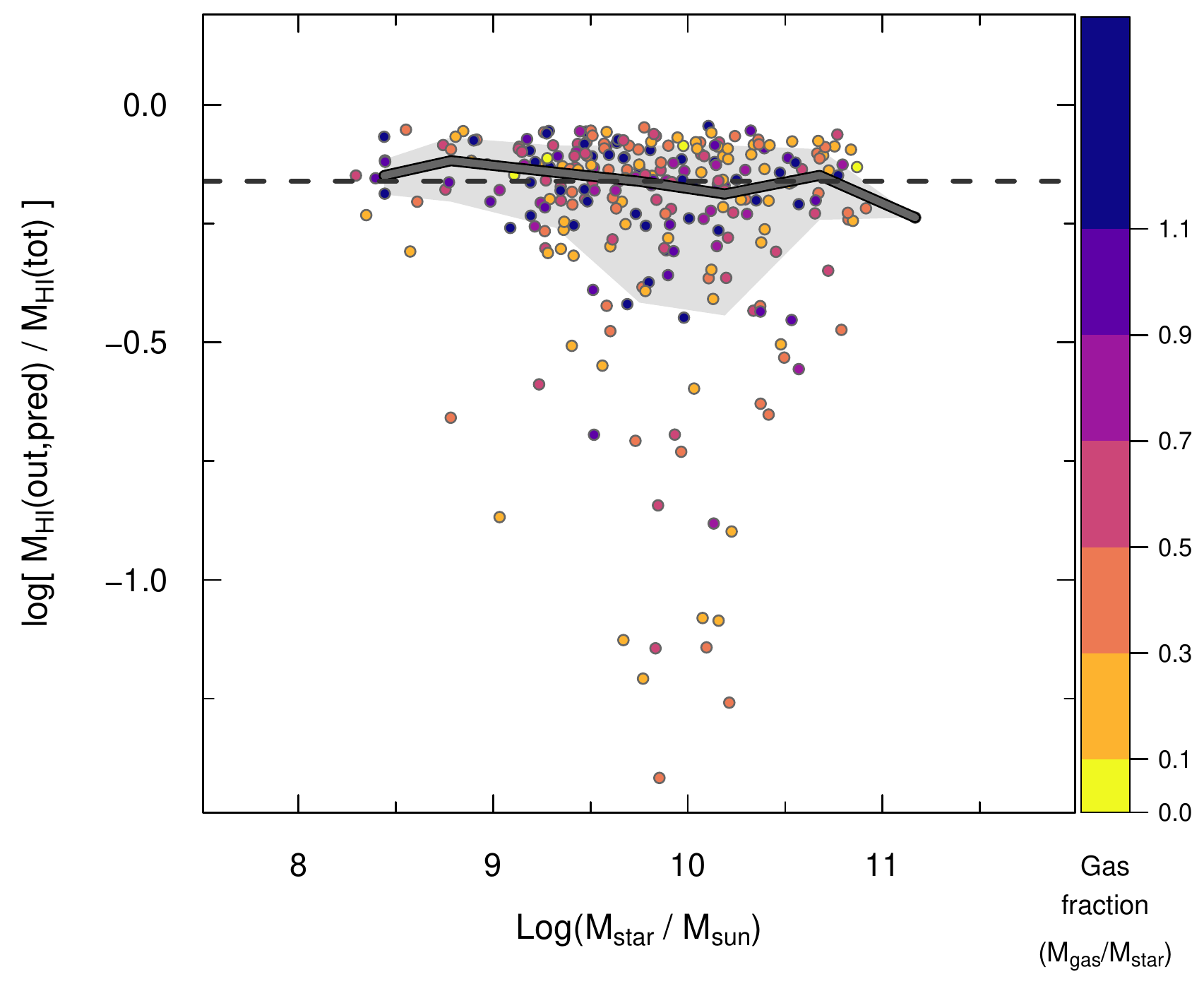}
\caption{Upper panel: ratio of \rwise\ and \rhi\ plotted against (log) \mstar.
These data are available for 268 galaxies.
Lower panel: fraction of \hi\ mass outside the galaxy stellar disk, as defined by \rwise.
In both panels, data points are coded by gas fraction as in previous figures.
\label{fig:hiradii}
}
\end{figure}

Our sample relies only on global \hi\ measurements to estimate \hi\ mass,
so nothing is known about the spatial distribution of the atomic gas.
However, \hi\ properties in nearby galaxies are governed by two interesting relations.
The first is the form of the radial profiles in star-forming disks; 
when normalized to a characteristic \hi\ radius, \rhi, typically defined
by the semi-major axis of a 1 \msunpc\ isophote, 
the radial distributions of \hi\ are self-similar in the outer regions
\citep[e.g.,][]{swaters02,wang14,wang16}.
The second is a partial consequence of the first, namely \hi\ mass \mhi\
and \rhi\ are tightly related to within $\sim$0.06\,dex
\citep[e.g.,][]{broeils97,swaters02,martinsson16,wang16}.
These two properties of \hi\ content can be combined to estimate the amount of
\hi\ within \rhi\ and beyond it. 

\begin{figure}[!ht]
\begin{center}
\includegraphics[width=0.40\textwidth]{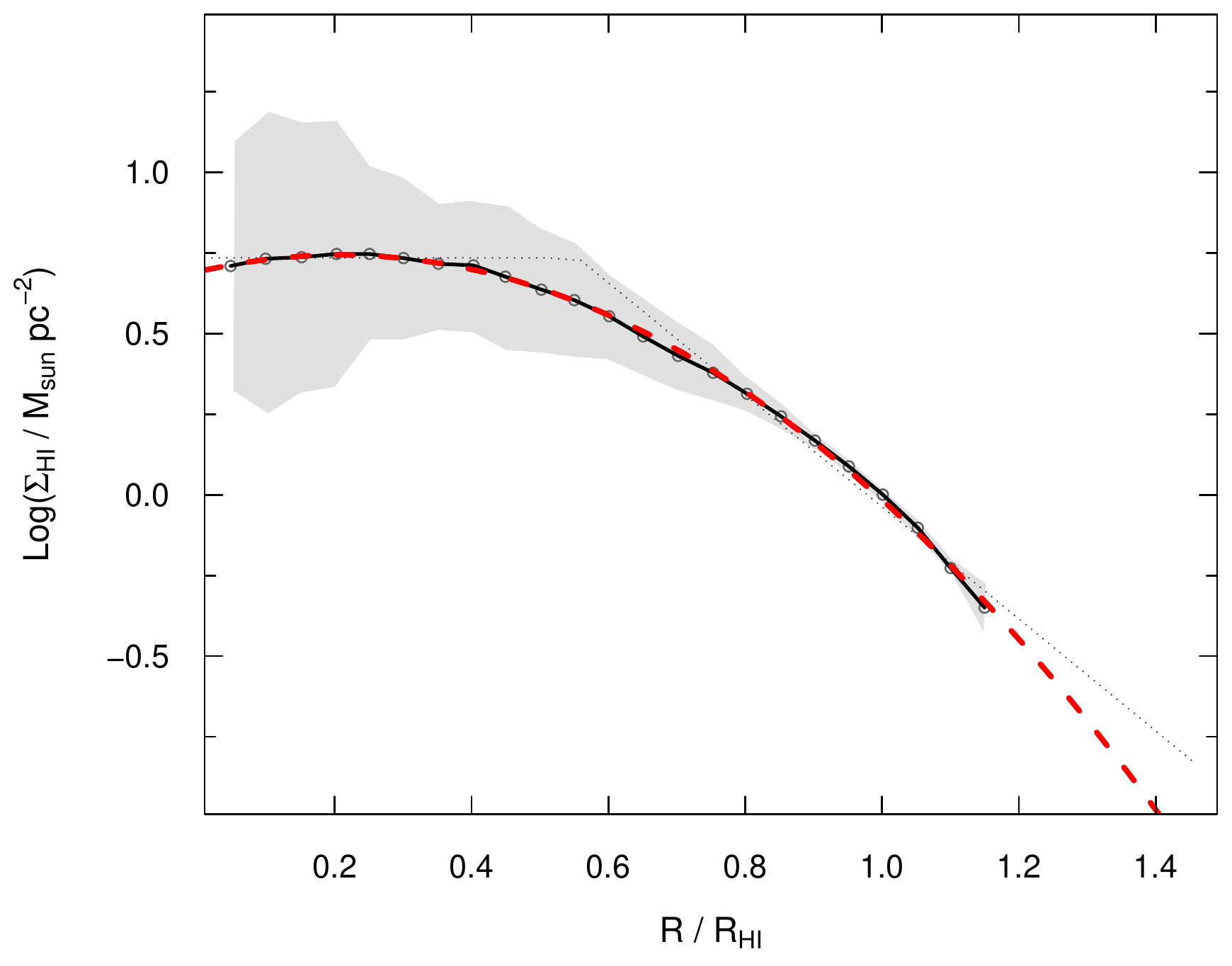}
\caption{``Standard'' self-similar \hi\ surface brightness profile taken
from \citet{wang20,wang14,wang16}. The solid (black) curve connects the
data points, while the gray shaded region shows the uncertainties as presented
by \citet{wang20}.
The dashed (red) curve shows the quadratic fit, and the dotted line corresponds
to a broken power-law that is however not a good approximation to the data.
\label{fig:histdprofile}
}
\end{center}
\end{figure}

Following the approach of \citet{wang20}, we have estimated the
size of the stellar disk using the semi-major axes measured by WISE, 
\rwise,
in the same photometric band, W1, used to infer stellar masses (see 
Sect. \ref{sec:magma}, \pone).
These are available for 268 of the 392 MAGMA galaxies, and 
have been converted from the apparent sky radii to kpc.
We then estimated \rhi\ according to \citet{wang16} \citep[see also][]{broeils97}.
The upper panel of Fig. \ref{fig:hiradii} shows 
\rwise/\rhi, where the data points are colored by gas fractions as in previous figures.
These ratios are typically less than unity with a median \rwise/\rhi\,=\,0.39,
as shown by the horizontal dashed line.
Figure \ref{fig:hiradii} also shows the running median for the MAGMA sample binned in \mstar.
There is a weak trend for most massive galaxies to have \rwise/\rhi\ larger than the median,
by $\la$0.1\,dex.

Figure \ref{fig:histdprofile} shows the ``standard'' self-similar profile
\hi\ profile given by \citet{wang20}, and shown in previous papers \citep{wang14,wang16}.
We fit with a quadratic polynomial (in log-lin space), and found that is gives 
a very good approximation (to within 0.005\,dex along the y axis).
Thus, after scaling to \rhi,
it is possible to simply integrate analytically to obtain
the \hi\ mass outside \rwise;
dividing by \mhi\ (global) gives the fraction
of the total \hi\ mass outside the stellar disk.
We have used the same limits of integration as \citet{wang20}, namely
\rwise/\rhi\ and 1.5, in order not to exceed the radial region over which the
profile has been computed.
This integration has been performed only for those galaxies for which \rwise/\rhi\,$\leq$1.2,
again following \citet{wang20}.
The results are shown in the lower panel of Fig. \ref{fig:hiradii}.
There is virtually no trend with \mstar\
and the median fraction of \hi\ outside the optical/stellar disk is 0.69,
as shown by a horizontal dashed line.
Interestingly, despite the different estimates of stellar disk dimensions,
these two conclusions are entirely consistent with \citet{wang20},
who found no trend with \mstar\ and a median fraction of \hi\ inside the stellar disk of 
$\sim$\,dex($-0.5$), or $\sim$0.32.
Thus, we conclude that there is a significant fraction of \hi\ gas outside the stellar
disk in MAGMA galaxies, but this fraction does not change systematically with \mstar.

\bigskip
\section{Gas fractions and colors \label{app:colors} }

\citet{Kannappan2013} found that long-term fractional stellar mass growth rate (FSMGRLT) correlates
quite well with $U-J$ \citep[see also][]{kannappan04}.
They formulated FSMGRLT as the ratio of the mass of stars formed over the last Gyr relative
to the pre-existing stellar mass, employing detailed optical-NIR SED fitting to define FSMGRLT.
Here we do not have access to analogous SED fitting for MAGMA galaxies, but exploit the correlation
of FSMGRLT with ultraviolet and NIR colors found by \citet{Kannappan2013}.
We use GALEX NUV and W1 (3.4\,\micron), two of the measurements used in \pone\ to estimate SFR
and \mstar; the underlying assumption is that \nuvw\ is a proxy for FSMGRLT. 

\begin{figure}[!ht]
\includegraphics[width=0.48\textwidth]{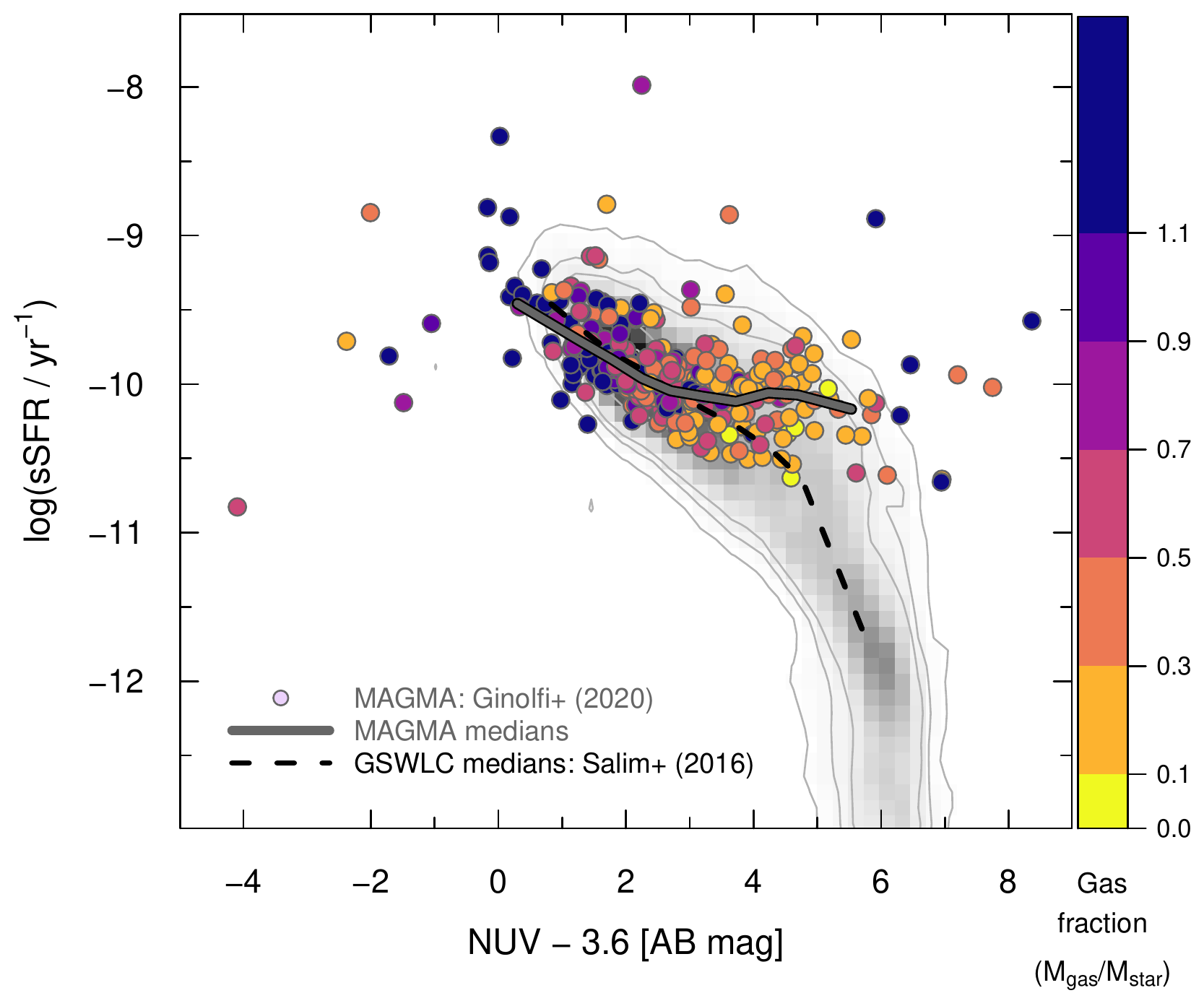}
\caption{sSFR (logarithm) plotted against NUV-W1 (AB) color. 
The gray region and contours correspond to the GSWLC \citep[][]{Salim2016}, with
the GSWLC medians over the same range as for MAGMA given by a black dashed line.
MAGMA medians are shown by a thick gray line with black borders as in previous figures.
As described in the text, the SFRs and stellar masses from the GSWLC are compatible with those for MAGMA.
\label{fig:uv36ssfr}
}
\end{figure}

The two quantities, sSFR and FSMGRLT, do not generally trace the same timescales, as illustrated in
Fig. \ref{fig:uv36ssfr} where sSFR is plotted against \nuvw\ (and by extension FSMGRLT).
Also shown in Fig. \ref{fig:uv36ssfr} as underlying contours and gray scale is the
GSWLC consisting of $\sim 700\,000$ galaxies at $z \la 0.3$ \citep{Salim2016}.
The \mstar\ and SFR scales in the GSWLC were the basis of our formulation for SFR, and
also with the \mstar\ scale as described in \pone.
Thus the comparison of MAGMA and GSWLC is expected to be self-consistent; here we have selected
only GSWLC galaxies in the redshift range $0.015\,\leq\,z\,\leq\,0.06$.
The power-law slope for sSFR with \nuvw\ flattens for MAGMA galaxies at \nuvw$\sim 2$
and log(sSFR)$\sim -10.2$, similar to the 
change in sSFR slope of the GBS and the GRS discussed in Sect. \ref{sec:alphaco} \citep[see also][]{bouquin15}.
The MAGMA trend also follows very closely the GSWLC for these blue colors and relatively
high sSFR.
However, the GSWLC power-law slope steepens at redder colors, \nuvw$\ga 5$;
for red colors and low sSFR there is little or no correlation between \nuvw\ and sSFR.
This slope inflection for GSWLC is similar to the behavior of the NUV$-$r color found by \citet{kaviraj07}
and \citet{salim14alone}, and roughly 
delineates the separation of the ``green valley" from the ``red sequence'' and ``blue cloud'' of SDSS galaxies 
\citep[e.g.,][]{schawinski14}. 

Ultimately, the differences between sSFR and FSMGRLT depend on the timescale of the star-formation tracer
used to define SFR.
By combining UV and mid-infrared (MIR), as advocated by \citet{Leroy2019} and as we have done in \pone\ for MAGMA, the timescales are blurred;
UV is expected to trace the $\ga$ 100--200\,Myr scales, while the MIR traces the warm dust emitted by \hii\ regions
over shorter time scale, 10-20\,Myr \citep[e.g.,][]{calzetti05}.
In Fig. \ref{fig:uv36ssfr}, sSFR for MAGMA (thus SFR over fairly short timescales) 
is higher than the longer-term SFR reflected by \nuvw.
This behavior of MAGMA with a flatter power-law slope of sSFR for red \nuvw, relative to the GSWLC, is related to our requirement
of \hi\ and \htwo\ detections.
As discussed above, at a given high \mstar\ (here traced by red \nuvw),
MAGMA galaxies contain more gas than a typical SFMS galaxy. 
Thus, sSFR is also somewhat higher than for a typical galaxy
in a mass-selected sample with similar mass and color.

\begin{figure*}[!ht]
\includegraphics[width=0.48\textwidth]{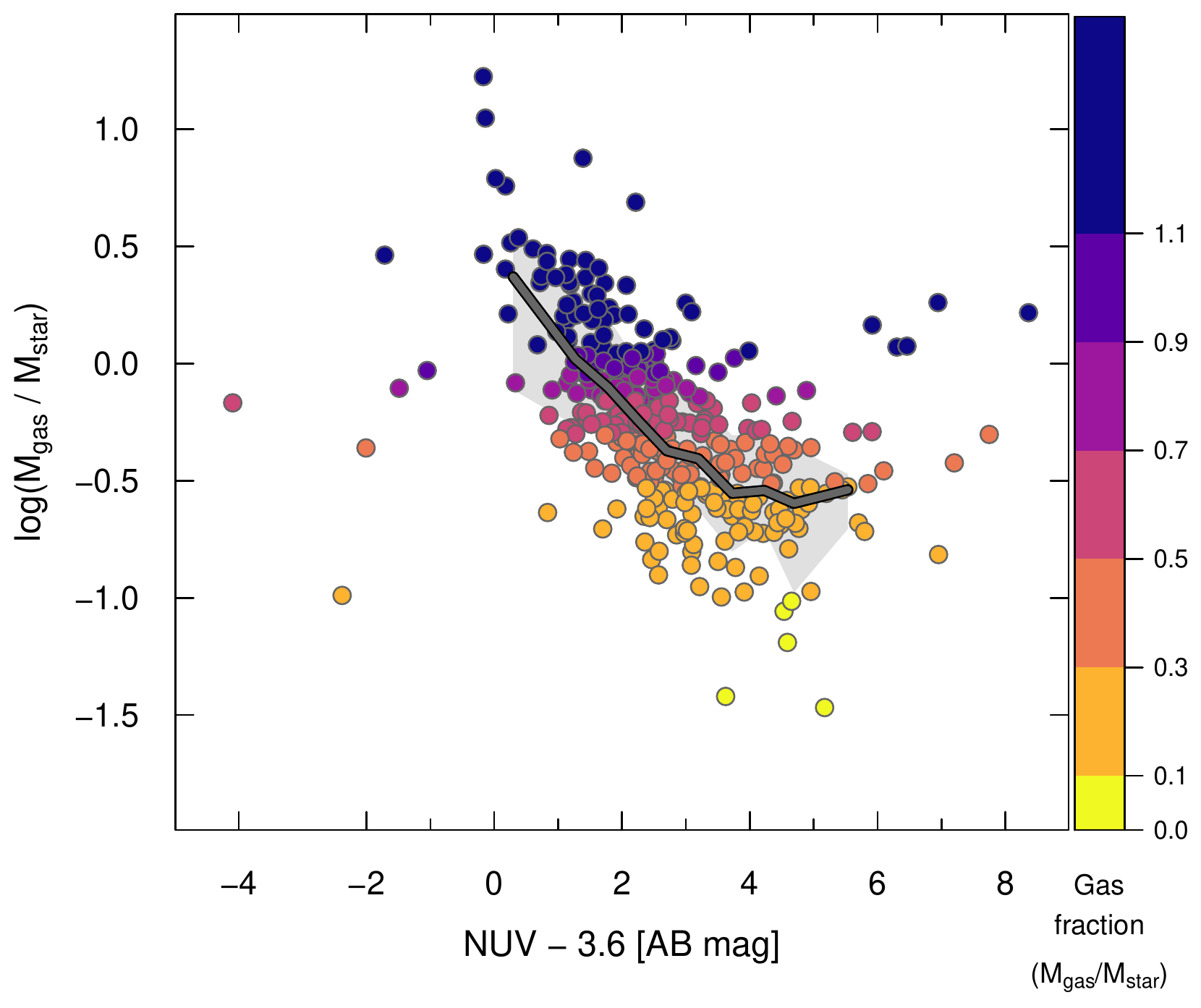}
\hspace{0.02\textwidth}
\includegraphics[width=0.48\textwidth]{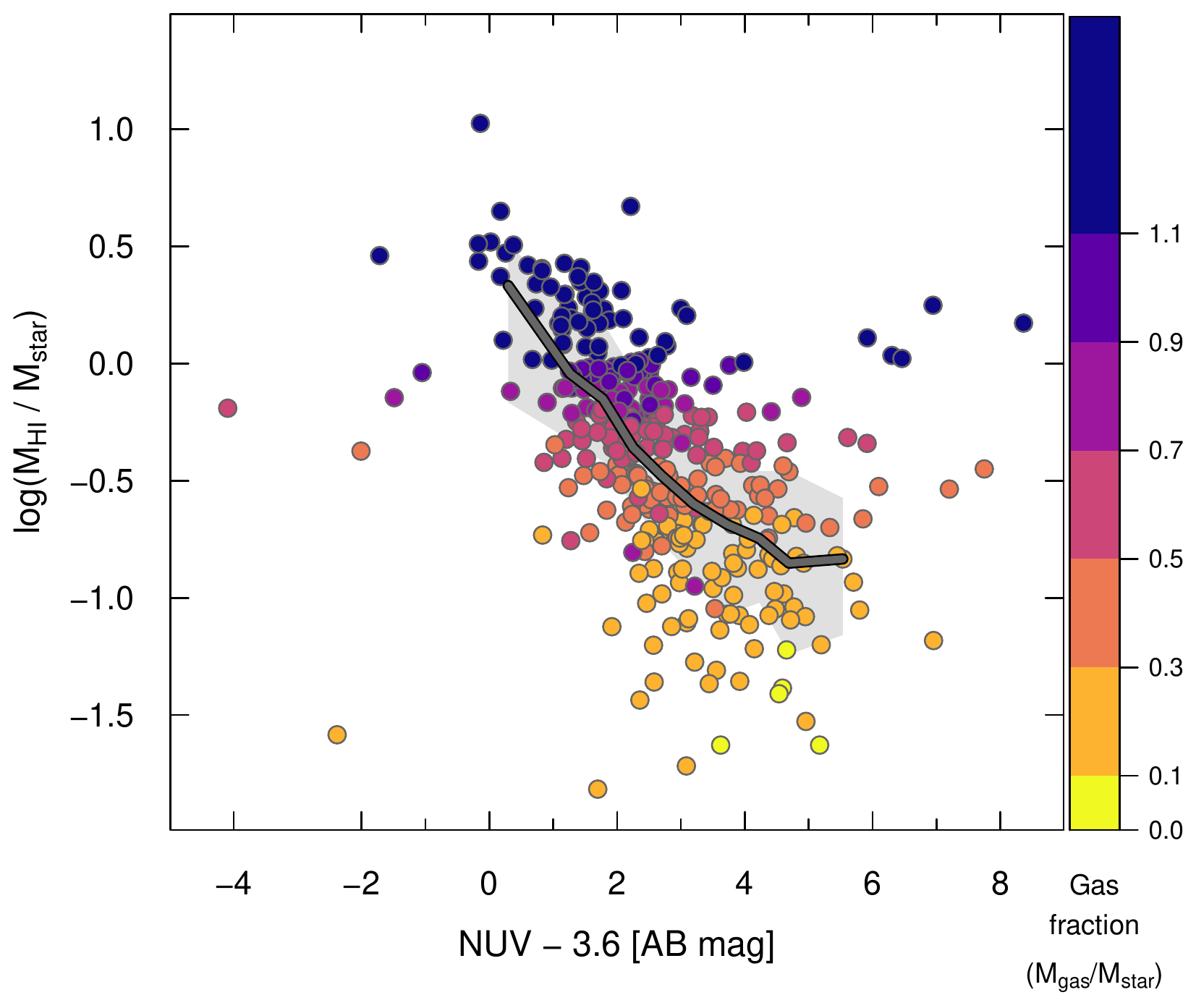}
\caption{Total gas fraction (\mgas/\mstar) plotted against the NUV-W1 (AB) magnitude (left panel) and 
\hi\ fraction (\mhi/\mstar) vs. NUV-W1 (right).
Both magnitudes have been corrected for Galactic extinction as described in Sect. \ref{sec:alphaco}.
The MAGMA medians are shown as a heavy gray line, and the light-gray region
corresponds to $\pm\,1\sigma$ in the distributions.
The contribution of \htwo\ to \mgas\ is evident in 
the upturn in the \mgas\ fraction for red NUV-W1 colors in the left panel, compared to the right.
\label{fig:uv36gas}
}
\end{figure*}

Following \citet{Kannappan2013}, we now compare gas fractions with \nuvw. 
Figure \ref{fig:uv36gas} shows the correlation of total gas fraction with \nuvw\ (\mgas/\mstar, left panel) and 
and \hi\ gas fraction with \nuvw\ (\mhi/\mstar, right).
The correlation of \fgas\ and \nuvw\ shown in the left panel extends only to \nuvw$\sim 4$;
for redder colors \fgas\ is roughly constant over the color range of the MAGMA sample.
This result is in disagreement with the trends found by \citet{Kannappan2013}, since they
found a good correlation for \fgas\ over the entire color range ($U-J$) probed by their sample.
However, they had only a handful of \htwo\ measurements, so equated total gas content with \mhi.
If, instead of \mgas/\mstar, we compare \mhi/\mstar\ with \nuvw\ (right panel of Fig. \ref{fig:uv36gas}), 
the MAGMA trends are in good agreement.
In this case, the correlation is present over roughly the entire range of \nuvw\
probed by MAGMA, with only a slight flattening at the reddest colors, \nuvw$\ga 5$.
The flattening at red colors (high \mstar) is also noticeable in Fig. \ref{fig:h2hibin_sfr}, where \hi\ content
is roughly constant for the most massive galaxies, while \htwo\ tends to increase with SFR.

If we continue with our assumption that \nuvw\ is tracing FSMGRLT, that is to say star
formation over the last $\sim$Gyr, Fig. \ref{fig:uv36gas} illustrates
that the \hi\ content is setting the stage for star formation over these timescales.
As also observed in the right panel of Fig. \ref{fig:h2hibin_sfr},
the galaxies with the lowest masses, highest sSFR, and bluest \nuvw\ also have the highest
\mhi/\mstar. 
Over a range of more than a factor of 10 in sSFR, the \htwo\ fraction \mhtwo/\mstar\
is roughly constant, while the \hi\ fraction is more than a factor of 10 higher than \mhtwo/\mstar\
at high sSFR.

At lower sSFR, higher \mstar, and redder \nuvw,
\hi\ plays a lesser role over long $\sim$Gyr timescales, as shown 
by the flattening of the power-law slope in the right panel of Fig. \ref{fig:uv36gas}.
Instead, for the reddest, most massive, galaxies,
the dominant role is played by \htwo.
This is also seen from the behavior of \mgas/\mstar\ in the left panel of Fig. \ref{fig:uv36gas},
and in Fig. \ref{fig:h2hibin_sfr}.

\end{document}